\documentclass[preprint,letterpaper,prd,nofootinbib,tightenlines,floatfix]{revtex4}
\usepackage{graphicx}
\usepackage{amssymb}
\usepackage{bm}
\usepackage{comment}
\usepackage{color}

\newcommand{\be}{\begin{equation}}
\newcommand{\ee}{\end{equation}}
\newcommand{\bea}{\begin{eqnarray}}
\newcommand{\eea}{\end{eqnarray}}

\def\Tr{\mathop{\rm Tr}}

\usepackage{dcolumn}
\newcolumntype{q}{D{.}{.}{1.2}}
\newcolumntype{d}{D{.}{.}{2.5}}
\newcolumntype{s}{D{.}{.}{4.8}}

\begin{document}

\title{Dark Nuclei II: Nuclear Spectroscopy in Two-Colour QCD}
 
\author{William~Detmold} \affiliation{ Center for Theoretical Physics,
  Massachusetts Institute of Technology, Cambridge, MA 02139, USA}
\author{Matthew McCullough} \affiliation{ Center for Theoretical
  Physics, Massachusetts Institute of Technology, Cambridge, MA 02139,
  USA} \author{Andrew Pochinsky} \affiliation{ Center for Theoretical
  Physics, Massachusetts Institute of Technology, Cambridge, MA 02139,
  USA}

\date{\today}

\begin{abstract}
  We consider two-colour QCD with two flavours of quarks as a possible
  theory of composite dark matter and use lattice field theory methods
  to investigate nuclear spectroscopy in the spin $J=0$ and $J=1$
  multi-baryon sectors.  We find compelling evidence that $J=1$
  systems with baryon number $B=2,3$ (and their mixed meson--baryon
  counterparts) are bound states -- the analogues of nuclei in this
  theory. In addition, we estimate the $\sigma$-terms of the $J=0$ and
  $J=1$ single baryon states which are important for the coupling of
  the theory to scalar currents that may mediate interactions with the
  visible sector.
\end{abstract}

\preprint{MIT-CTP {4555}}

\maketitle

\section{Introduction}

The matter that we are made of and observe in our environment exhibits
remarkably intricate structure.  From the underlying, relatively
simple rules of the Standard Model, enormous complexity emerges at low
energies, first at the level of hadrons where many different meson and
baryon states and resonances are observed. Beyond this, a further
layer of complexity emerges as baryons interact strongly to form
nuclei and hypernuclei in which a vast variety of physical phenomena
manifest. While certain features of the physical world may be rather
specific (for example, the anomalously large scattering lengths of the
nucleon--nucleon interaction), the existence of complex nuclear
structure has recently been shown to persist at unphysical values of
the quark masses through numerical lattice QCD calculations
\cite{Beane:2010hg,Beane:2011iw,Beane:2012vq,Yamazaki:2012hi,Beane:2013kca}. In
this regard, it is interesting in and of itself to consider more
general variations of QCD and other strongly coupled gauge theories
and ask what form of complexity emerges.  In the context of composite
models of dark matter, such investigations also have phenomenological
implications. Indeed, it seems quite reasonable that if the dark matter 
sector is strongly interacting, it could be at least as complex as the
visible sector, but it is important to test such assumptions.

In this paper, we investigate one of the simplest strongly interacting
theories and address the central question of what level of complexity
arises therein. To be specific, we consider SU($N_c=2$) gauge theory
with $N_f=2$ flavours of fermions in the fundamental representation
and focus on fermion masses in the range where the mass of the vector
meson is less than twice the mass of the pseudoscalar meson.  Using
lattice field theory methods developed for QCD, we investigate the
spectrum of multi-baryon\footnote{Because of the symmetries of the
  $N_{c}=2$ theory, mesons and baryons occur in the same multiplets
  and multi-baryon states have degenerate multi-meson and
  multi-meson--multi-baryon partner states as will be discussed
  below.} states that exist in this theory and find that bound-state
analogues of nuclei form which we refer to in this context as {\it
  dark nuclei}. Specifically, we find compelling evidence that the
baryon number $B=2$ and $3$ states with angular momentum and parity
$J^P=1^+$ are stable against breakup into their constituent baryons.
The occurrence of nuclear bindings gives rise to a new energy scale,
$E_B/B$, the binding energy per baryon, in the theory that can differ
substantially from the natural scale of the dark strong interaction,
$\Lambda_{\rm QC_2D}$.  In nature, the ratio $E_B/B M_{p}$ (where $M_p$ is the proton mass)
is ${\cal O}$(0.2 -- 0.7\%), but in QCD with large quark masses, more
significant bindings are seen
\cite{Beane:2010hg,Beane:2011iw,Beane:2012vq,Yamazaki:2012hi,Beane:2013kca},
with $E_B/BM_{p}\sim{\cal O}$(1--3\%). In the two-colour
case considered here, we find that similar ratios are possible.  The
presence of nuclear bindings, and of nuclear reactions, in this theory
engenders a plethora of phenomenological considerations for strongly
interacting dark matter that we explore in a companion article
\cite{Detmold:2014qqa}.

In the context of phenomenology in the dark sector, it is also
interesting to investigate the ways in which such a strongly
interacting dark sector could interact with the Standard Model or with
additional dark sector particles, for example, other weakly coupled 
dark gauge dynamics.  There are many possible types of
interactions to consider and to begin to address this in the
particular model considered here, we calculate the $\sigma$-terms of
the dark baryons that would govern the couplings to scalar
currents. We find the dimensionless couplings for a quark flavour $q$
in a hadron $H$, $f_{q}^{(H)} \equiv \langle H | m_q
\overline{q}q|H\rangle/M_H \sim 0.15$--$0.3$ at the quark masses that
we work.

The structure of this article is as follows. In Section
\ref{sec:lattice}, we introduce the lattice formulation of the theory
and discuss details of the implementation.  Section \ref{sec:single}
discusses the single hadron spectroscopy and the determination of the
scale of the theory through the pseudoscalar decay constant, while
Section \ref{sec:multi} presents our investigations of multi-hadron
spectroscopy and the extractions of nuclear binding energies. Section
\ref{sec:properties} focusses on the types of nuclear properties and
processes that could be extracted using lattice field theory methods,
and in it, we compute the $\sigma$-terms of the single hadron states
approximately using partially quenched methods. Finally in Section
\ref{sec:discussion}, we discuss our results and other recent
investigations as examples of ``nuclear physics'' in a more general
context.
We discuss phenomenological considerations of our results in the context of
strongly interacting dark matter  in a companion article \cite{Detmold:2014qqa}.

\section{The lattice calculation}
\label{sec:lattice}
We consider the strongly interacting SU($N_c=2$) gauge theory with
$N_f=2$ flavours of fermions in the fundamental representation.
Two-colour QCD is perhaps the simplest candidate for an interesting
theory of strongly interacting dark matter and is a natural place for
investigations using lattice field theory methods as calculations are
computationally less demanding than for other possible theories.  In
this context, this theory has been considered recently in
Refs.~\cite{Lewis:2011zb,Buckley:2012ky,Hietanen:2014xca}.  The
pseudo-real nature of representations of SU(2), leads to a
colour-singlet spectrum of mesons (quark--anti-quark states) and
baryons (di-quark states) and a larger global symmetry whereby the
left- and right-handed fundamental matter fields can be embedded in
representations of a SU($2N_{f}$) symmetry. Under this enlarged
symmetry, composite mesons, baryons and anti-baryons occur in the same
multiplets. A further consequence of pseudo-reality, is that the
theory can be studied in the presence of non-zero quark chemical
potential
\cite{Kogut:1991ts,Skullerud:2003yc,Akemann:2006xn,Hands:2011ye,Takahashi:2013mja}.

We follow the formulation of
Refs.~\cite{Lewis:2011zb,Hietanen:2014xca}, and use standard Wilson
gauge and fermion actions with two flavours of mass-degenerate quarks,
$\psi_f(x)$ for $f=u,d$. The Euclidean-space lattice action is
constructed in terms of SU(2)-valued gauge link variables,
$U_{\mu}(x)$, and is given by 
\bea
S_{\rm Wilson}&=&\frac{\beta}{2}\sum_{x}\sum_{\mu,\nu}\left[1-\frac{1}{2}\Re\Tr\left(U_\mu(x)U_\nu(x+\hat\mu)U_\mu^\dagger(x+\hat\nu)U_\nu^\dagger(x)\right)\right]  \nonumber \\
&&-\frac{1}{2}\sum_{f=u,d}\sum_x\sum_\mu\left[\overline\psi_f(x)(1-\gamma_\mu)U_\mu(x)\psi_f(x+\hat\mu)+\overline\psi_f(x+\hat\mu )(1+\gamma_\mu)U^\dagger_\mu(x)\psi_f(x)\right]  \nonumber\\
&&+\sum_{f=u,d}\sum_x(4+m_{0,f})\overline\psi_f(x)\psi_f(x)\,, 
\eea
where $\beta$ is the gauge coupling and $m_{0,f}$ are the bare fermion
masses that we choose to be degenerate, $m_{0,u}=m_{0,d}\equiv
m_{0}$. The calculations presented herein make use of field
configurations that were generated using the hybrid Monte-Carlo (HMC)
algorithm implemented in a modified version of the {\tt chroma}
library for lattice field theory calculations \cite{Edwards:2004sx}.
The calculations were performed using a double precision version of
the codes and fermion inversions were run to a residual of
$10^{-10}$. In addition to the gauge coupling and bare mass, the
individual simulations depend on the geometry of the lattice which is
taken to be $L^3\times T$, with spatial and temporal extents of $L$
and $T$, respectively. In order to understand the parameter dependence
of the theory a large set of choices of $\{\beta,m_{0},L,T\}$ have
been studied. At certain parameter values, a direct comparison with
the results (specifically, plaquette values and pion masses) of
Refs.~\cite{Lewis:2011zb,Hietanen:2014xca} has been made; these works
use different software bases and the agreement that is found provides
a useful validation of the simulations.

For the primary studies presented in this work, we investigate the
theory in parameter ranges where it is computationally feasible (as a
model for dark matter, there is no strong preference for particular
values of the fermion masses).  While the regime of very light quark
masses compared to the scale of the theory is interesting
\cite{Buckley:2012ky}, it is not viable to perform quantitative
studies at this point without using large scale computational
resources that are of similar magnitude to those used in $N_c=3$
QCD phenomenology. We focus on somewhat heavier masses that are also of
phenomenological interest and aim separately to explore the $\beta$
and $m_0$ dependence for a range of different spatial and temporal
extents. The lattice spacing and single hadron spectroscopy are
primarily determined (up to exponentially small corrections) by
$\beta$ and $m_0$ provided that the lattice volume is sufficiently
large, and the correspondence between lattice parameters and physical
parameters can be made in the single hadron sector alone. Once this
has been accomplished, multi-body spectroscopy requires careful
analysis of volume dependence and hence is more computationally
expensive. The parameters of the primary simulations are shown in
Table \ref{paramtab}.

\begin{table}
  \begin{center}
    \begin{ruledtabular}
      \begin{tabular}{ccccr}
        Label & $\beta$ & $m_0$ & $L^3\times T$ & $N_{\rm traj}$   \\
        \hline
        $A$ & 1.8 & $-1.0890$ & $12^3\times72$ & 5,000 \\
        &  & & $16^3\times72$ & 4,120 \\
        &  & & $20^3\times72$ & 3,250 \\
        \hline
        $B$ & 2.0 & $-0.9490$ & $12^3\times48$ & 10,000 \\
        &  & & $16^3\times48$ & 4,000 \\
        &  & & $20^3\times48$ & 3,840 \\
        &  & & $24^3\times48$ & 2,930 \\
        \hline
        $C$ & 2.0 & $-0.9200$ & $12^3\times48$ & 10,000 \\
        &  & & $16^3\times48$ & 9,780 \\
        &  & & $20^3\times48$ & 10,000 \\
        \hline
        $D$ & 2.0 & $-0.8500$ & $12^3\times48$ & 9,990 \\
        &  & & $16^3\times48$ & 5,040 \\
        &  & & $16^3\times72$ & 5,000 \\
        &  & & $20^3\times48$ & 5,000 \\
        &  & & $24^3\times48$ & 5,050\\
        \hline
        $E$ & 2.1 & $-0.7700$ & $12^3\times72$ & 5,000 \\
        &  & & $16^3\times72$ & 5,000 \\
        &  & & $20^3\times72$ & 4,300 \\
        \hline
        $F$ & 2.2 & $-0.6000$ & $12^3\times72$ & 5,000 \\
        &  & & $16^3\times72$ & 5,000 \\
        &  & & $20^3\times72$ & 5,000 \\
        &  & & $24^3\times72$ & 5,070 
      \end{tabular}
    \end{ruledtabular}
    \caption{ The parameters of the main ensembles used in this work.}
    \label{paramtab}
  \end{center}
\end{table}     

For each ensemble, we run the Monte-Carlo evolution for a large number
of trajectories as shown in Table \ref{paramtab}.  After allowing
${\cal O}(400)$ trajectories for thermalisation, we use every 10th
trajectory for measurements.

\section{Single hadron spectroscopy and properties}
\label{sec:single}

On each configuration we generate 8 smeared sources (after 10
repetitions of stout smearing of the gauge links
\cite{Morningstar:2003gk} with smearing factor 0.08, we perform 80
iterations of APE \cite{Albanese:1987ds} smearing of width $4.0$),
equally separated in time, but randomly placed in space. For each
source, we solve the Dirac equation using either the conjugate
gradient (CG) or stabilised biCG algorithms, demanding convergence to
a residual of $10^{-10}$ in the resulting quark propagator (we have
checked on a subset of measurements that solving to machine precision
does not change our results). On a subset of ensembles, we also
generate propagators from point sources at the same locations in order
to enable the extraction of the pseudoscalar decay constant.

In order to study the single hadron spectrum, we use the propagators
computed on each ensemble of gauge configurations to measure
correlation functions with the quantum numbers of the various states
we are interested in. Because of the relations between meson
($\overline{\psi}\psi$) and baryon ($\psi\psi$) systems, we focus on
isovector mesonic operators 
\be 
{\cal O}_{\{S,P,V_\mu,A_\mu\},s}({\bf
  x},t) = \overline{\psi}_{u}({\bf
  x},t)\{1,\gamma_5,\gamma_\mu,\gamma_\mu\gamma_5\}\psi_{d}({\bf
  x},t)\,.  
  \ee 
  The subscript $s=\{{\cal P},{\cal S}\}$ on the
operators corresponds to whether it is constructed from local (${\cal P}$) or
smeared (${\cal S}$) quark fields.  From these interpolating operators, we build
correlation functions
 \bea
C^{s,s^{\prime}}_{X,Y}(t,T;{\bf p}) &=& \Tr\left[ e^{-HT}  \sum_{\bf x}e^{i{\bf p}\cdot {\bf x}}{\cal O}_{X,s}({\bf x},t) {\cal O}^\dagger_{Y,s^{\prime}}(0,0) \right] \nonumber \\
&=&\sum_{n}\sum_{\bf x}\langle n |e^{i{\bf p}\cdot {\bf x}} e^{-HT}
{\cal O}_X({\bf x},t) {\cal O}^\dagger_Y(0,0) | n \rangle\,,
\label{eq:singlecorr}
\eea 
for the various combinations of $X,Y= S,P,V_\mu,A_\mu$ and
choices of smearing of source and sink, $s$ and $s^{\prime}$. The sum
over $n$ is a sum over a complete set of states $\{|n\rangle\}$. In
the limit of large temporal extent of the lattice geometry,
$T\to\infty$, the vacuum state, $|\Omega\rangle$, dominates the
correlation function, but we are careful to explore the effects of the
finite temporal extent that allow additional contributions to
multi-hadron correlation functions in particular
\cite{Detmold:2008fn}.

Correlators for scalar and axial-vector baryons (di-quarks) are
similarly constructed from operators 
\be
\label{eq:BaryonInterpolators} {\cal O}_{\{N,\Delta_i\},s}({\bf x},t)
= {\psi}^\top_{u}({\bf
  x},t)(-i\sigma_2)C\{1,\gamma_i\gamma_5\}\psi_{d}({\bf x},t)\,, 
  \ee
where $(-i\sigma_{2})$ is the antisymmetric tensor of SU($N_{c}=2$)
and $C$ is the charge conjugation matrix.  However, as mentioned above
these baryons are degenerate with pseudoscalar and vector mesons and
so these correlators contain no new information.

In the limit of large temporal extent, these correlators decay with
time dependence that is characterised by the energies of the
eigenstates of the appropriate quantum numbers and by computing that
dependence, the eigen-energies can be extracted. That is, assuming $X$
and $Y$ are commensurate, 
\be
 C^{s,s^{\prime}}_{X,Y}(t,T;{\bf 0})
\stackrel{T\to\infty}{\longrightarrow} \sum_n
Z^{(n)\dagger}_{X,s}Z^{(n)}_{Y,s^{\prime}} e^{-E_n t}\,, 
\ee 
where the
$Z^{(n)}_{Y,s}= \langle n | {\cal O}^{s\dagger}_Y |\Omega\rangle$ are
overlap factors of the corresponding source and sink interpolating
operators onto the eigenstates enumerated by $n$ such that $E_n\leq
E_m$ for $n<m$, where $E_{n}$ is the energy of the eigenstate $|n\rangle$. 
For the finite temporal extent of the calculations we
perform, the effects of propagation of hadronic states around the
temporal boundary can also be important and modify the above
expression from exponential time dependence to a more complicated form
(for example,
$C_{P,P}^{s,s}\sim\sum_{n}|Z_{P,s}^{{(n)}}|^{2} \cosh[E_n(t-\frac{T}{2})]$). This
is taken in to account in our analysis where appropriate, and will be
returned to in the discussion of multi-hadron systems.

In order to extract the eigen-energies and related quantities, we
perform correlated $\chi^2$-minimising fits to our numerical data.
Results from the multiple different source locations on each
configuration are averaged, after appropriate translation, before any
further analysis proceeds. Results on consecutive configurations are
further averaged into blocked measurements, with a typical block size
consisting of 5--10 configurations (50--100 trajectories) depending on
the ensemble. Correlators are also reflected around the midpoint of
the temporal extent, averaging $\frac{1}{2}[C(t)\pm C(T-t)]$, to
reduce fluctuations. Statistical uncertainties are determined by using
the measurements on a given blocked ensemble to define multiple bootstrap
ensembles on which separate fits are performed using a globally
defined covariance matrix. The number of bootstrap ensembles is
typically 400. The systematic uncertainties in fits to single hadron
correlation functions are estimated by considering a large range of
fitting windows, $[t_{\rm min},t_{\rm max}]$ and using the width of
the variation with $t_{\rm min}$ and $t_{\rm max}$ over the range to
define the uncertainty.

\subsection{Quark masses}

\begin{figure}
  \includegraphics[width=\columnwidth]{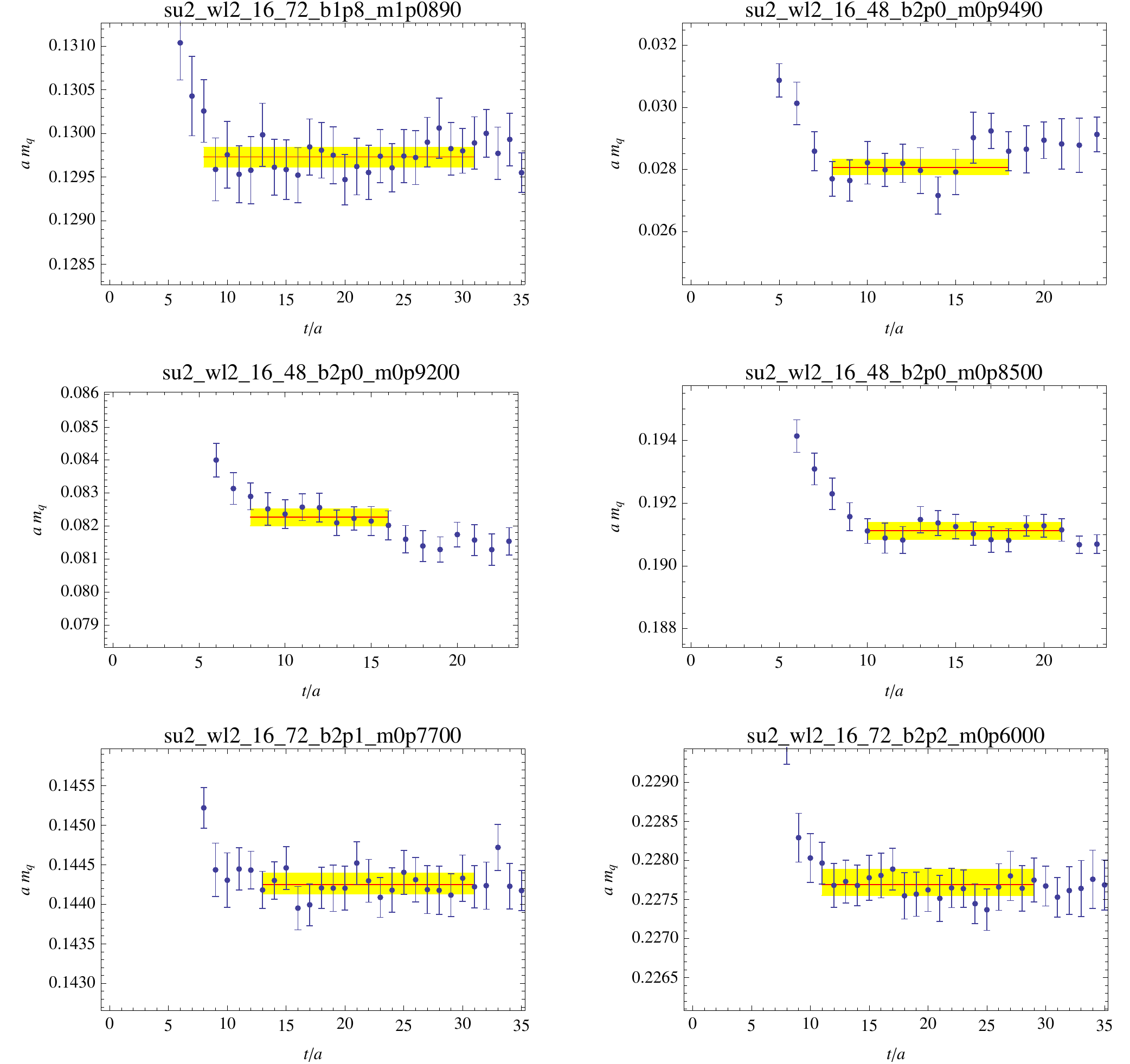}
  \caption{Ratios from which the PCAC quark mass can be extracted for
    $L=16$ ensembles. The shaded bands correspond to the extracted
    value of the ratio along with statistical and systematic
    uncertainties combined in quadrature.}
  \label{fig:mqfits}
\end{figure}

The PCAC quark mass can be extracted from the ratio of certain
combinations of the axial vector and pseudoscalar correlation
functions, namely 
\be 
a\, m_q^{PCAC} = \lim_{t,T\to\infty}
\frac{C^{{\cal P},{\cal P}}_{A_4,P}(t+1,T;{\bf 0})- C^{{\cal P},{\cal
      P}}_{A_4,P}(t-1,T;{\bf 0})}{4C^{{\cal P},{\cal
      P}}_{P,P}(t,T;{\bf 0})}\,, 
\ee 
which we access in a correlated
manner for each ensemble using the bootstrap procedure. These ratios,
and the associated constant in time fits, are shown for exemplary
ensembles in Fig.~\ref{fig:mqfits} and the values are listed in Table
\ref{masstab} for the $L=16$ ensembles (volume effects are seen to be
very small).
\begin{table}
  \begin{center}
    \begin{ruledtabular}
      \begin{tabular}{dddsssss} %%zxqdcolumn
        \multicolumn{1}{c}{Ensemble} & \multicolumn{1}{c}{$\beta$} & \multicolumn{1}{c}{$m_0$} &   \multicolumn{1}{c}{$a\ m_q$} & \multicolumn{1}{c}{$a\ f_\pi$} & \multicolumn{1}{c}{$a\ m_\pi$} & 
        \multicolumn{1}{c}{$m_\pi/m_\rho$}
        \\
        \hline
        A & 1.8 & -1.0890 & 0.1299(1)(1) & 0.259(1)(1) & 0.8281(8)(5) &  0.844(1)(2) 
        \\
        B & 2.0 & -0.9490 & 0.0280(2)(4) & 0.101(3)(5) & 0.347(6)(13)&  0.663(9)(10)
        \\
        C & 2.0 & -0.9200 & 0.0823(3)(3) & 0.159(2)(4) & 0.609(3)(4) & 0.826(3)(4)
        \\
        D & 2.0 & -0.8500 & 0.1911(3)(2) &0.2156(16)(11) & 0.9151(13)(6)& 0.910(2)(2)
        \\
        E & 2.1 & -0.7700 & 0.1442(1)(1) & 0.1582(1)(1)& 0.7450(9)(7) & 0.904(2)(2)
        \\
        F & 2.2 & -0.6000 & 0.2277(2)(1)& 0.1525(5)(7) & 0.8805(7)(5)& 0.951(3)(3)
      \end{tabular}
    \end{ruledtabular}
    \caption{ The PCAC quark masses, pion decay constants, pion masses
      and pion to $\rho$ mass ratio computed on $L=16$ ensembles.}
    \label{masstab}
  \end{center}
\end{table}

\subsection{Pion decay constant and scale setting}

Setting the overall scale in this theory is arbitrary as there is no
physical quantity to match to. We choose the scale through the
pion decay constant, $f_\pi$, defined through 
\be 
\langle 0 | J_{\mu
  5}^a | \pi^b(p) \rangle \equiv i f_\pi \delta^{ab} p_\mu \,, 
  \ee
where $J_{\mu 5}^a$ is the axial current. The value of $\Lambda=4\pi
f_\pi$ is a proxy for the typical scale in the theory, although other
quantities such as the $\rho$ meson mass could also be used.

\begin{figure}
  \includegraphics[width=\columnwidth]{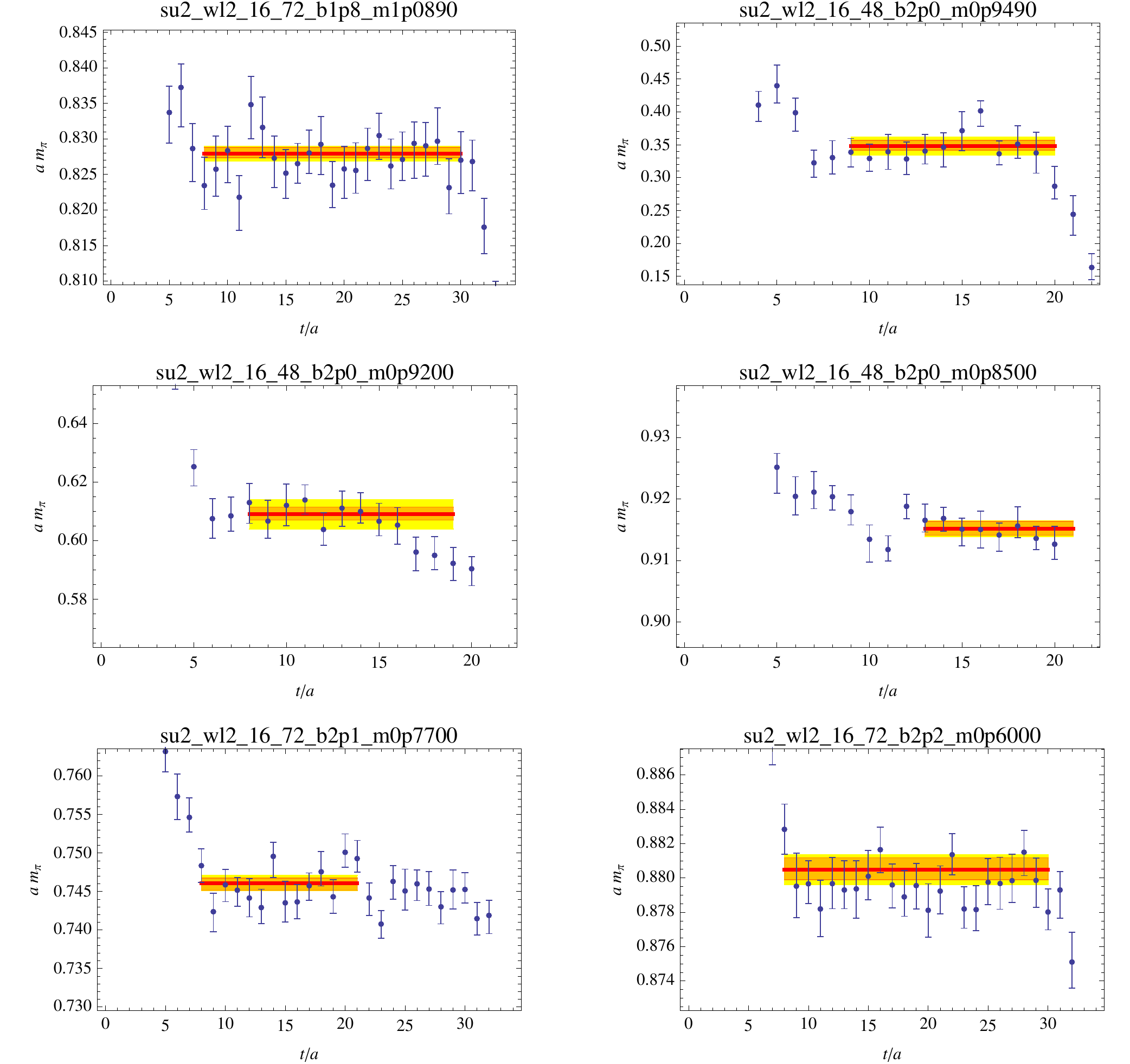}
  \caption{Effective mass plots for the $C_{P,P}^{{\cal P},{\cal P}}$
    correlators for $L=16$ ensembles that are used for the extraction
    of $f_\pi$ and $m_\pi$. The shaded bands correspond to the
    extracted value of the pion mass along with the statistical
    uncertainties (inner band) and the statistical and systematic
    uncertainties combined in quadrature (outer band).}
  \label{fig:mpifits}
\end{figure}

The axial current used in the lattice calculation differs from that
defined in the continuum 
\be 
J_{\mu 5}^{a,\rm latt}={\cal Z}_A J_{\mu
  5}^{a,\rm QCD} \,,
\ee
and there is a finite renormalisation that must
be undertaken to convert to the continuum. We use the one loop
perturbative determination of this renormalisation constant, ${\cal
  Z}_A=1- g_0^2 C_F d_A(1)$ where $C_F=(N_c^2-1)/2N_c = 3/4$ and
$d_A(1)=0.100030(2)$ \cite{Groot:1983ng}.  Using this, the pion decay
constant can be determined from fits to two-point correlation
functions.  We follow Ref.~\cite{Aoki:2010dy}
% (the normalisation of
%$f_{\pi}$ used here differs from that of
%Refs.~\cite{Lewis:2011zb,Hietanen:2014xca}) 
and used the various
different correlators, $C_{X,Y}^{s,s^{\prime}}$ for
$\{X,Y\}=\{P,A_{4}\}$ and $\{s,s^{\prime}\}={{\cal P},{\cal S}}$ to
extract $m_{\pi}$, $Z^{(0)}_{P,{\cal P}}$, $Z^{(0)}_{P,{\cal S}}$,
$Z^{(0)}_{A_{4},{\cal P}}$, and $Z^{(0)}_{A_{4},{\cal S}}$. Having
determined these, the decay constant in lattices units is given by 
\be
a\ f_{\pi}= \frac{{\cal Z}_{A}{Z}^{(0)}_{A_{4},{\cal P}}}{a m_{\pi}}\,.
\ee
The extracted values of the decay constant and pion mass are shown
in Table \ref{masstab} in lattice units and effective mass plots of
some of the correlators that enter the fits are shown in
Fig.~\ref{fig:mpifits}.

To set the scale, we require $f_{\pi}=246$ GeV on each ensemble at a
common value of $m_\pi/ m_\rho=0.9$ (this value of $f_{\pi}$ is motivated in the
context of strongly coupled theories of electroweak symmetry breaking,
but is a somewhat more arbitrary choice for a theory of the dark sector).  That is,
\be
a=\frac{ a f_\pi(m_\pi/m_\rho=0.9)}{246}\ {\rm GeV}^{-1}.
\ee
Our calculations are not performed exactly at bare quark masses
corresponding to $m_\pi/m_\rho=0.9$ for each value of beta, so we
interpolate our data to that value assuming $f_\pi= f_0 + f_1 \beta +
f_2 \beta^2 + f_m m_\pi^{2}/m_\rho^2$ (adding higher order terms in
either $\beta$ or $m_\pi/m_\rho$ does not alter the extraction
significantly).  The resulting lattice spacings for the different
$\beta$ values that we use are shown in Table \ref{tab:aVsBeta}.  This
approach differs somewhat from Ref.~\cite{Hietanen:2014xca}, where the
scale setting is performed using $f_{\pi}$ after it has been
extrapolated to the chiral limit. Since we do not explore the chiral
regime, this approach is not practical in the current study. If the pion decay constant is
linearly dependent on the PCAC mass (or equivalently, quadratically on
$m_\pi$) as one may expect in the heavy quark regime, the difference
amounts to an overall rescaling, leaving ratios of lattice spacings
unaltered. Since we could equally well have demanded $f_{\pi}=2.46$
GeV, such a scaling is irrelevant and we find that the ratio of
lattice spacings at $\beta=2.0$ and $\beta=2.2$ determined here is
similar to that in Ref.~~\cite{Hietanen:2014xca}.
\begin{table}
  \begin{center}
    \begin{ruledtabular}
      \begin{tabular}{cc}
	$\beta$ & $a\ [10^{-3}\ {\rm fm}]$ \\
	\hline
	1.8 & 0.35(2) \\
	2.0 & 0.24(1) \\
	2.1 & 0.19(1)    \\
	2.2 & 0.14(2) \\
      \end{tabular}
    \end{ruledtabular}
    \caption{ The lattice spacings for the various gauge couplings as
      defined in the main text at the common mass ratio
      $m_\pi/m_\rho=0.9$.}
    \label{tab:aVsBeta}
  \end{center}
\end{table}

\subsection{Meson and baryon masses}

The enhanced symmetries of the $N_c=2$ theory mean that mesons and
baryons occur in degenerate multiplets.  The pseudoscalar mesons,
which we refer to as the $\pi^{\pm,0}$, belong to the 5-dimensional,
fundamental representation of the residual Sp(4)$\sim$SO(5) symmetry
with the other elements being a scalar, isoscalar baryon $N\sim ud$
(which we refer to as the nucleon) and a conjugate anti-baryon,
$\overline{N}$. While the masses of all of these states are expected
to vanish in the chiral limit, there are significant explicit symmetry
breaking terms in the Lagrangian for the parameters we consider and so large deviations from
the expectations of SU(4)$\to$Sp(4) chiral dynamics
\cite{Bijnens:2009qm,Bijnens:2011fm} are expected.  Choosing $X=Y=V_{1,2,3} $, we
have also investigated the isovector, vector mesons and the
corresponding ``$\Delta$'' ($J^P =1^+$ isoscalar axial-vector baryon)
and ``anti-$\Delta$'' baryons of the theory in detail. States with
other quantum numbers are discussed below.

\begin{figure}
  \includegraphics[width=\columnwidth]{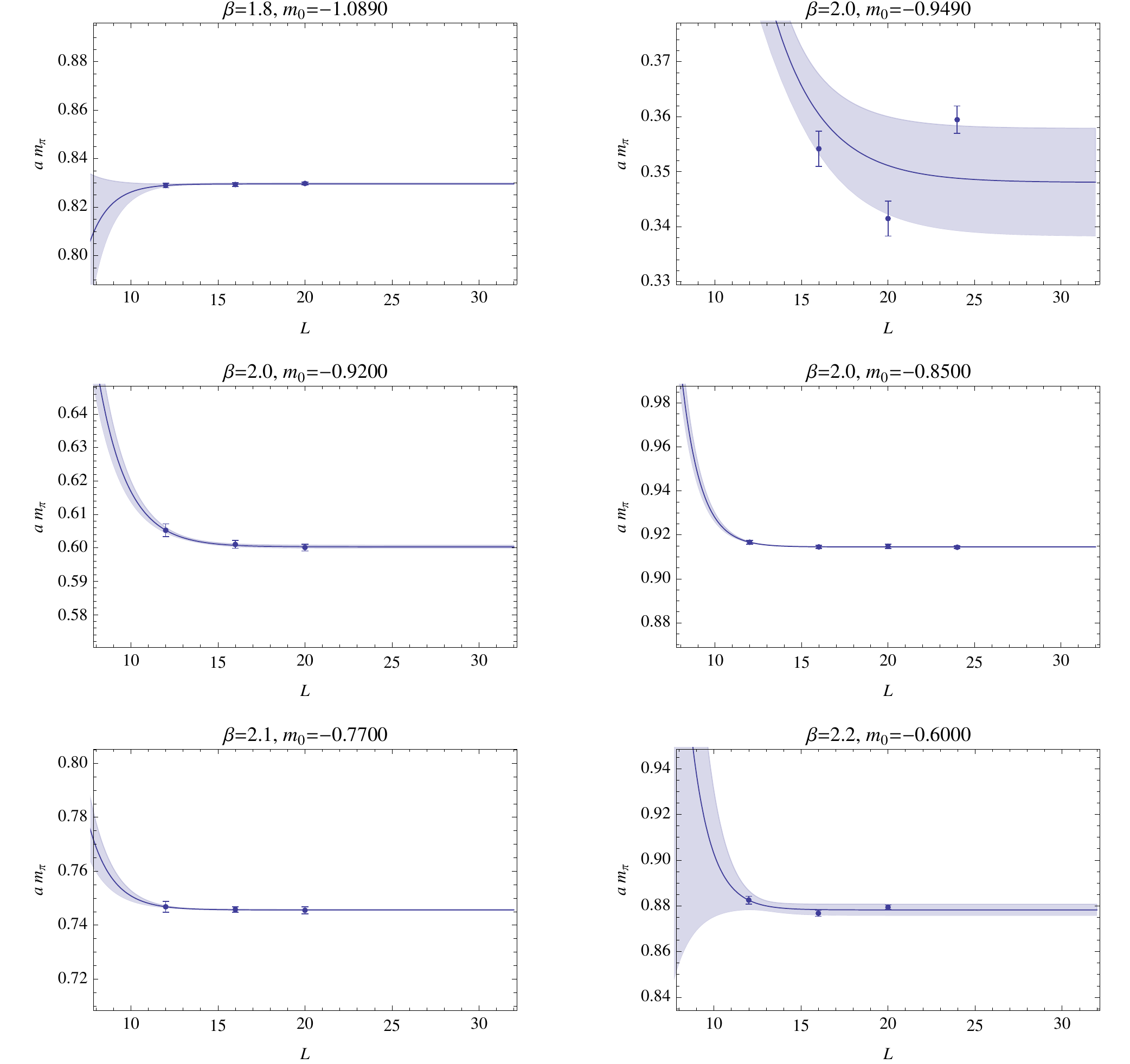}
  \caption{Volume dependence of the pion (equivalently, nucleon) mass
    for each setting of $\beta$ and $m_0$.  The curves and shaded
    regions correspond to fits of the form $m_\pi(L)=m_\pi(\infty) +
    \delta_L e^{-m_\pi L}$.}
  \label{mpivsL}
\end{figure}

\begin{figure}
  \includegraphics[width=\columnwidth]{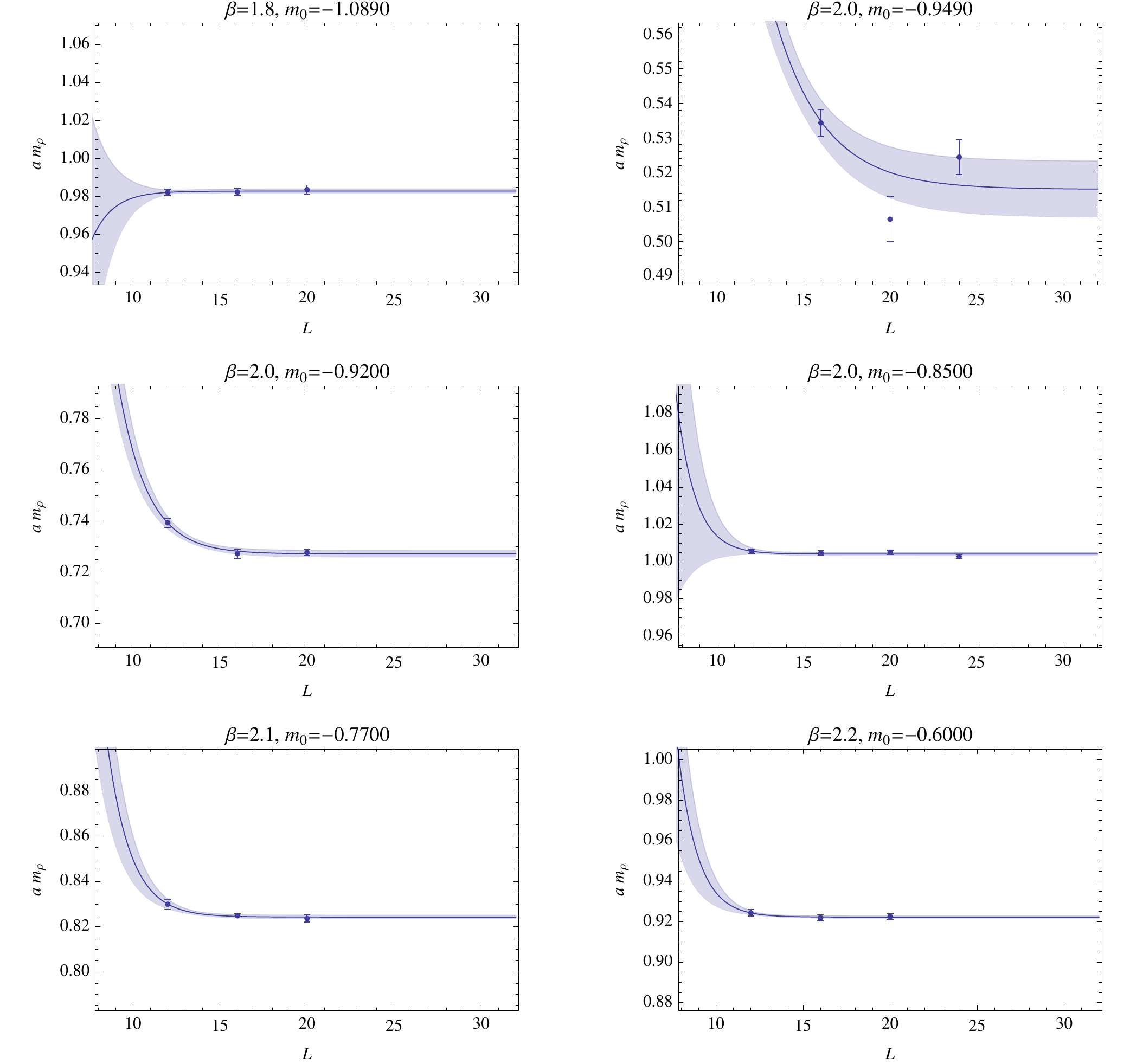}
  \caption{Volume dependence of the $\rho$ (equivalently, $\Delta$)
    mass for each setting of $\beta$ and $m_0$.  The curves and shaded
    regions correspond to fits of the form $m_\rho(L)=m_\rho(\infty) +
    \tilde\delta_L e^{-m_\pi L}$.}
  \label{mrhovsL}
\end{figure}

In figures \ref{mpivsL} and \ref{mrhovsL}, we show the volume
dependence of the baryon masses for each set of ensembles.  Typically
only very small volume dependence is observed, consisted with
exponential corrections from states propagating around the spatial
directions of the lattice geometry. We are confident that the minimal
dependence seen here will not pollute the extraction of multi-hadron
scattering parameters and binding energies discussed below, although
the $B$ ensemble with the lightest masses and $L=12$, where
$m_{\pi}L\sim5$, should be treated with caution.

\begin{figure}
  \includegraphics[width=0.7\columnwidth]{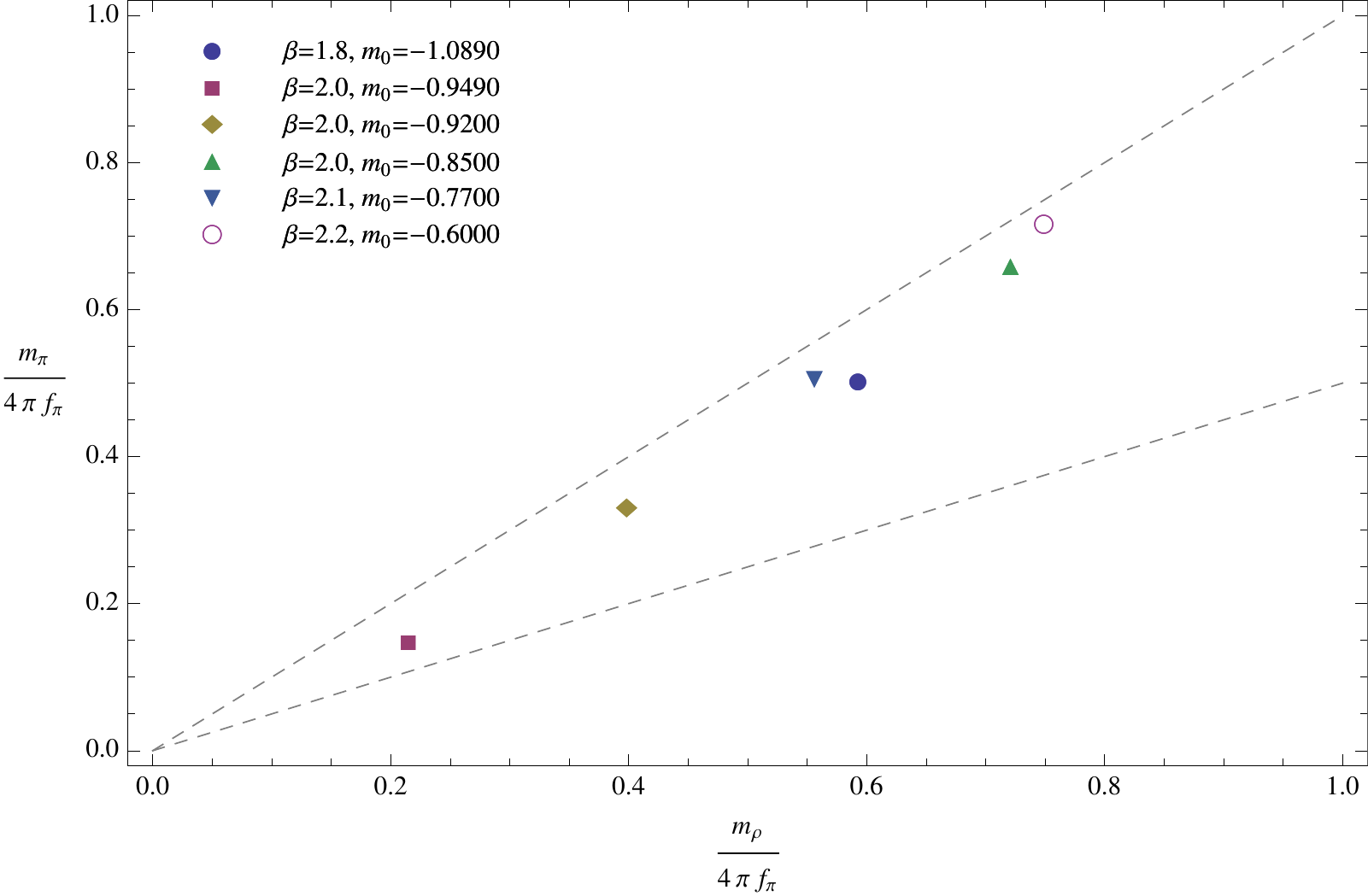}
  \caption{Relationship between the $\rho$ and pion masses for the
    quark masses used in this work. The dashed lines indicate where
    $m_{\rho}=m_{\pi}$ and $m_{\rho}=2m_{\pi}$, respectively.}
  \label{fig:mrhoVsmpi}
\end{figure}

Figure \ref{fig:mrhoVsmpi} shows the range of values
of $m_{\pi}/4\pi f_{\pi}$ that we realise as a function of
$m_\rho/4\pi f_{\pi}$. 
Note that the $\rho$ is stable against decay to multiple pions at the
masses we have chosen and that remains so over quite a range of
masses.  Similarly, the $\Delta$ is stable against decay to a nucleon
plus multiple pions.  However, this is only true in the mass range
considered here; for vanishingly light quarks, the nucleon and pion
become massless but the $\rho$ and $\Delta$ remain massive with
$m_\rho=m_\Delta\sim \Lambda_{\rm QC_2D}$.  Hence, there is a value of
the quark masses at which the $\rho$ and $\Delta$ become unstable to
$\pi\pi$ and $\pi N$ decays, respectively.

The isovector axial-vector meson states (isoscalar vector baryons) can
be similarly investigated, although we do not pursue calculations
here.  Refs.~\cite{Lewis:2011zb,Hietanen:2014xca} have performed this
investigation and find that these states are somewhat heavier than the
vector mesons over the wide range of quark masses that were
studied. Isoscalar mesons are more difficult to investigate, but may
be interesting for phenomenological reasons. We leave investigations
of this sector to future work.

\subsection{Single particle dispersion relations}

\begin{figure}
  \includegraphics[width=\columnwidth]{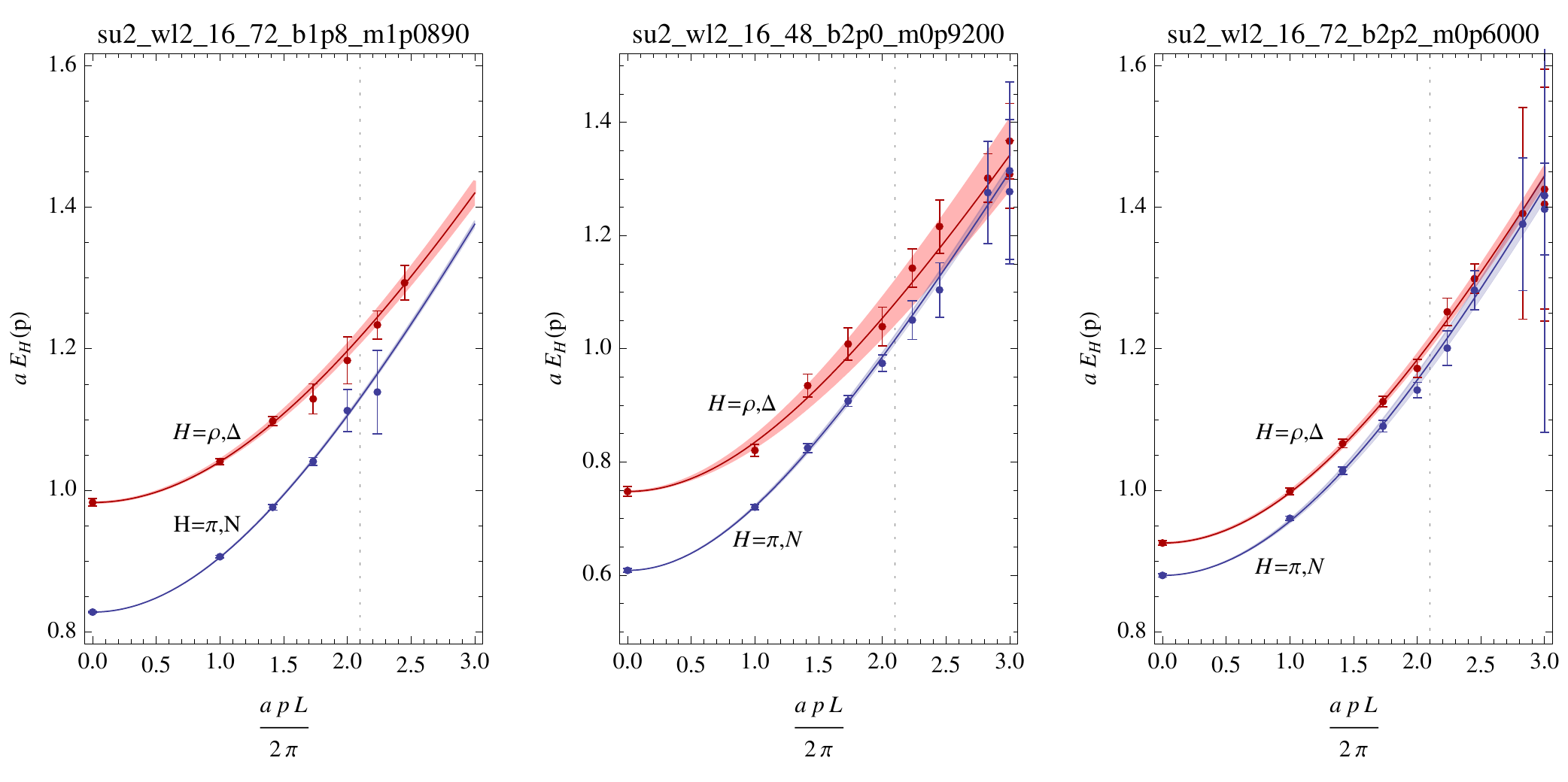}
  \caption{The pion(nucleon) and $\rho$($\Delta$)  dispersion relations on three of
    the ensembles. The bands show fits to the data to the left of the
    dashed line as discussed in the text. The shaded bands correspond
    to the 90\% confidence regions of the fits.}
  \label{fig:disprel}
\end{figure}

In order to control systematics in our discussion of the
multi-particle spectrum, we also investigate the dispersion relations
of the single pion and $\rho$ meson at non-zero lattice spacing by
measuring the correlators in Eq.~(\ref{eq:singlecorr}) for all lattice
momenta with $\left|\frac{a L{\bf
      p}}{2\pi}\right|^{2}\leq9$. Fig.~\ref{fig:disprel} shows the
extracted $\pi$ and $\rho$ energies as a function of momentum for
three of the ensembles.  Fits to the low momentum region of the
dispersion relations using the continuum motivated form 
\be
E_{H}(p)=\sqrt{M_{H}^{2}+c_{H}^{2}p^{2}} 
\ee
 indicate that
discretisation effects are relatively mild.  The extracted values of
the ``speed of light'', $c_{H}$, for the pseudoscalar and vector
states are shown in Table \ref{disptab} for each ensemble.  We focus
on the $C_{P,P}^{{\cal P},{\cal P}}(t,T;{\bf p})$ and
$C_{V_{3},V_{3}}^{{\cal P},{\cal P}}(t,T;{\bf p})$ correlators which
have the best interpolating fields for the moving mesons, and fits are
performed to the data for $\left[\frac{a p L}{2\pi}\right]^{2}<4$. The
uncertainty on the speed of light is determined from the parameter
confidence interval of the fit and from the variation of the value
between correlators with different types of source and sink smearing,
specifically $C_{X,X}^{{\cal P},{\cal P}}$ and $C_{X,X}^{{\cal
    S},{\cal S}}$.  Comparing the different ensembles, it is clear
that the values for larger $\beta$ tend towards unity and even for
$\beta=2.0$, the deviations from the continuum expectation are
$\sim$7\%. Consequently, we proceed to use the continuum dispersion in
analysis of the multi-hadron spectrum.
\begin{table}
  \begin{center}
    \begin{ruledtabular}
      \begin{tabular}{cccccc}
        Ensemble & $L^{3}\times T$ & $\beta$ & $m_0$ &$c_{\pi}$ & $c_{\rho}$ \\
        \hline
        $A$ & $16^{3}\times 72$  & 1.8 & $-1.0890$ &0.93(1)& 0.87(4)\\
        $B$ & $16^{3}\times 48$ & 2.0 & $-0.9490$ & 0.92(5) & 0.97(5) \\
        $C$ & $16^{3}\times 48$ & 2.0 & $-0.9200$ & 0.99(2) & 0.94(1)\\
        $D$ & $16^{3}\times 48$ & 2.0 & $-0.8500$ & 0.94(2) & 0.92(3)\\
        $E$ & $16^{3}\times 72$ & 2.1 & $-0.7700$ & 0.95(1)& 0.93(3)\\
        $F$ & $16^{3}\times 72$ & 2.2 & $-0.6000$ &0.96(2)& 0.94(1)\\
      \end{tabular}
    \end{ruledtabular}
    \caption{ The speeds of light extracted from fits to the $\pi$ and
      $\rho$ dispersion relations, $c_\pi$ and $c_\rho$, respectively.
    }
    \label{disptab}
  \end{center}
\end{table}     

\section{Nuclear/multi-hadron spectroscopy}
\label{sec:multi}

\begin{figure}
  \includegraphics[width=0.6\columnwidth]{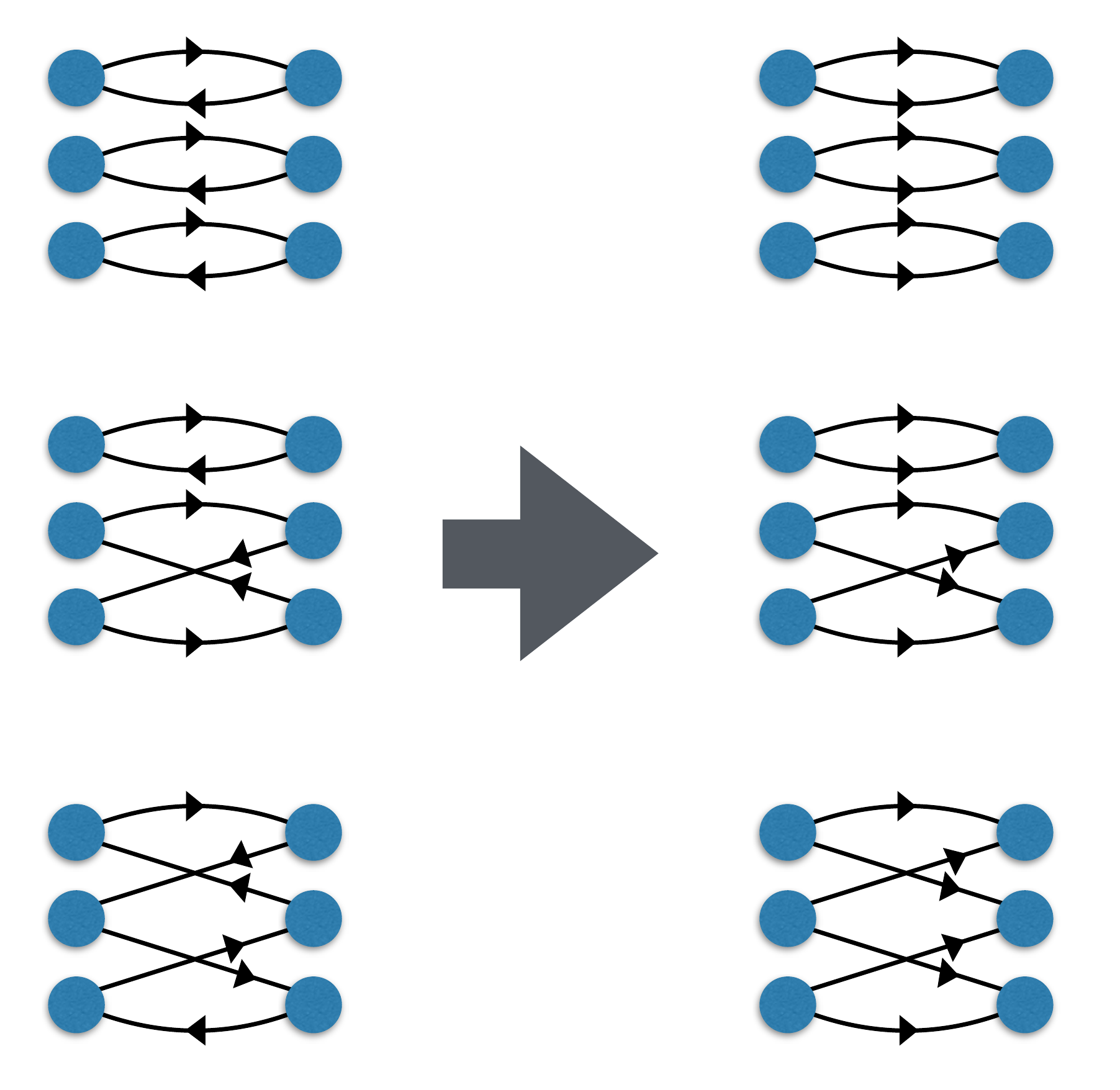}
  \caption{Relationship between $|I|=I_z=n$ multi-meson contractions
    and $B=n$ multi-baryon contractions for $n=3$. On the left, we
    consider the three topologies of quark contractions that
    contribute to the $|I|=I_z=n=3$ multi-meson correlator, with lines
    with arrows point right(left) corresponding to up quark (anti-down
    quark) propagators. On the right, the contractions that result
    from replacing the anti-down propagators by down quark propagators
    which correspond to the contractions for $n$ baryons of opposite
    parity to the mesons.}
  \label{fig:multimesTomultibaryon}
\end{figure}

Just as there are degeneracies between the various meson and baryon
states in the single hadron sector, the multi-hadron systems also fall
into multiplets containing multi-meson, multi-baryon, and
multi-meson--multi-baryon systems. In what follows, we will focus on
the computationally simplest to access systems that contain the
$(\overline{u} d)^n$ multi-meson systems with maximal
isospin\footnote{Isospin refers to a global SU(2) subgroup of SU(4)
  that is preserved through the symmetry breaking SU(4)$\to$Sp(4) with
  generators ${\bf T}={ \left(\begin{array}{cc}{\bm \tau} & 0 \\ 0 &
        {\bm \tau} \end{array}\right)}$, and baryon number refers to a
  global U(1) symmetry.}, $|I|=I_z=n$. By consideration of the quark
contractions that these systems require, it is clear that these meson
states are degenerate with corresponding multi-baryon states as
indicated in Fig.~\ref{fig:multimesTomultibaryon}. These systems have
no disconnected/annihilation type contractions and so have
corresponding degenerate multi-baryon partners. The relation is made
exact by using the identities for the quark propagator
\cite{Lewis:2011zb} 
\bea 
S(y,x)&=& C^{\dagger}
(-i\sigma_{2})^{\dagger}S(x,y)^{T}
(-i\sigma_{2})C\,, \\
S(y,x)&=&\gamma_{5}S^{\dagger}(x,y)\gamma_{5} 
\eea
where $(-i\sigma_{2})$ is the antisymmetric tensor of SU($N_{c}=2$) and $C$
is the charge conjugation matrix (the first relation is specific to
the two colour theory).  Multiple applications of these relations
replace the multi-meson correlator by the multi-baryon correlator for
baryons that are of opposite parity to the mesons.

Group theoretically, we consider the $n^{\rm th}$ tensor product of
fundamental representations of Sp(4)$\sim$SO(5) and consider only
states in the totally symmetric flavour irrep., which form multiplets
of size $(n+1)(n+2)(2n+3)/6$ \cite{hamermesh1989group}.  In what
follows, we will refer to the $I=0$, $B=n$ component of each
multiplet, noting that it is degenerate with states with baryon number
$-n\le B \le n$ of varying multiplicities.  We will focus on angular
momentum $J=0$ and $J=1$ systems which can be thought of as $nN$ and
$(n-1)N\Delta$ states, respectively (more properly, the eigenstates
have a Fock component of this form).

In order to construct two-point correlators of the appropriate quantum
numbers efficiently, we make use of the methods developed in
Refs.~\cite{Beane:2007es,Detmold:2008fn, Detmold:2008yn,
  Detmold:2011kw,Detmold:2012wc} for the study of multi-meson systems
in $N_c=3$ QCD. These directly translate to the current situation
because of the $N_c=2$ specific identification of the $n\ N$
correlator with the $n\ \pi^+$ correlator and the $(n-1)\ N\Delta$
correlator with the $(n-1)\ \pi^+\rho^+$ correlator, as discussed
above.  To this end, we study the correlators 
\be
\label{eq:Cn}
C_{nN}(t)=\left\langle 0 \left| \left(\sum_{\bf x}{\cal O}_N^{{\cal
          P}}({\bf x},t)\right)^n \left({\cal O}_N^{{\cal
          S}\dagger}({\bf x}_0,t_0)\right)^n\right|0\right\rangle\,,
\ee
 and 
 \be 
\label{eq:CnD}
 C^{(i,j)}_{nN,\Delta}(t)=\left\langle 0 \left|
    \left(\sum_{\bf x}{\cal O}_N^{{\cal P}}({\bf
        x},t)\right)^n\sum_{\bf x} {\cal O}_{\Delta_j}^{{\cal P}}({\bf
      x},t) \left({\cal O}_N^{{\cal S}\dagger}({\bf x}_0,t_0)\right)^n
    {\cal O}_{\Delta_i}^{{\cal S}\dagger}({\bf
      x}_0,t_0)\right|0\right\rangle\,, 
\ee
 where $({\bf x}_0,t_0)$ is
the chosen source location and ${\cal O}_{N,\Delta_j}^s$ are the
interpolating operators for the nucleon and $\Delta$ states defined in
Eq.~(\ref{eq:BaryonInterpolators}). In our study, we average over all
polarisations of the $\Delta$ correlators.  In the limit of very large
time separations and with an infinite temporal extent of the lattice
geometry, these correlators are dominated by the energies of the $nN$ 
and $nN\Delta$ ground states, $E_{nN}$ and $E_{nN,\Delta}$,
respectively.  The factorially large numbers of contractions that
these correlators encompass are performed using the methods of
Refs.~\cite{Detmold:2008fn,Detmold:2011kw,Detmold:2012wc}.  Since the
number of quark degrees of freedom that can be sourced at a single
space-time point is $N_s N_c=8$, the construction of propagators from
a single source limits our calculations to $n\le8$ in the present
calculation.

As we are interested in the hadronic interactions, it is also useful
to define the ratios 
\be
\label{eq:ratios}
R_{nN}(t)\equiv \frac{C_{nN}(t)}{[C_{1N}(t)]^n},\qquad
R_{nN,\Delta}(t)\equiv
\frac{\sum_{i}C^{(i,i)}_{nN,\Delta}(t)}{[C_{1N}(t)]^n\sum_i
  C^{(i,i)}_{0N,\Delta}(t)}, 
\ee
 that fall off at late times with
characteristic exponential dependence on the energy shifts, $\Delta
E_{nN} = E_{nN} - n E_{N}$ and $\Delta E_{nN,\Delta} = E_{nN,\Delta} -
n E_{N} - E_{\Delta}$, respectively. Provided we consider Euclidean
times large enough that the numerators and denominators in these
ratios have been separately saturated by their ground states, these
ratios potentially allow us to take advantage of correlations between the
different terms in extracting the energy shifts.

Since the systems that we are interested in easily factorise into
multiple colour singlet states, the finite temporal extent of the
lattice geometries that we work with has an important consequence
\cite{Detmold:2008yn,Detmold:2011kw}. The interpolating operators that
we use are designed to produce a particular set of quantum numbers
propagating over the time-slices that separate the source and the
sink. However, they can also produce the same overall quantum numbers
by having some part of the system propagate around the temporal
boundary. The expected forms of the $J=0$ correlators are then 
\be
\label{eq:fitform}
C_{nN}(t) = \sum_{m=0}^n Z_{n;m} \cosh\left( \delta E_{n;m} t_T
\right) + Z_{n;\frac{n}{2}} \delta_{n \bmod{2},0} + \cdots \,, 
\ee
where $t_T = t-T/2$, $\delta E_{n;m} = E_{(n-m)N}-E_{mN}$ and $m$
counts the number of forward going $N$'s (more precisely the forward
going baryon number) and the ellipsis denotes excited state
contributions either in the forward going signal, or in the thermal
contributions.  The second term in Eq.~(\ref{eq:fitform}) appears for
even $n$ and is a time independent contribution resulting from half
the system propagating forward in time and half propagating backward
in time.  For the $J=1$ correlators, the analogous expression is
\cite{Detmold:2011kw} 
\be
\label{eq:fitformDelta}
C_{nN,\Delta}(t) = \sum_{m=0}^n \sum_{j=0}^1 Z_{n;m,j} \cosh\left(
  \delta E_{n;m,j} t_T \right) + \cdots \,, 
\ee
 where $m(j)$ counts
the number of forward going $N$'s ($\Delta$'s) in a given term,
$\delta E_{n;m,j} = E_{(n-m)N,(1-j)\Delta}-E_{mN,j\Delta}$, and the
ellipsis denotes excited state contributions.

In order to extract the energies, and thereby the energy shifts,
  we use these forms to fit to the various correlators. 
  We consider  a number of different approaches for dealing with the thermal 
  effects. One strategy is to consider most of time extent of the lattice 
  geometry, $[t_{\rm min},T-t_{\rm min}]$, excluding only the  region where excited states 
  contaminate the signal, and use the  thermal state behaviour
  of Eqs.~(\ref{eq:fitform}) and (\ref{eq:fitformDelta}). This can be
  done most efficiently in a cascading fashion, first fitting the one
  nucleon energy from $C_{1N}(t)$ and then using this value in the
  fit of $C_{2N}(t)$ where the main goal is to extract
  $E_{2N}$ and so on (a similar approach was previously used in
  Ref.~\cite{Detmold:2011kw}). The $Z$ factors are linear parameters
  and are eliminated using variable projection
  \cite{golub1973differentiation} and thus the minimisation at each
  $n$ is with respect to a single parameter. Alternatively, we can consider 
  fits that only treat the dominant forward 
  and backward going states over a restricted time range, $[t_{\rm min},t_{\rm max}]\cup[T- t_{\rm max},T-t_{\rm min}]$, omitting the regions where either thermal- or excited-states are
  relevant. Finally, We can also directly analyse the ratios in Eq.~(\ref{eq:ratios}). All methods lead to  extractions of the energies that are consistent 
  for most states.  For the main discussion, we opt for the simplest 
  approach, analysing the correlators themselves without thermal effects, and use effective mass plots to identify the time-slices where 
  thermal effects are negligible. The statistical
  uncertainties are estimated using the bootstrap method and we choose time ranges 
  conservatively such that adding a time-slice does not alter the results significantly.

\begin{figure}
  \includegraphics[width=0.8\columnwidth]{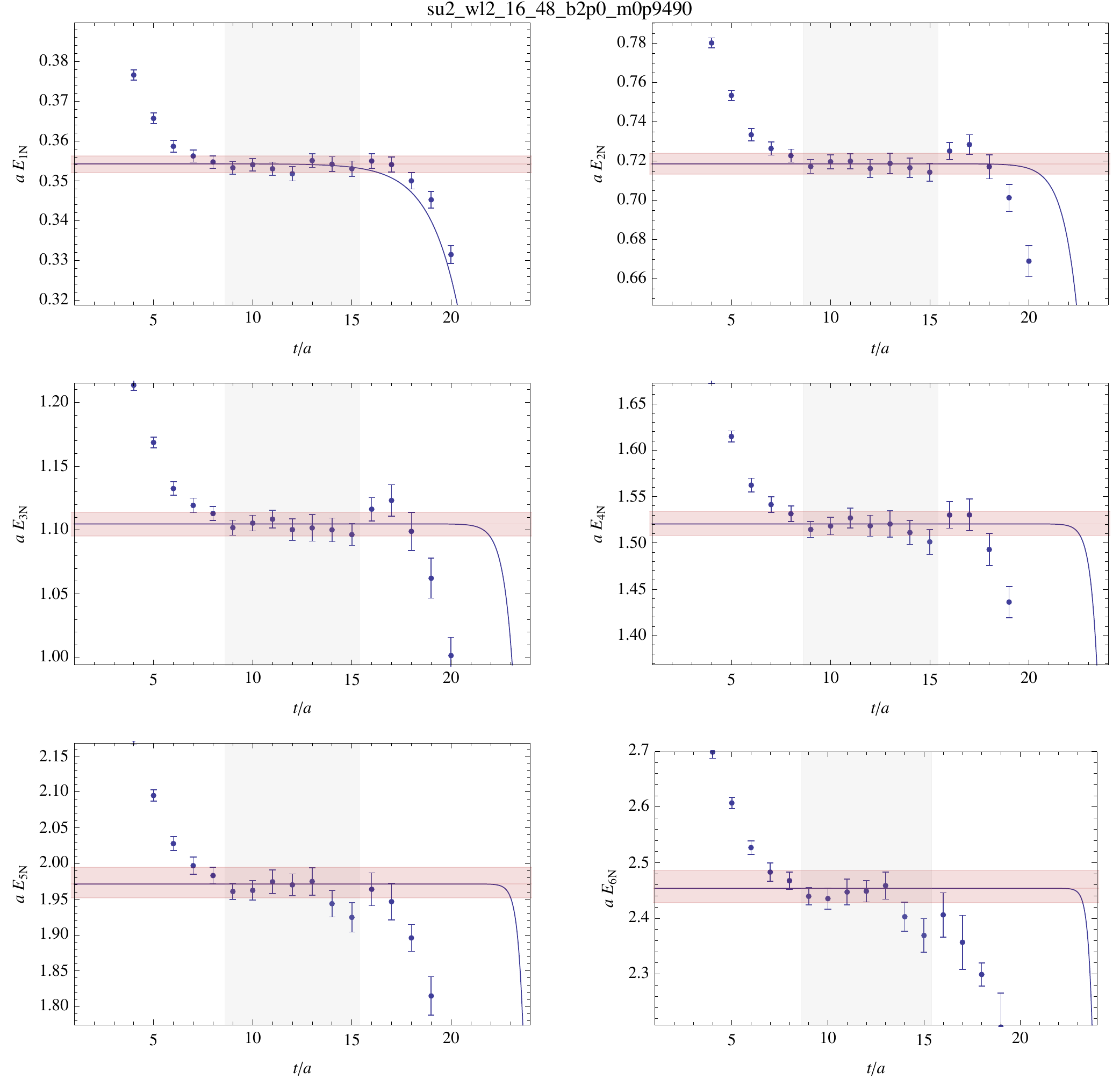}
  \caption{Effective mass plots for the $J=0$ correlators of
    Eq.~(\ref{eq:Cn}) for the $16^{3}\times 48$ $B$ ensemble for
    $n=1,\ldots,6$ nucleons. The horizontal band shows the energy extracted 
    from fits to the correlator, while the vertical band indicates the range of 
    time-slices used in the fits.}
  \label{fig:effmassnucleiA}
\end{figure}

\begin{figure}
  \includegraphics[width=0.8\columnwidth]{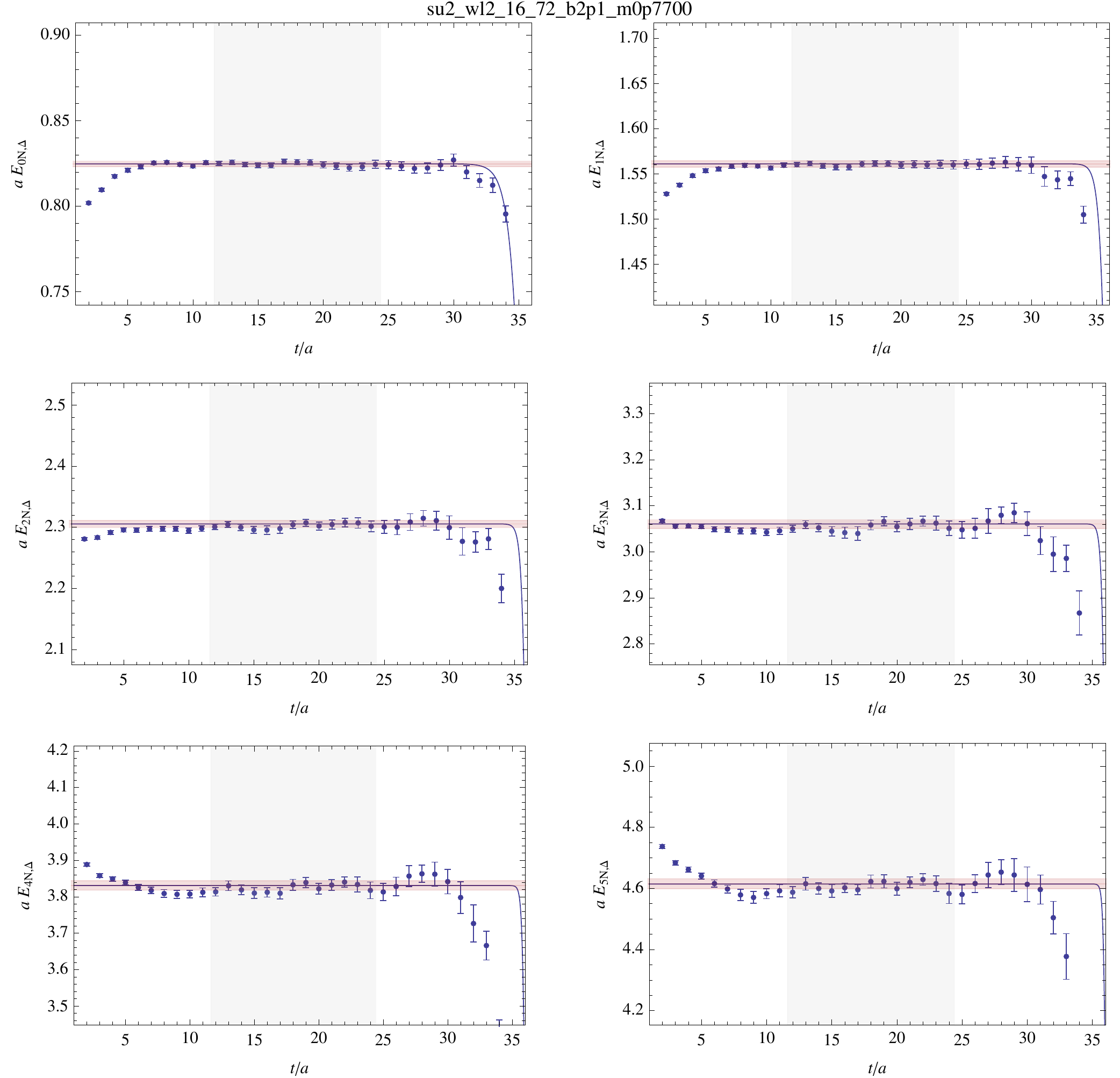}
  \caption{Effective mass plots for the $J=1$ correlators of
    Eq.~(\ref{eq:CnD}) for the $16^{3}\times 72$ $E$ ensemble for a
    single $\Delta$ and $n=0,\ldots,5$ nucleons. The horizontal band shows the energy extracted 
    from fits to the correlator, while the vertical band indicates the range of 
    time-slices used in the fits.}
  \label{fig:effmassnucleiB}
\end{figure}

In Figs.~\ref{fig:effmassnucleiA} and \ref{fig:effmassnucleiB}, the
effective masses of the correlators are shown for
two of the ensembles for  baryon numbers from 2 to 7, along with
the resulting energies extracted from fits to the
correlators (horizontal band). The vertical shaded regions show the 
fit ranges that are used. For the lighter mass ensembles, strong thermal effects are clearly visible,
as seen in Fig.~\ref{fig:effmassnucleiA} in particular.
The energies extracted from the fits on each of the ensembles are
tabulated in tables in Appendix \ref{fitresults} and summarised in
Figs.~\ref{fig:energyShiftExtractionsA} -- \ref{fig:energyShiftExtractionsF}.
\begin{figure}
  \includegraphics[width=\columnwidth]{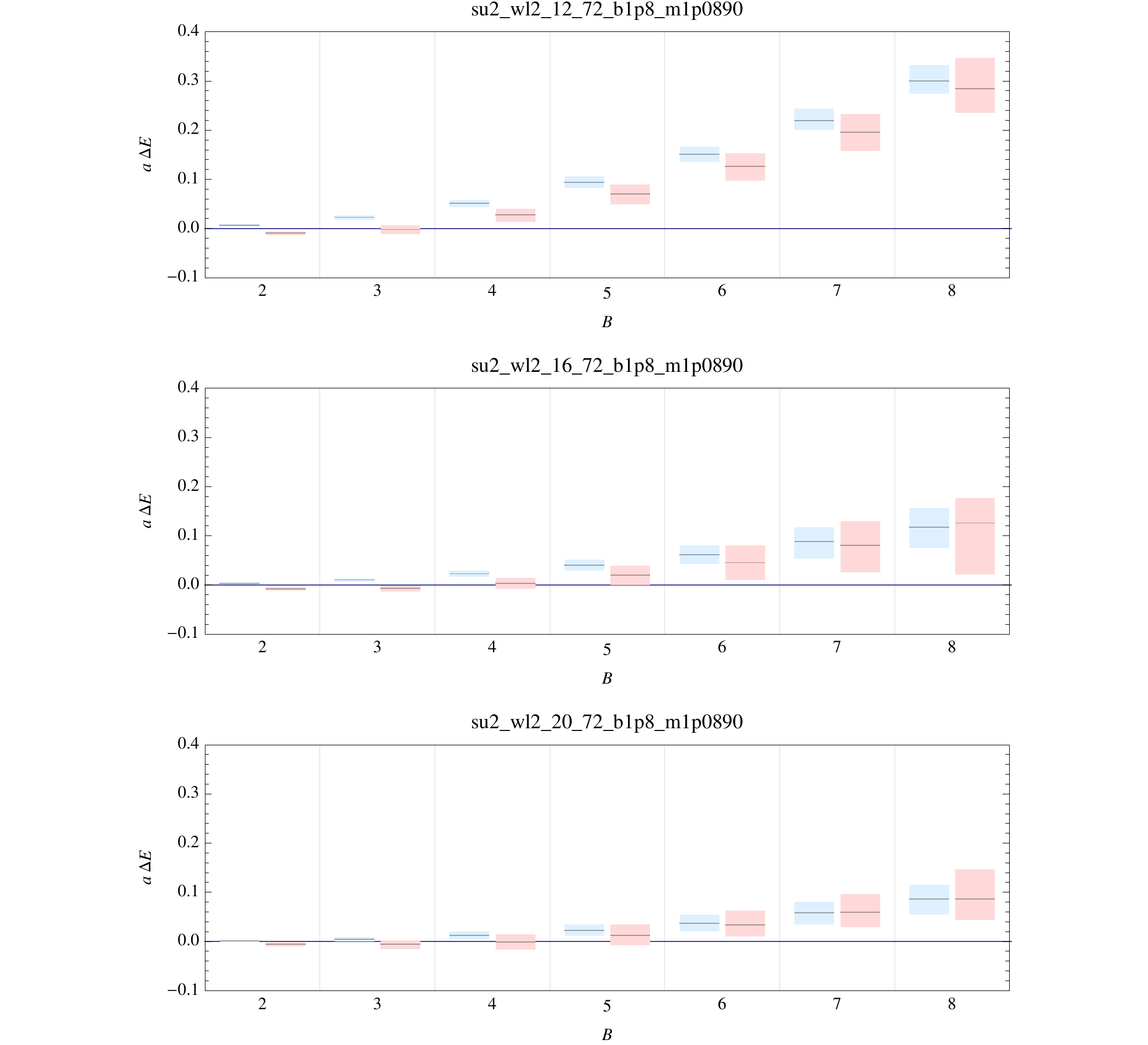}
  \caption{Extracted energy shifts for the $J=0,1$ systems for the $A$
    ensembles. For each baryon number, $B$, the left (blue) region
    corresponds to the $J=0$ system and the right (red) region
    corresponds to the $J=1$ system.}
  \label{fig:energyShiftExtractionsA}
\end{figure}
\begin{figure}
  \includegraphics[width=\columnwidth]{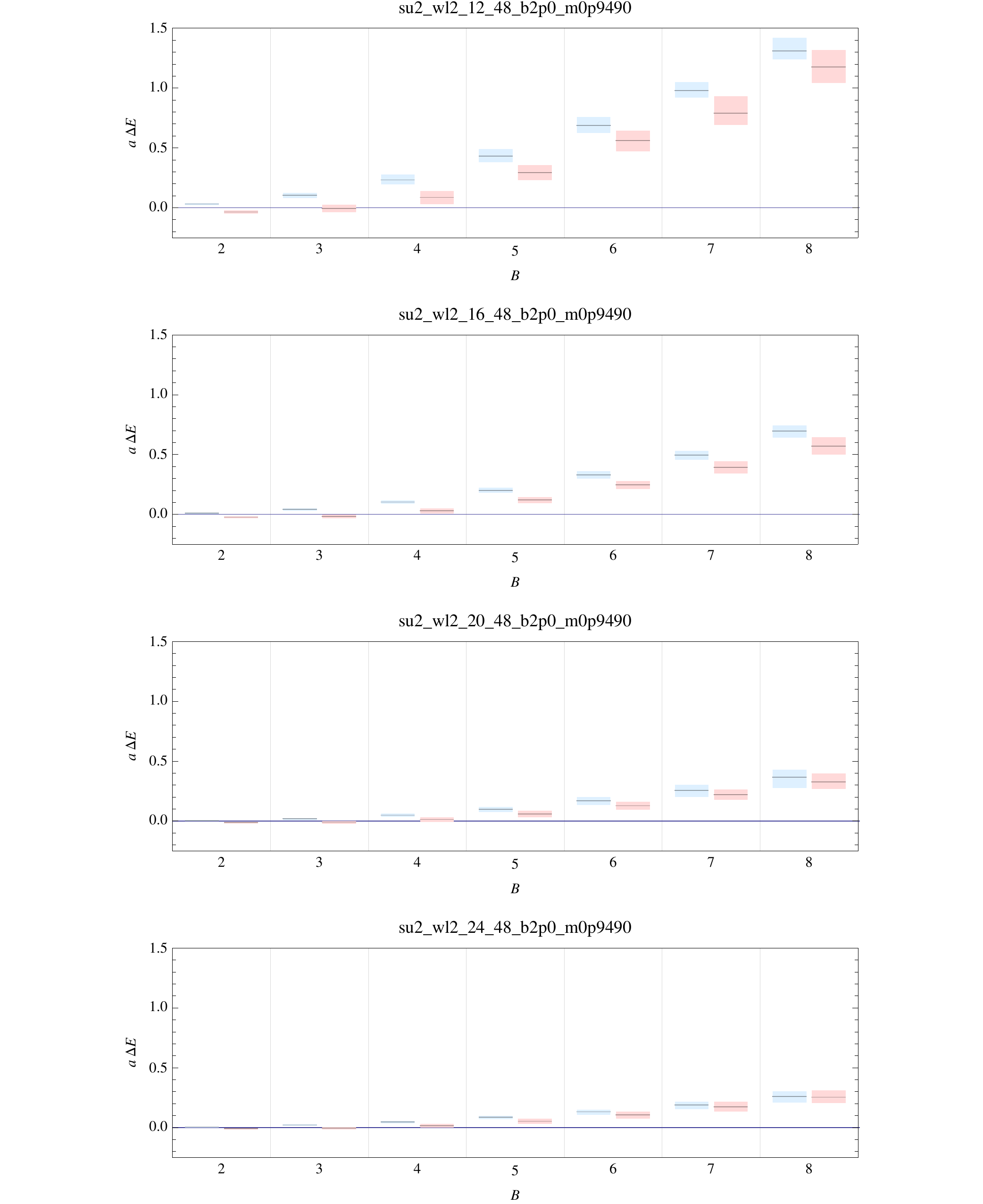}
  \caption{Same as Fig.~\protect{\ref{fig:energyShiftExtractionsA}}
    for the $B$ ensembles.}
  \label{fig:energyShiftExtractionsB}
\end{figure}
\begin{figure}
  \includegraphics[width=\columnwidth]{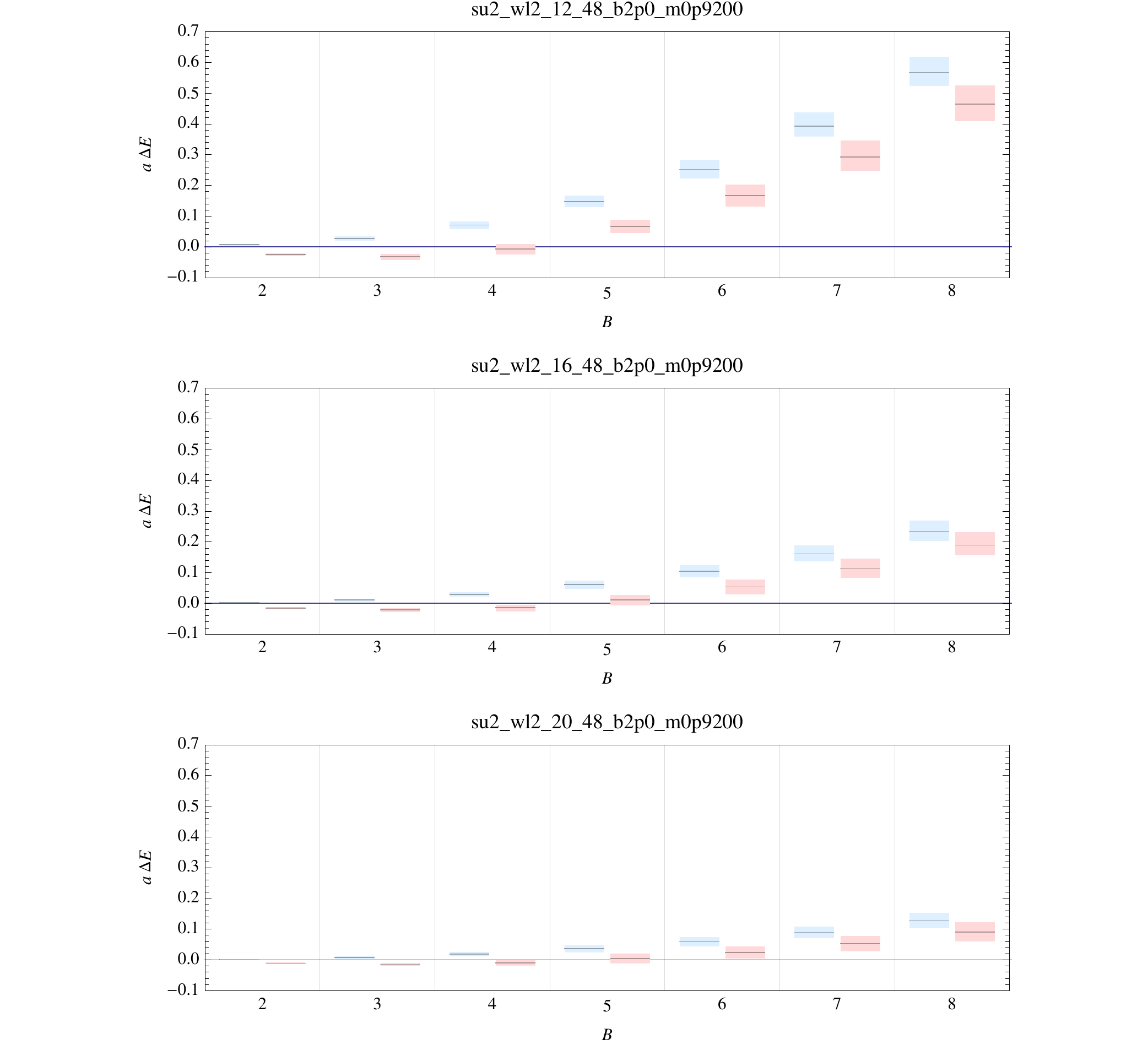}
  \caption{Same as Fig.~\protect{\ref{fig:energyShiftExtractionsA}}
    for the $C$ ensembles.}
  \label{fig:energyShiftExtractionsC}
\end{figure}
\begin{figure}
  \includegraphics[width=\columnwidth]{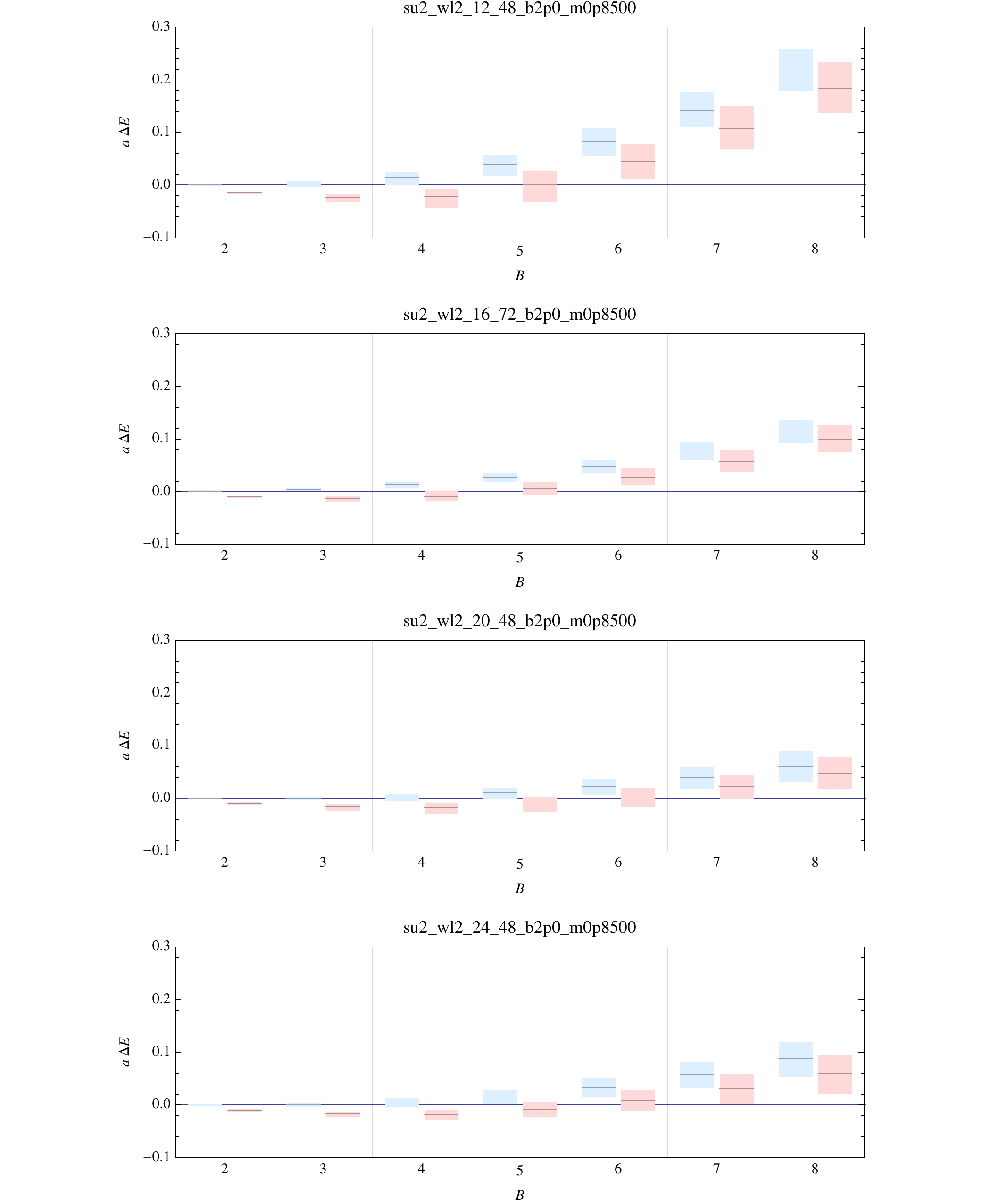}
  \caption{Same as Fig.~\protect{\ref{fig:energyShiftExtractionsA}}
    for the $D$ ensembles.}
  \label{fig:energyShiftExtractionsD}
\end{figure}
\begin{figure}
  \includegraphics[width=\columnwidth]{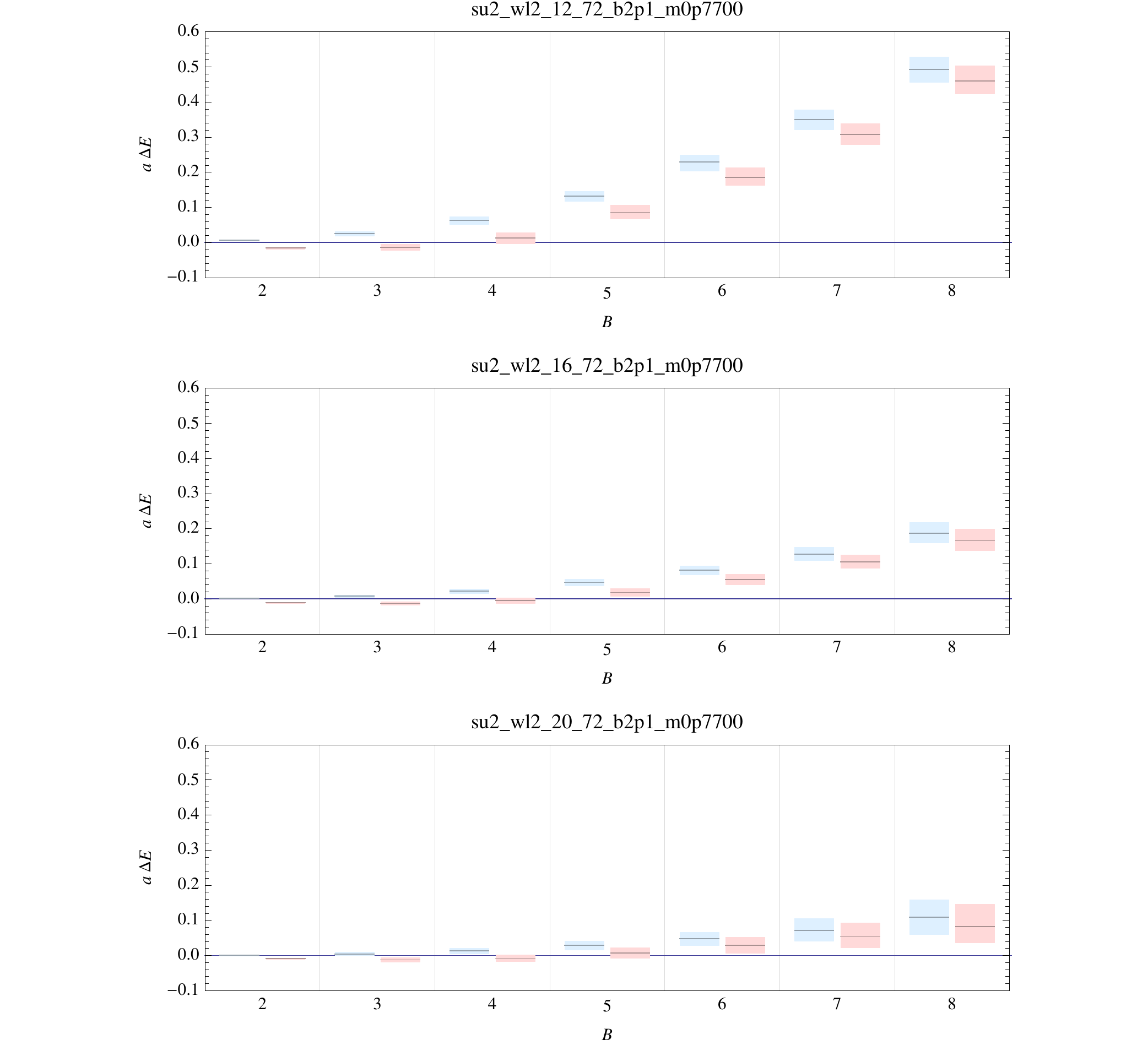}
  \caption{Same as Fig.~\protect{\ref{fig:energyShiftExtractionsA}}
    for the $E$ ensembles.}
  \label{fig:energyShiftExtractionsE}
\end{figure}
\begin{figure}
  \includegraphics[width=\columnwidth]{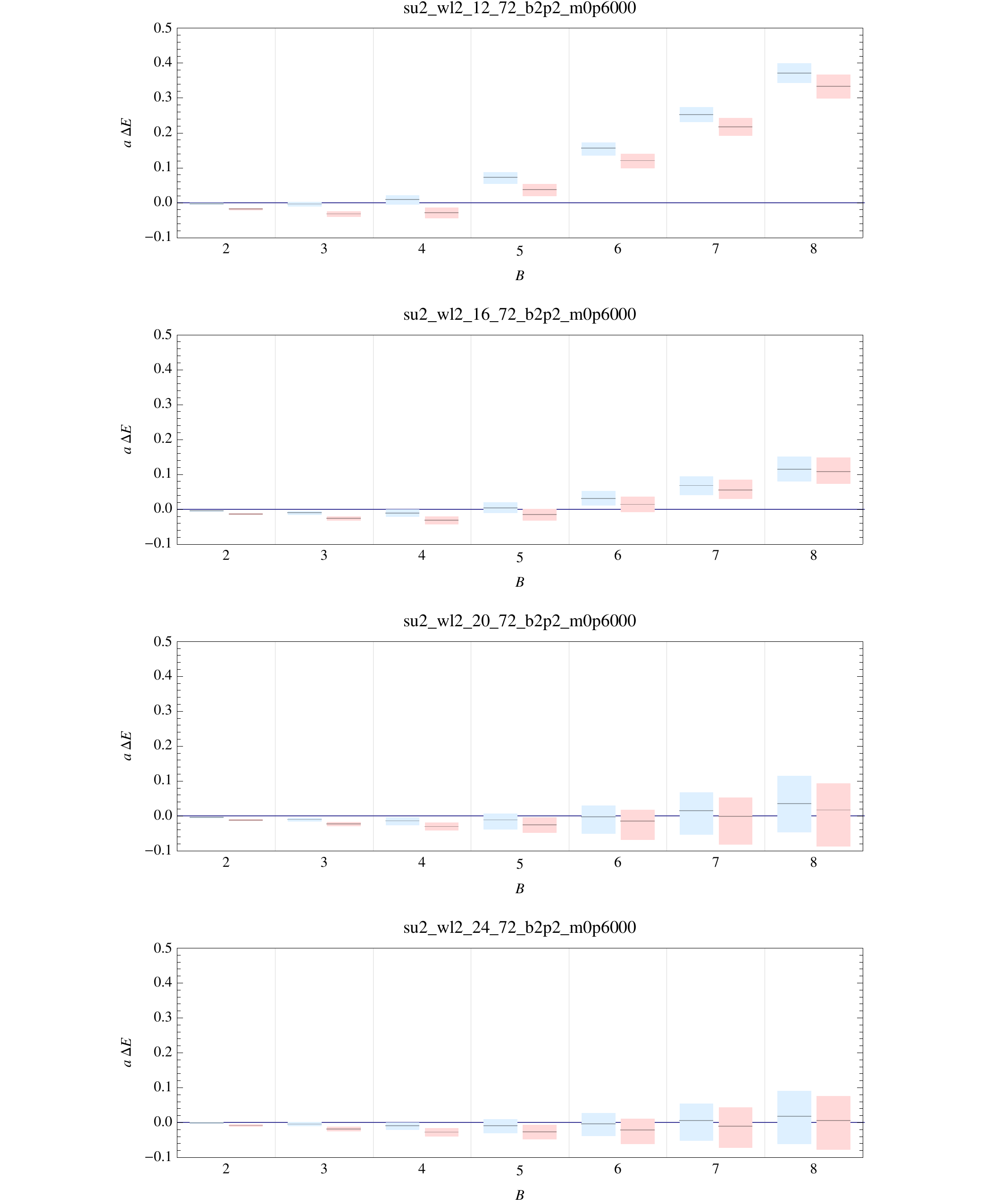}
  \caption{Same as Fig.~\protect{\ref{fig:energyShiftExtractionsA}}
    for the $F$ ensembles.}
  \label{fig:energyShiftExtractionsF}
\end{figure}

\subsection{Bound versus scattering states}

After studying the spectrum at multiple different volumes, we can
investigate whether states are bound states or scattering states. For
the case of two-hadron \cite{Luscher:1986pf,Luscher:1990ux} and
three-hadron \cite{Polejaeva:2012ut,Briceno:2012rv,Hansen:2013dla}
systems, or for weakly interacting $n$ boson systems
\cite{Beane:2007qr,Detmold:2008gh}, the expected dependence of
scattering states on the volume is known and is determined by the two-
and three- body interactions. In the limit of small interactions and
large volumes, the expectation is that the energies of these systems
will scale with $1/L^3$ if they are unbound. For two-body bound states, the volume
dependence is exponential, $e^{-\gamma L}/L$, with the exponent
determined by the binding momentum, $\gamma$ \cite{Beane:2003da}. There
is also a general expectation \cite{Kreuzer:2010ti} that higher-body,
deeply bound states will have localised wavefunctions and depend
exponentially on the volume for sufficiently large volumes, and if the
binding is arising dominantly from two-body interactions, a similar
scaling may be expected. Given this, we attempt to perform fits to the
volume dependence using two hypotheses with functional forms
corresponding to scattering and bound state systems. Specifically 
\bea
H_1:&\quad \Delta E_{\rm bound}(L) =& -\Delta E_{\infty}\left[1+ C\ \frac{e^{-\kappa
      L}}{L}\right]\,,
\\
H_2:&\quad \Delta E_{\rm scatter}(L) =& \frac{2\pi A}{\mu L^3}
{{\left(\begin{array}{c}n\\2\end{array}\right)}}
\left[1-\left(\frac{A}{\pi L}\right){\cal I} +\left(\frac{A}{\pi
      L}\right)^{2}[{\cal I}^{2}+(2n-5){\cal J}] \right]
+\frac{B}{L^{6}}\,, 
\eea 
where $A$, $B$, $C$, $\Delta E_{\infty}$ and
$\kappa$ are in general free parameters and the geometric constant
${\cal I}=-8.9136329$, ${\cal J}=16.532316$.  For two-body systems the
bound state hypothesis simplifies as $\Delta E_{\inf}=\frac{\gamma^2}{2\mu}$,
$\kappa=\gamma$ and $C=\frac{12}{\gamma} \hat{C}$, where
$\mu=\frac{m_1 m_2}{m_1+ m_2}$ is the reduced mass for a system
involving particles of masses $m_1$ and $m_2$ and $\gamma\equiv\sqrt{2\mu\; \Delta E_{\infty}}$ is the
infinite volume binding momentum \cite{Beane:2003da, Davoudi:2011md},
leaving two fit parameters, $\gamma$ and $\hat{C}$.  In order to allow
bound state hypothesis fits with only three volumes, we make the same
substitutions for higher body systems (the relationships between the parameters are now
assumptions), although this means that the conclusions
for $n>2$ are less definitive.  For the case of weakly-interacting $n$
body scattering states (unbound), the parameter $A$ corresponds to the
two-body scattering length and $B$ receives  contributions from effective
range corrections and three-body
interactions~\cite{Beane:2007qr,Detmold:2008gh}.

By analysing the performance of the two different models in fits to
data for multiple volumes, we can ascertain whether particular states
are likely bound states or finite volume scattering states for the
particular quark masses and lattice spacing under consideration.  To
assess this, we define the Bayes factor \cite{kass1995bayes} 
\be
\label{eq:Bayes}
K= \frac{ P(D|H_1)}{P(D|H_2)} = \frac{\int P(D|H_1,p_1)P(p_{1}|H_1) d
  p_1} {\int P(D|H_2,p_2)P(p_{2}|H_2) d p_2}\,, 
\ee 
which is the ratio
of likelihoods of the hypotheses given the data, $D$, each of which
can be computed as the integral over the parameters of the model, $p_i$,
of the likelihood of the data given the model for those parameters
weighted by the prior probability of the parameters given the
model. This last factor is input, and we choose Gaussian priors for $A$,
$B$ and $\hat{C}$ and an exponential distribution for $\gamma$ with
widths $10$, $10^5$, $0.1$, $10$, respectively (the extracted Bayes
factors are insensitive to these choices).  The likelihood function is
defined by 
\be 
\log P(D|H_i,p_i) = -\frac{1}{2}\sum_{j=1}^N
\frac{\left[d_j - H_i(x_j;p_i)\right]^2}{\sigma_j^2}\,, 
\ee
 for a set of $N$ data points, $D=\{(x_{1},d_1,\sigma_1),\ldots (x_{N},d_N,\sigma_N)\}$, 
 with coordinates, $x_{i}$, values, $d_{i}$, and uncertainties, $\sigma_i$.  
 The integrals defining the Bayes
factor, Eq.~(\ref{eq:Bayes}), are calculated as follows. The $H_1$
model is linear in $\hat C$, which allows the corresponding Gaussian
integral to be computed exactly. The remaining integral over $\gamma$
is computed numerically. Similarly, in the $H_2$ model, the integral
over $B$ is Gaussian, but the integral over $A$ requires numerical
computation.

Establishing an infinite volume binding is not the final result; to
extract physical information we then need to extrapolate to the
continuum limit\footnote{In principle, the continuum extrapolation
  should be performed for a number of fixed physical volumes, and only
  then should the resulting energy shifts be extrapolated to the
  infinite volume limit. However this would require extensive careful
  tuning of lattice geometries and lattice spacings and a more prosaic
  approach is adopted here. It would also be possible to perform a
  single coupled fit to the $a$, $L$ and $m_q$ dependence, but this is
  technically challenging.} and investigate the dependence on the
quark mass.

\subsection{$J^P=0^+$ and $J^P=1^+$ multi-baryon systems}

In the $J=0^+$, $n\ N$ systems, the scattering state fits are
typically strongly preferred; either all the extracted energy shifts
are positive for that set of ensembles\footnote{If there is a bound
  state in infinite volume, but all the extracted energies are above
  the threshold, then there must be a volume larger than the largest
  available where the system is at threshold and hence physically very
  extensive. It is thus subject to large volume effects that
  invalidate the bound state model as given.} or the Bayes factor is
very small indicating the scattering fit is preferred. For larger $n$,
the positive energy shifts get larger faster than multiple two-body
interactions would predict, thereby indicating the presence of
repulsive three-body interactions. As we are interested in bound
states, we do not pursue these states further in the present study.

\begin{figure}
  \includegraphics[width=0.45\columnwidth]{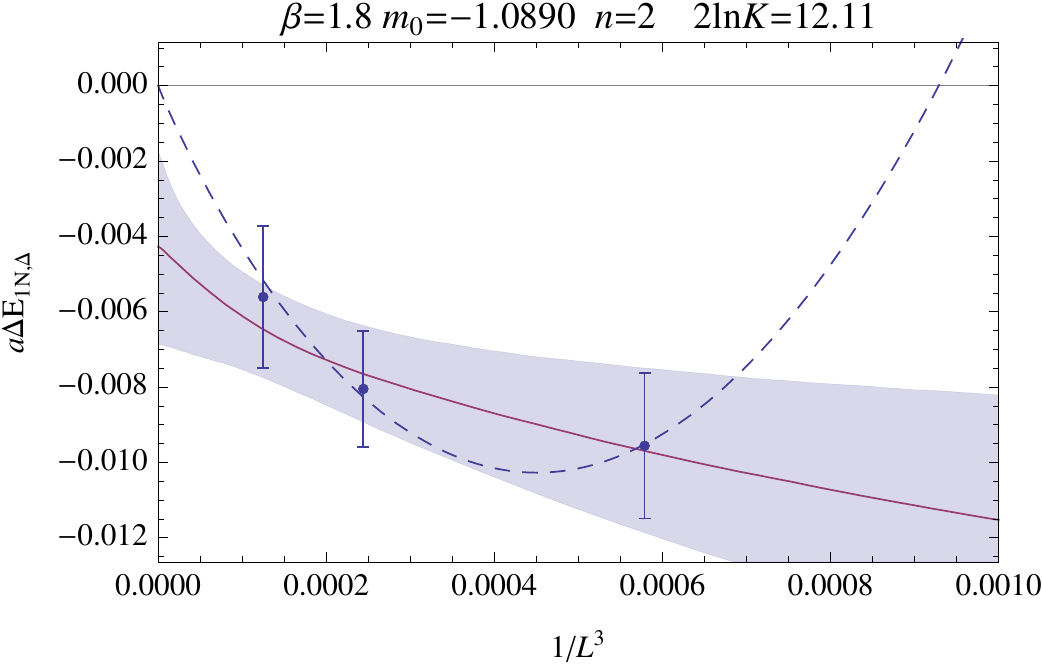}
  \qquad
  \includegraphics[width=0.45\columnwidth]{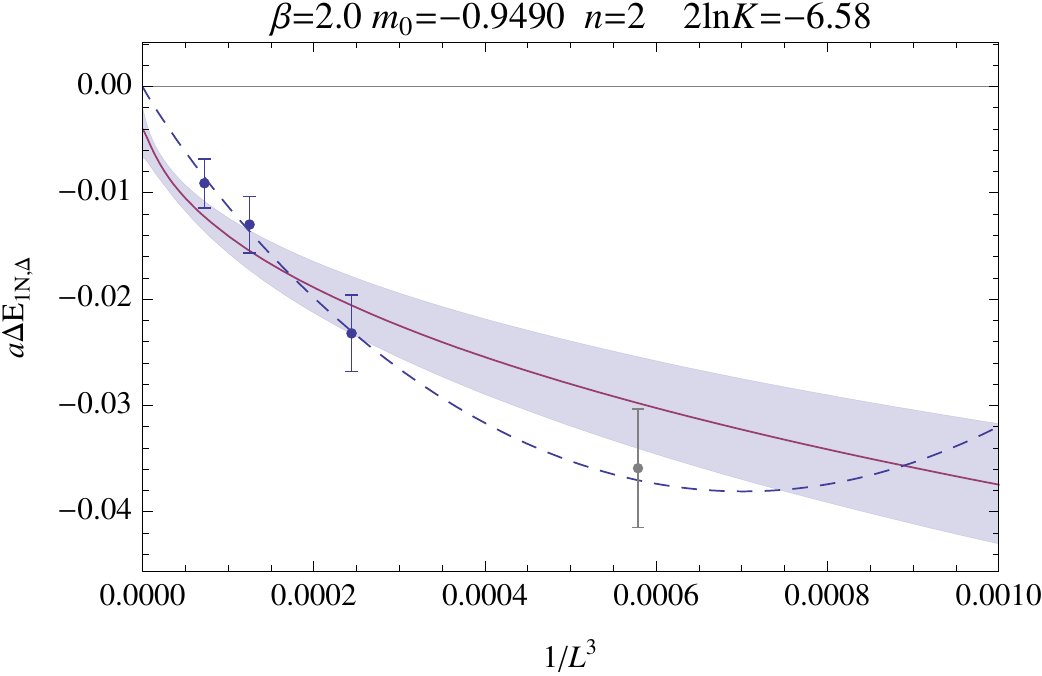}
  \\ \vspace*{4mm}
  \includegraphics[width=0.45\columnwidth]{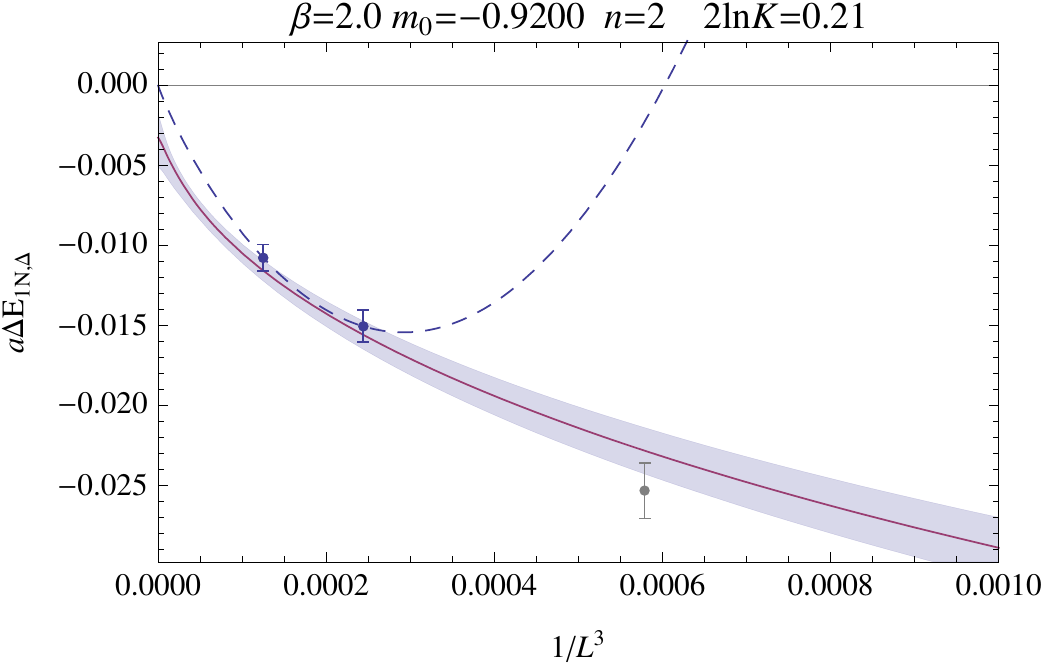}
  \qquad
  \includegraphics[width=0.45\columnwidth]{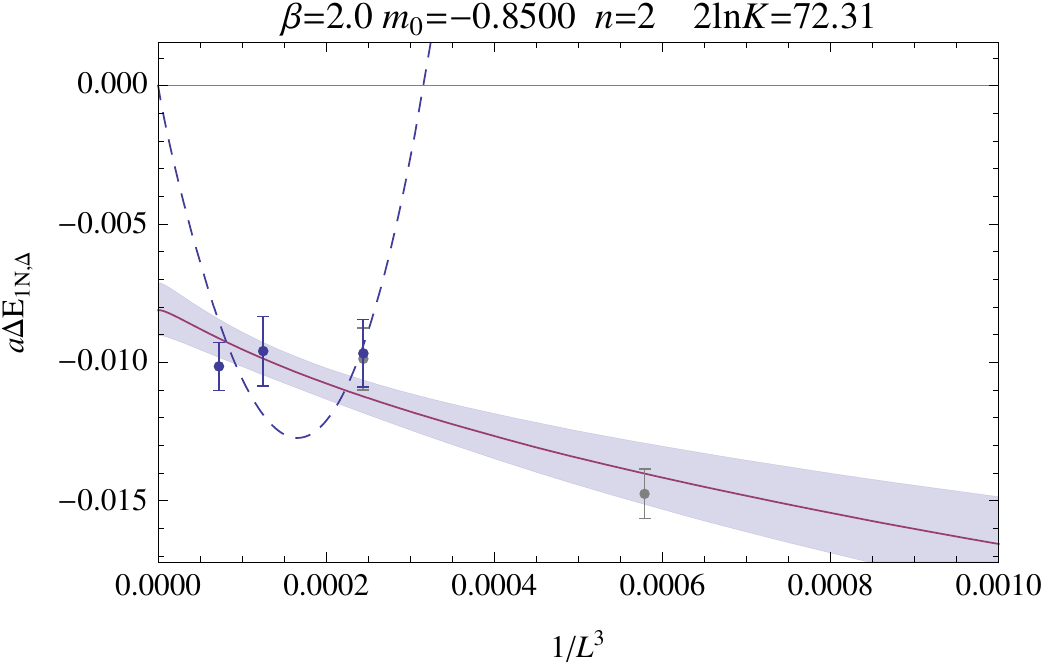}
  \\ \vspace*{4mm}
  \includegraphics[width=0.45\columnwidth]{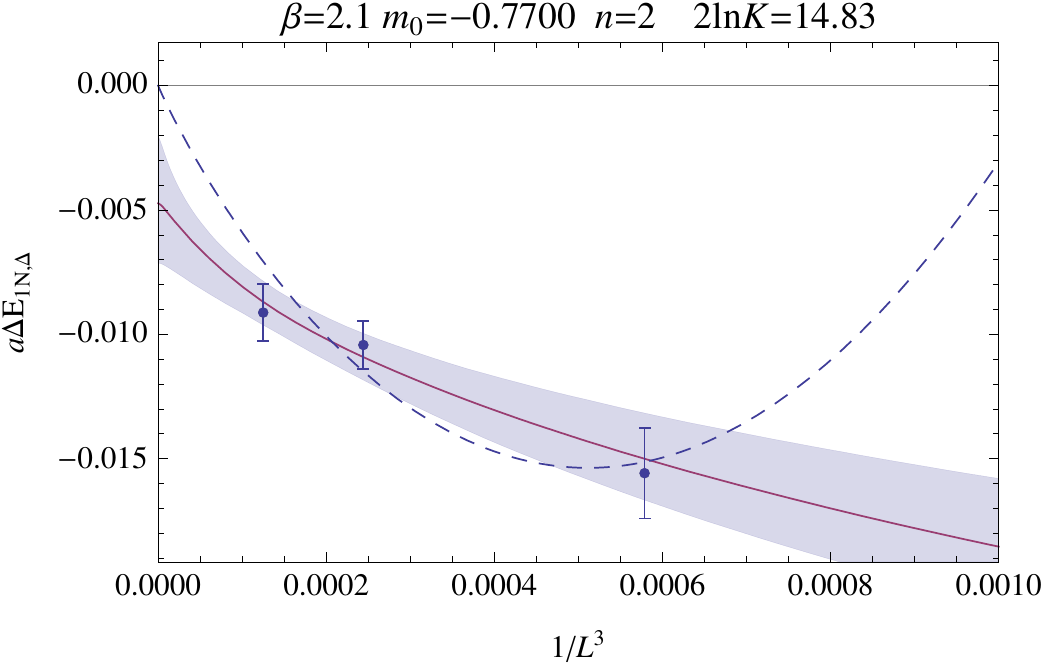}\qquad
  \includegraphics[width=0.45\columnwidth]{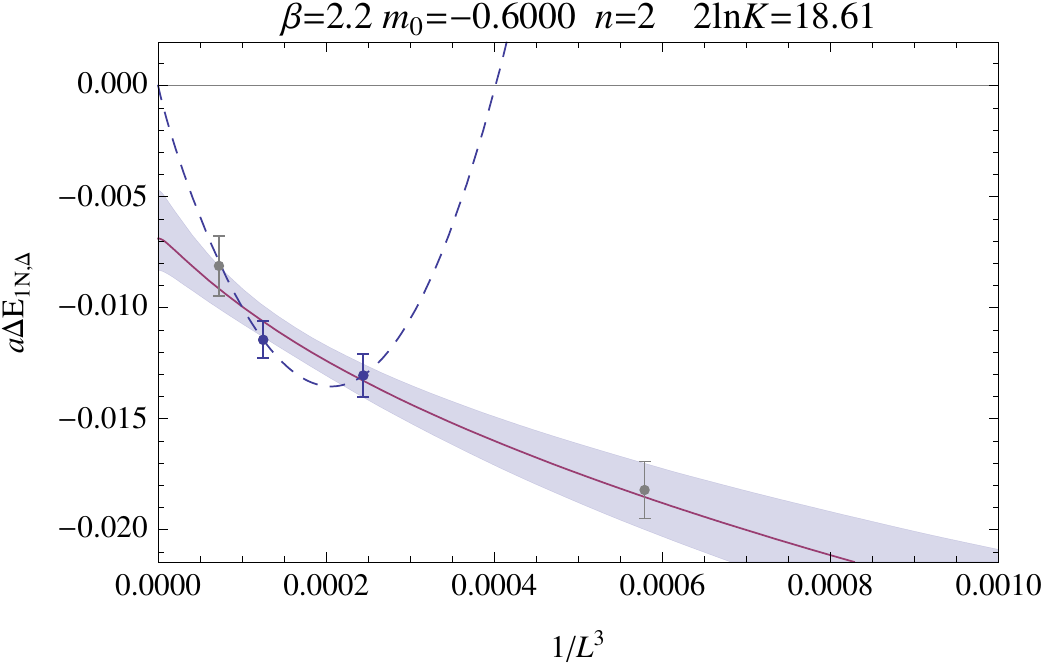}
  \caption{Infinite volume extrapolations of energy shifts for the
    $1N\Delta$ systems.}
  \label{fig:energyShiftExtrapolations2}
\end{figure}

\begin{figure}
  \includegraphics[width=0.45\columnwidth]{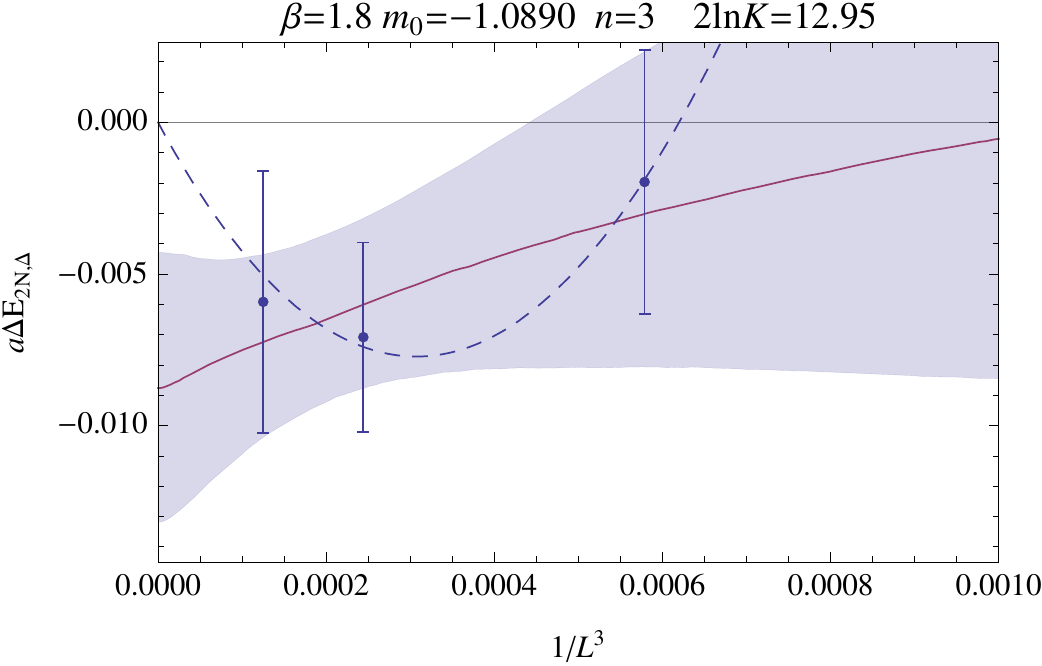}
  \qquad
  \includegraphics[width=0.45\columnwidth]{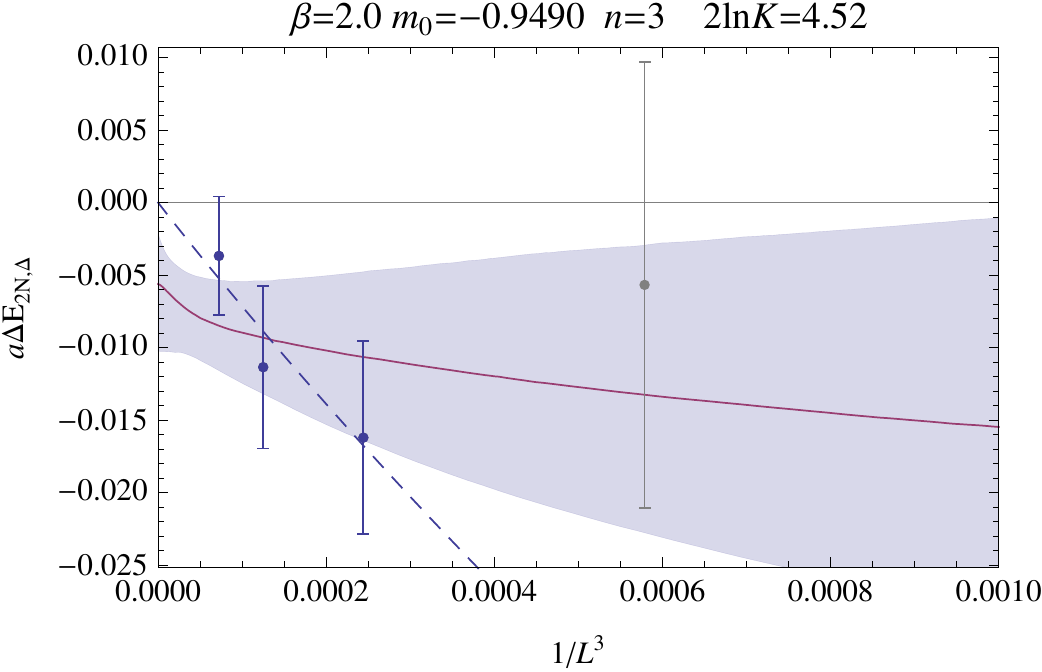}
  \\ \vspace*{4mm}
  \includegraphics[width=0.45\columnwidth]{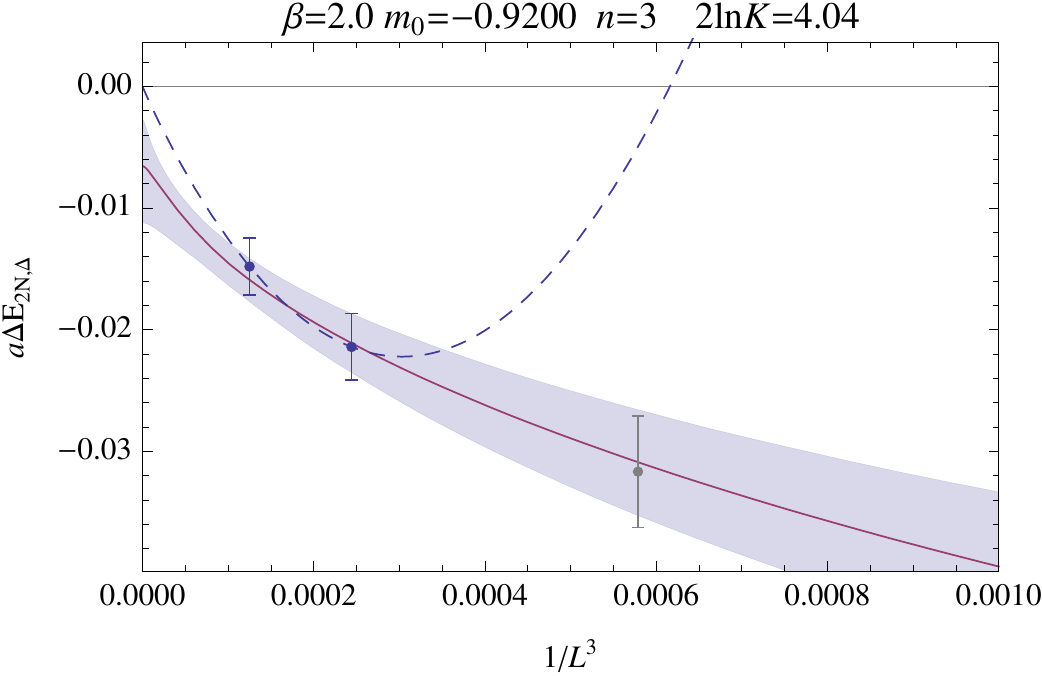}
  \qquad
  \includegraphics[width=0.45\columnwidth]{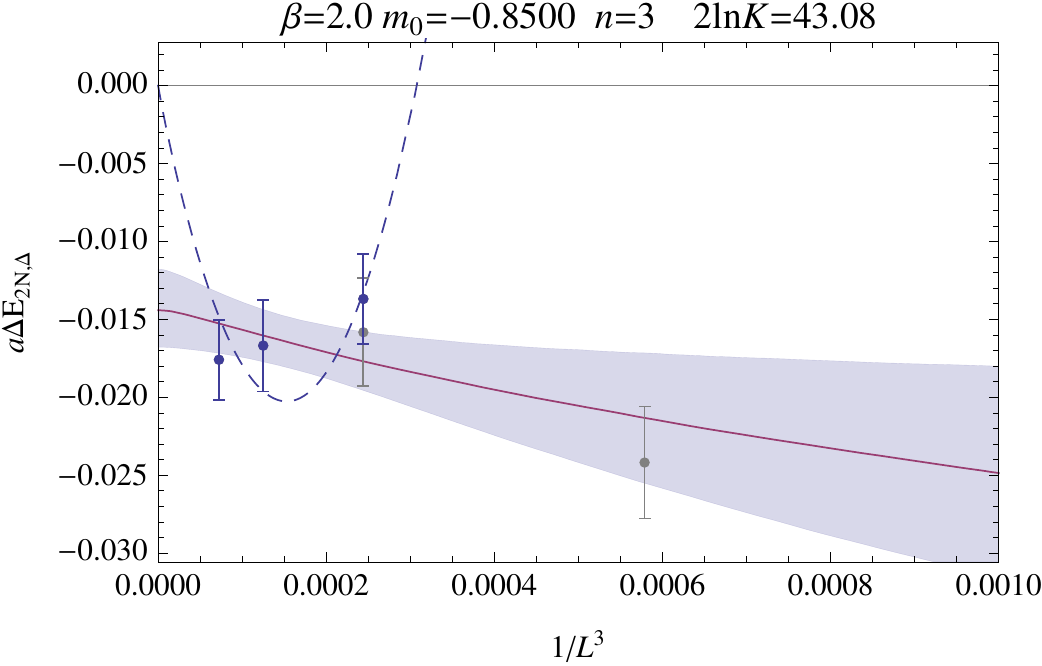}
  \\ \vspace*{4mm}
  \includegraphics[width=0.45\columnwidth]{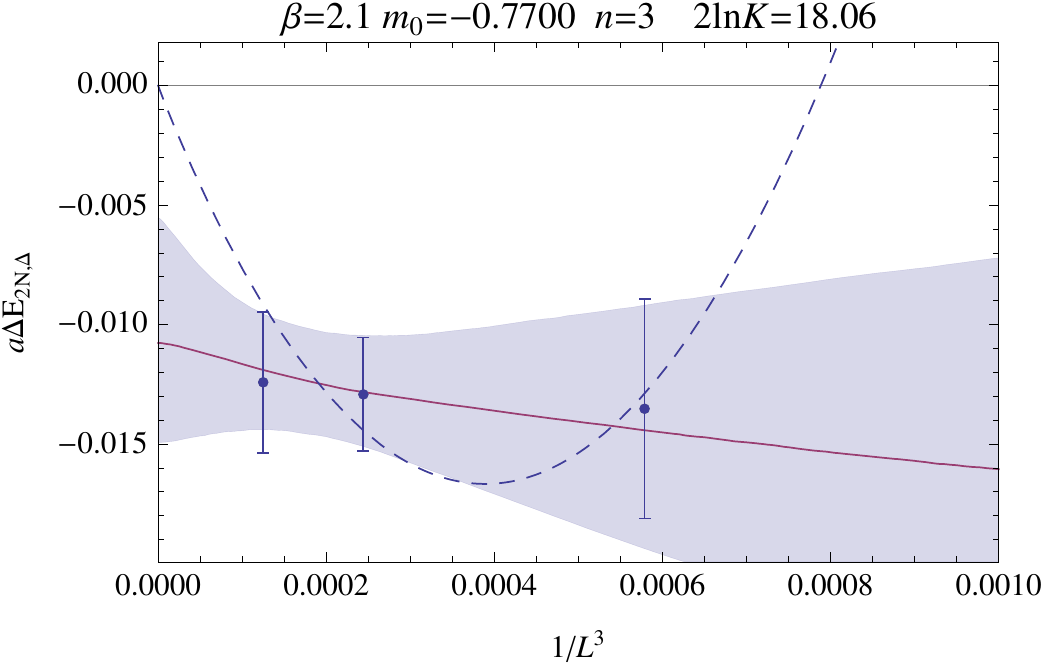}\qquad
  \includegraphics[width=0.45\columnwidth]{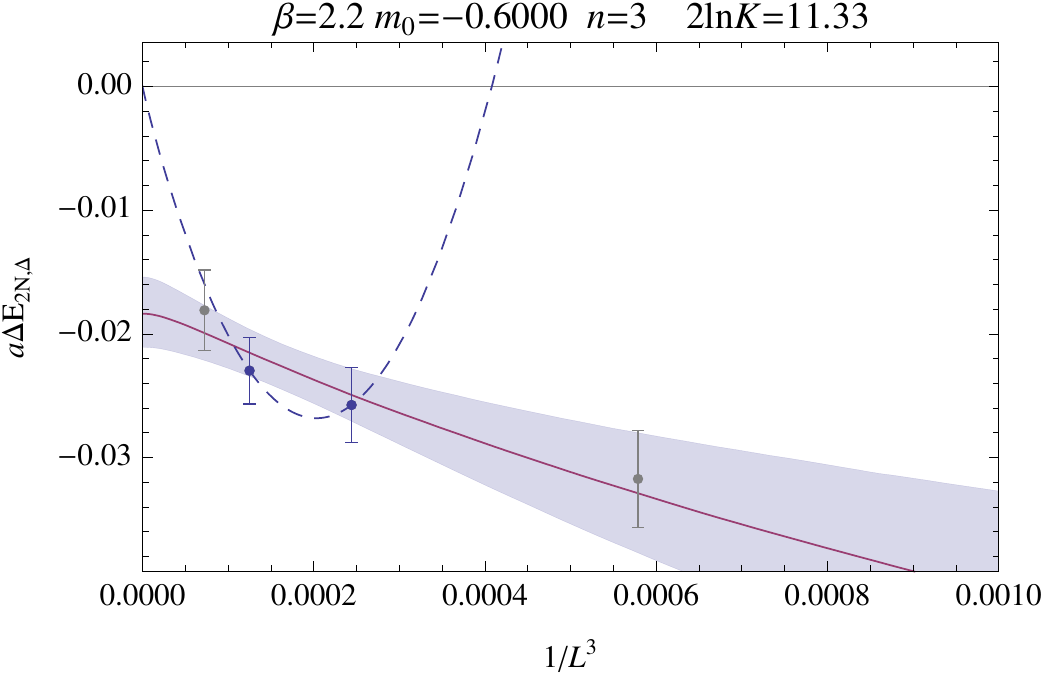}
  \caption{Infinite volume extrapolations of energy shifts for the
    $2N\Delta$ systems.}
  \label{fig:energyShiftExtrapolations3}
\end{figure}

\begin{figure}
  \includegraphics[width=0.45\columnwidth]{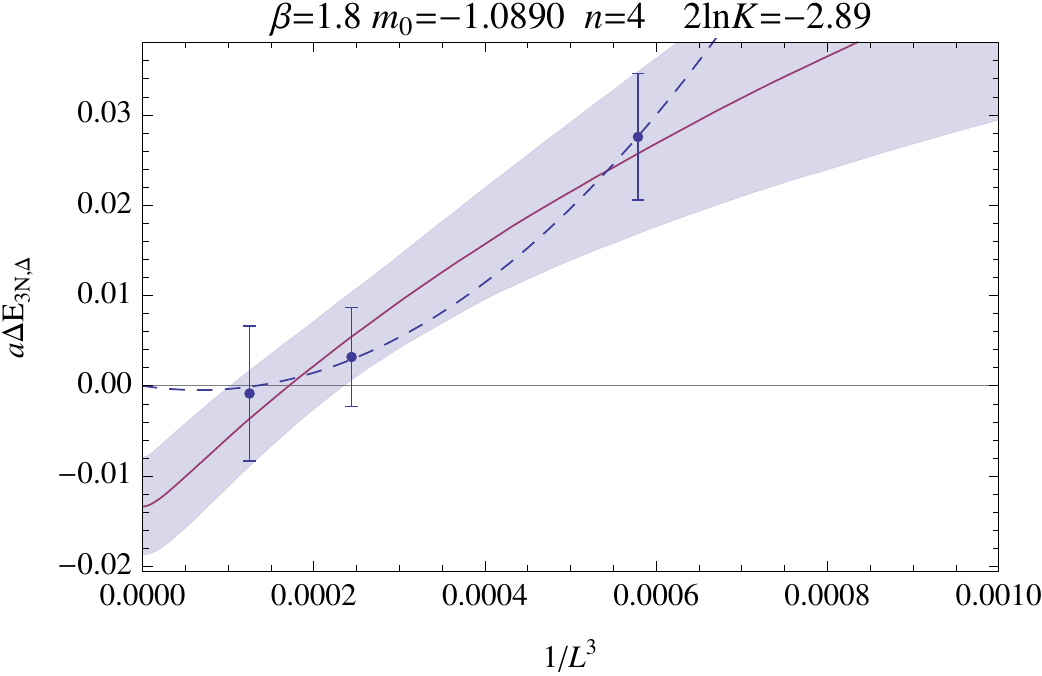}
  \qquad
  \includegraphics[width=0.45\columnwidth]{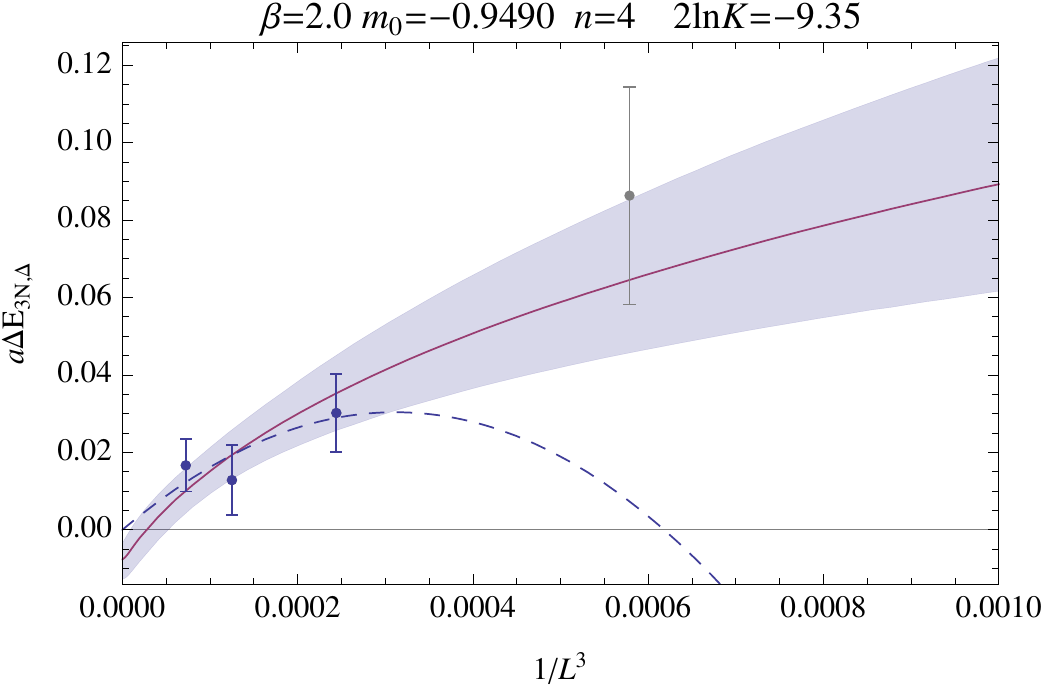}
  \\ \vspace*{4mm}
  \includegraphics[width=0.45\columnwidth]{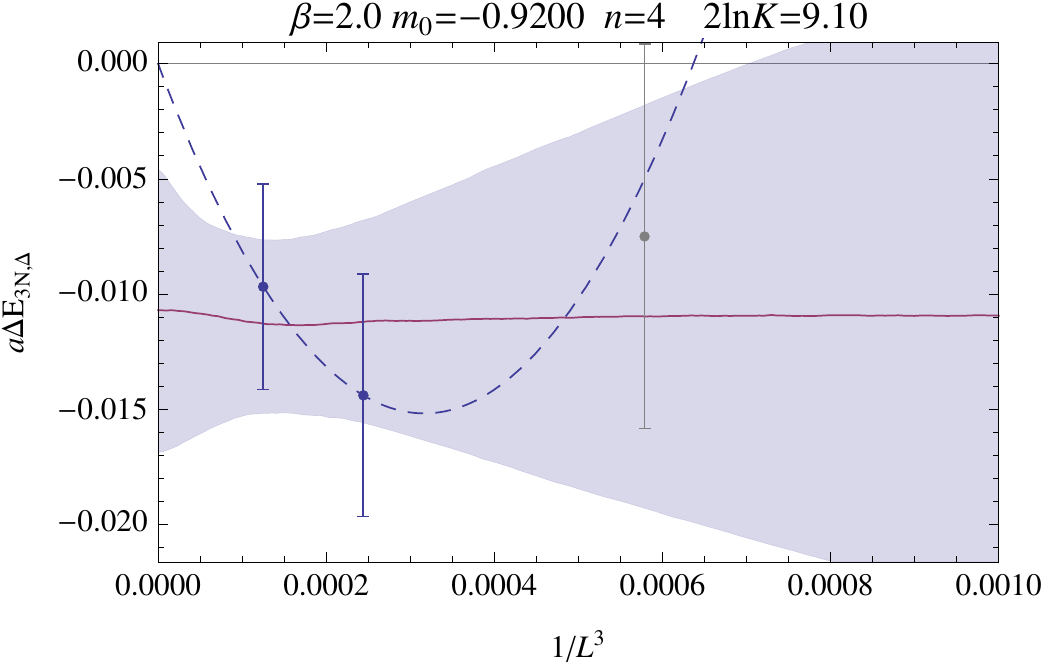}
  \qquad
  \includegraphics[width=0.45\columnwidth]{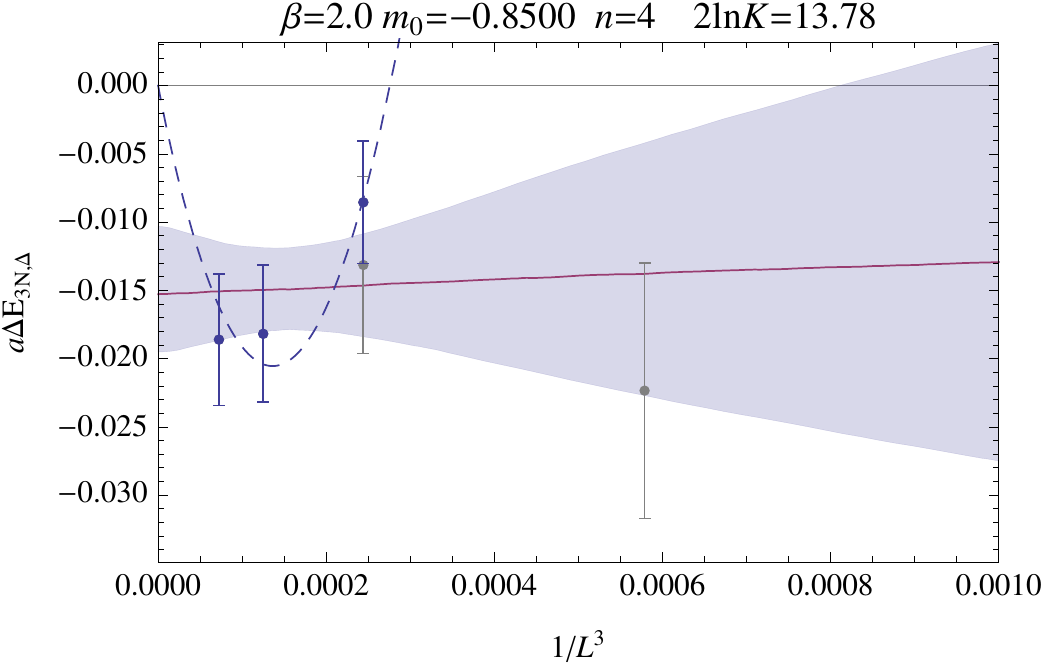}
  \\ \vspace*{4mm}
  \includegraphics[width=0.45\columnwidth]{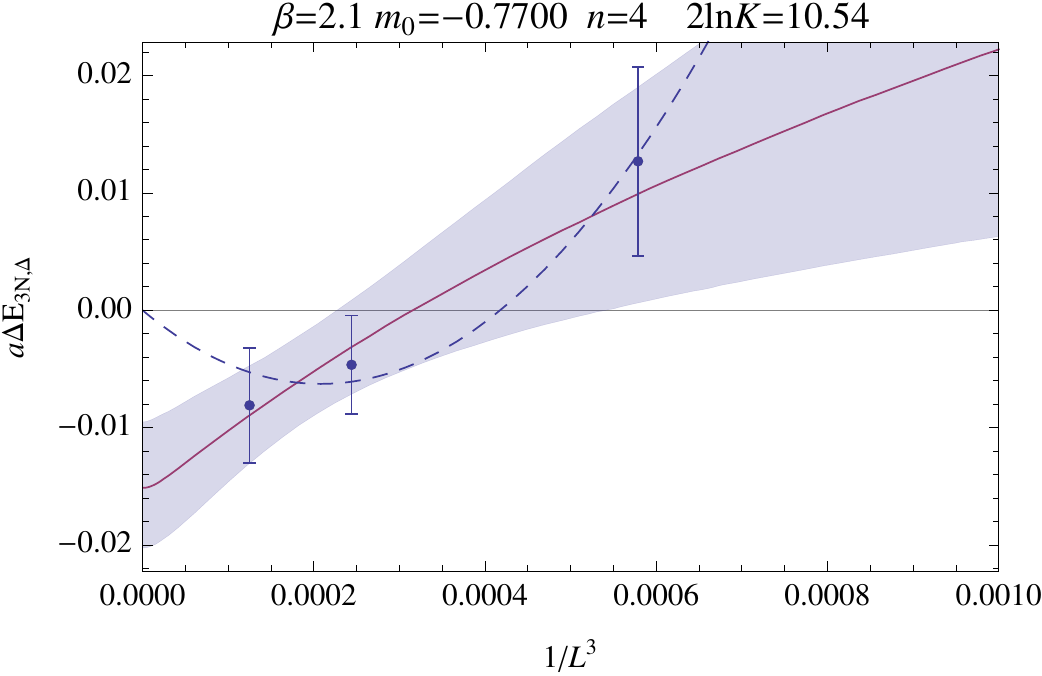}\qquad
  \includegraphics[width=0.45\columnwidth]{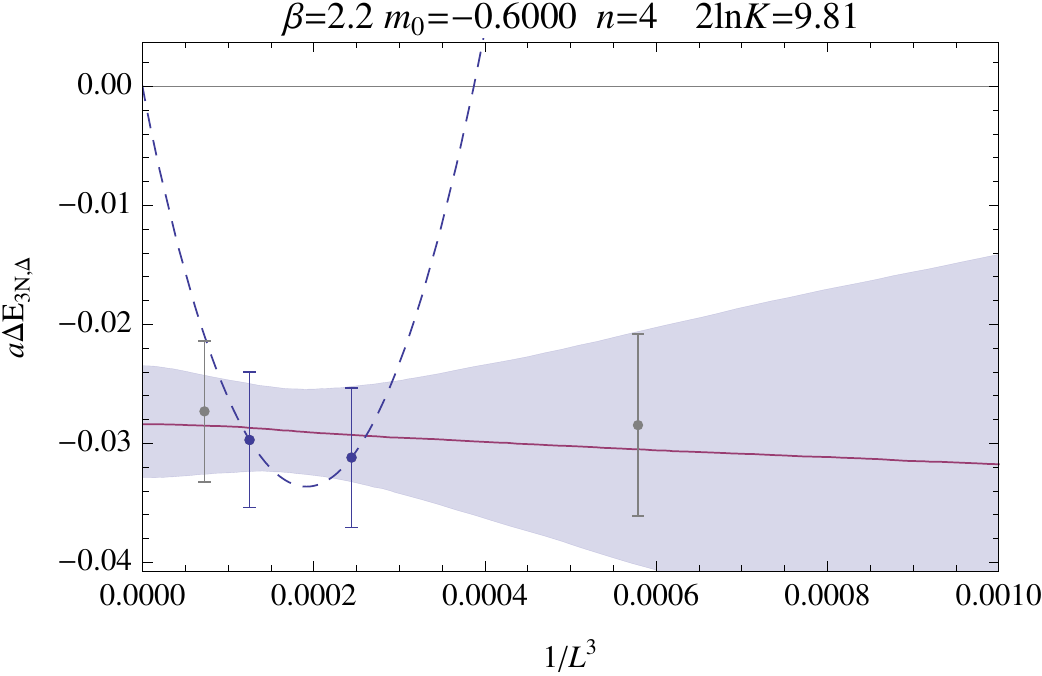}
  \caption{Infinite volume extrapolations of energy shifts for the
    $3N\Delta$ systems.}
  \label{fig:energyShiftExtrapolations4}
\end{figure}

\begin{figure}
  \includegraphics[width=0.45\columnwidth]{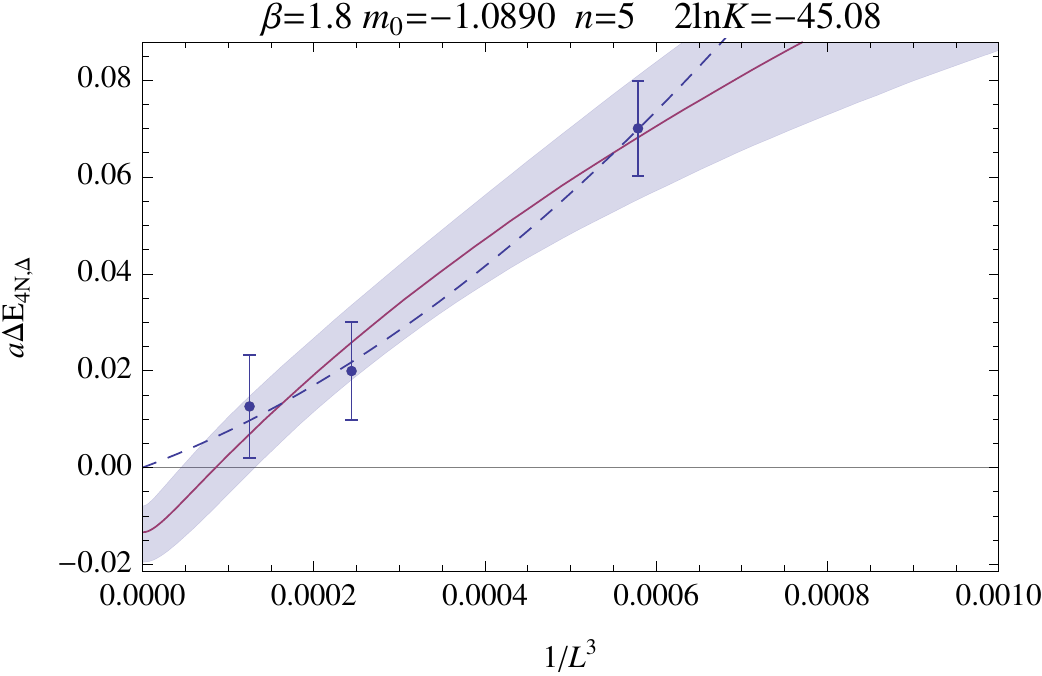}
  \qquad
  \includegraphics[width=0.45\columnwidth]{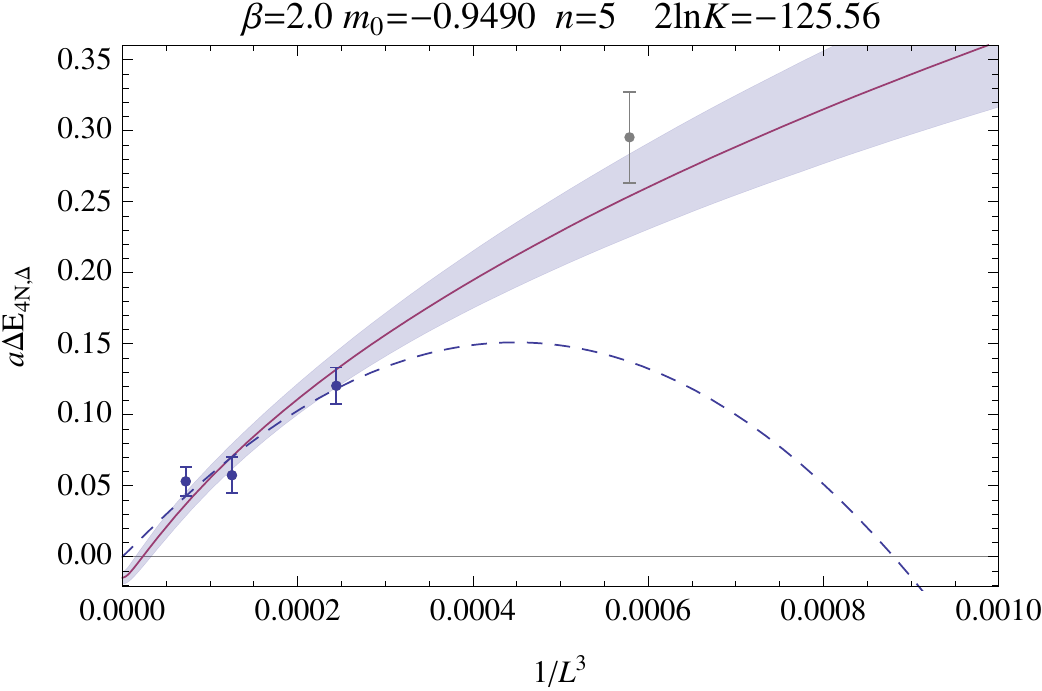}
  \\ \vspace*{4mm}
  \includegraphics[width=0.45\columnwidth]{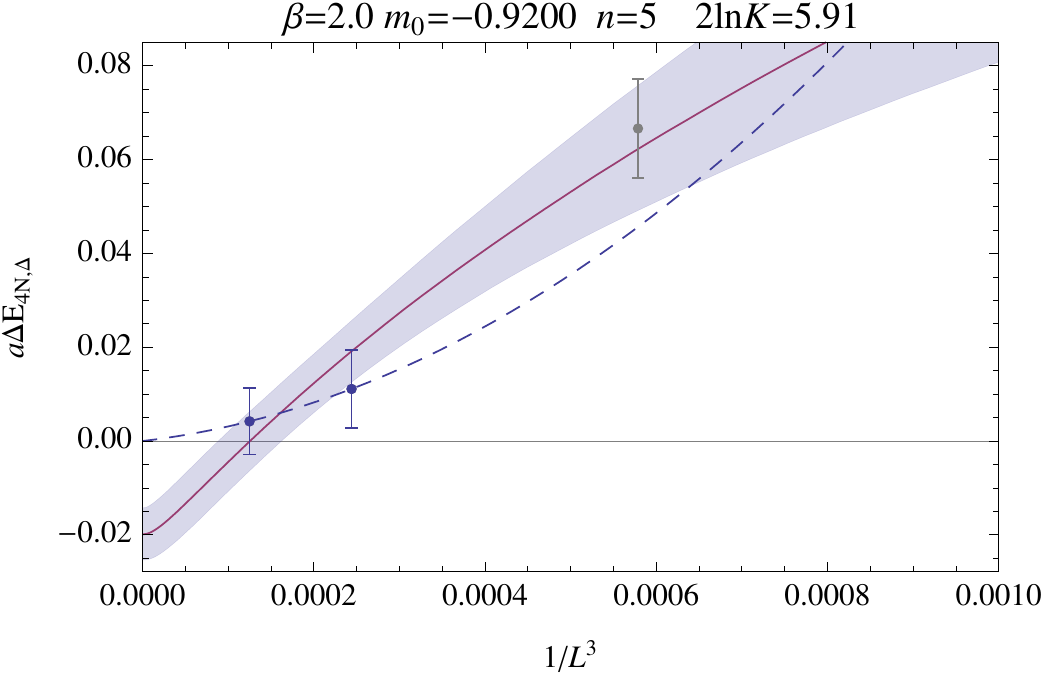}
  \qquad
  \includegraphics[width=0.45\columnwidth]{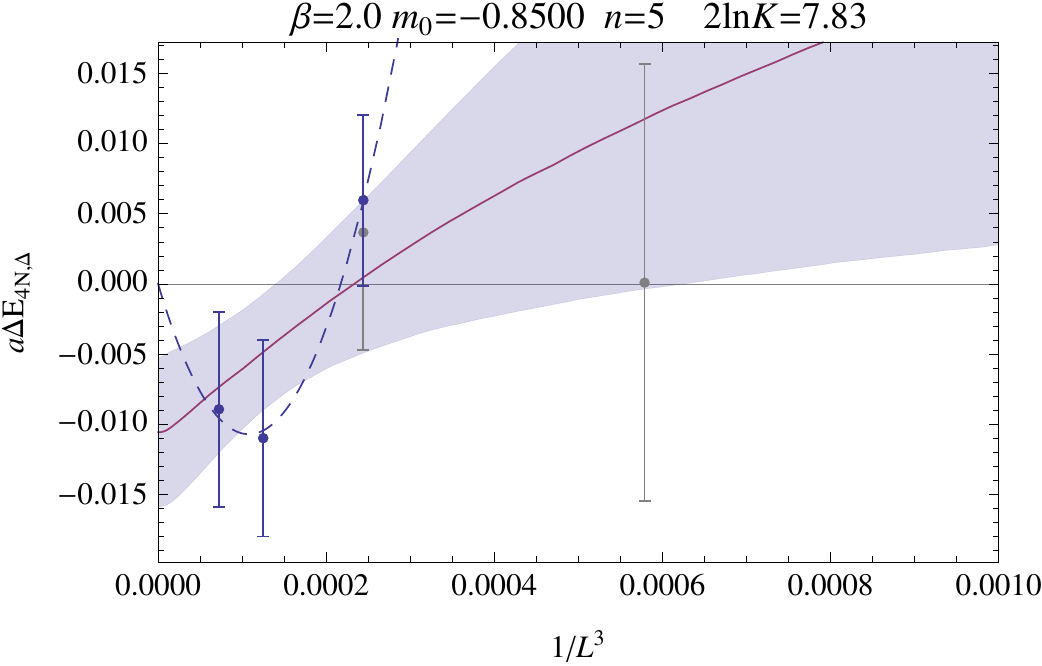}
  \\ \vspace*{4mm}
  \includegraphics[width=0.45\columnwidth]{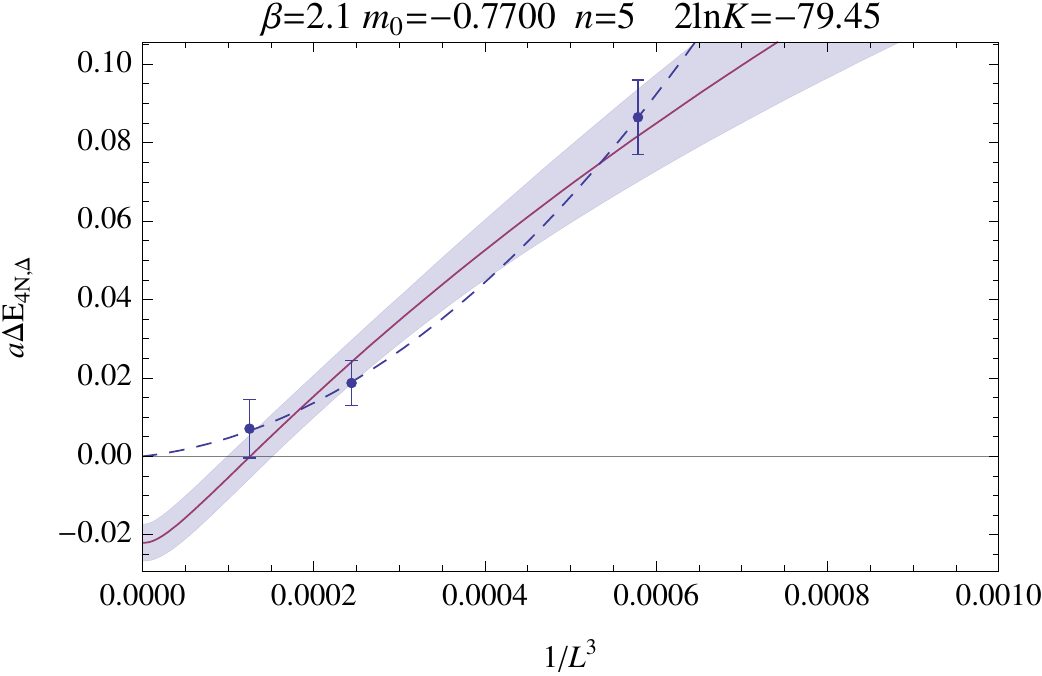}\qquad
  \includegraphics[width=0.45\columnwidth]{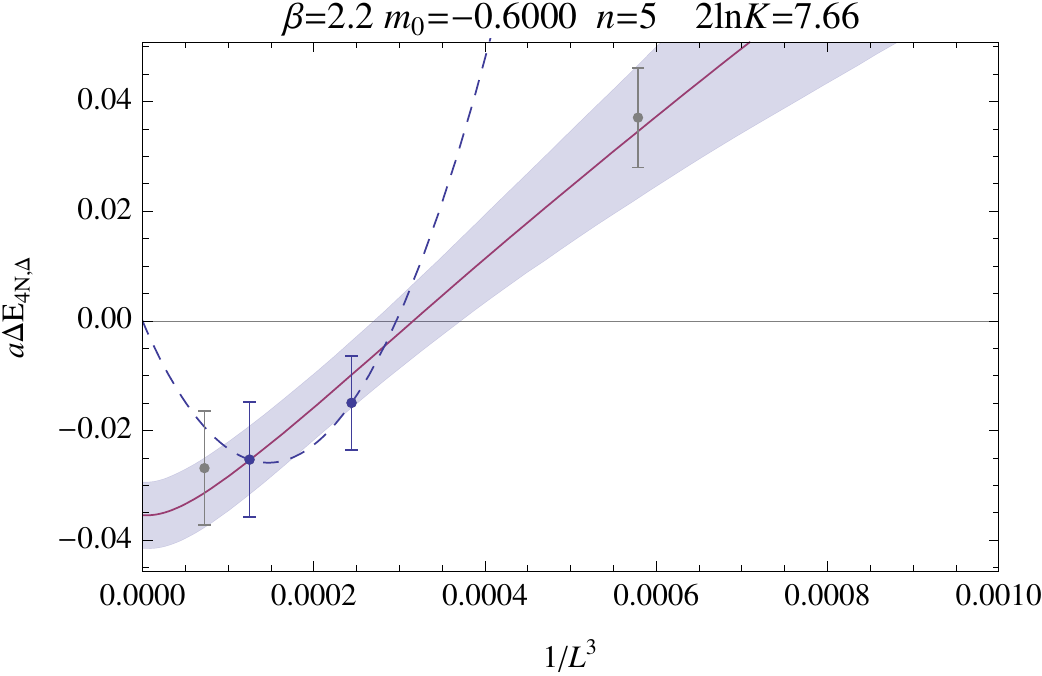}
  \caption{Infinite volume extrapolations of energy shifts for
    the $4N\Delta$ systems.}
  \label{fig:energyShiftExtrapolations5}
\end{figure}

For the $J=1^+$, $nN\Delta$ systems, we focus on cases where the
largest volume energy shift is negative and then compute the Bayes
factor to determine whether a bound state or an attractive scattering
state is preferred. The values of $2\ln[K]$ are shown in Table
\ref{ivrestab}; a value of $2\ln[K]>6$ is considered strong evidence
\cite{kass1995bayes} that hypothesis $H_1$ is preferred to $H_2$,
while $2\ln[K]>10$ is very strong evidence.  For states with a
positive value of $2\ln[K]$, we extract the 67\% credible interval on the
binding momentum, $\gamma$, and these values are also displayed in
Table \ref{ivrestab}.  In Figures
\ref{fig:energyShiftExtrapolations2}--\ref{fig:energyShiftExtrapolations5},
we show the resulting fits of the binding energies of the $(n-1)N\Delta$ 
systems for the various ensembles for $n=2,\ldots,5$. We show both the bound
state fit (solid line) and scattering fit (dashed line) and also
display the Bayes factor and the 67\% credible interval of the bound
state fit (the shaded region).  To assess systematics of these fits,
we remove the smallest volume ensembles from the analysis and
re-perform the fits. However there are only minor shifts in most cases
that are consistent with the extrapolation uncertainty.
One can speculate on causes of positive and negative values of $2\log
K$ listed in Table~\ref{ivrestab}.  It appears that data favour $H_2$
if either the system is unbound at both largest volumes, or if there
is only moderate curvature in the fit to the $H_2$ model as happens
for ($\beta=2.0$, $m_0=-0.9490$, $N=2$) ensemble. It is also worth
noting that for ($\beta=2.0$, $m_0=-0.9200$, $n=2$) we observe $2\log
K=0.21$, indicating close to even odds between $H_1$ and $H_2$. It it
is possible that more precise data would have lead to different
conclusions in both cases. The continuum limit fits discussed below also
indicate that the binding momenta are expected to be rather small on
these ensembles.
\begin{table}
  \begin{center}
    \begin{ruledtabular}
      \begin{tabular}{cccccc}
	Ensemble & $\beta$ & $m_0$ &  $B$ & $2\ln[K]$ & $a \gamma$ \\
        \hline
        $A$ &  1.8 & $-$1.0890 & 2 & 12.11 & 0.062(19) \\
        $B$ & 2.0 & $-$0.9490 & 2 & $-$6.58 & --- \\
        $C$ & 2.0 & $-$0.9200 & 2 & 0.21 & 0.046(12) \\
        $D$ & 2.0 & $-$0.8500 & 2 & 72.31 & 0.0881(52) \\
        $E$ & 2.1 & $-$0.7700 & 2 & 14.83 & 0.061(17) \\
        $F$ & 2.2 & $-$0.6000 & 2 & 18.61 & 0.079(11) \\
        \hline
        % 2NDelta
        $A$ &  1.8 & $-1.0890$ & 3 & 12.95 & 0.089(23) \\
        $B$ & 2.0 & $-0.9490$ & 3 & 4.52 & 0.049(18) \\
        $C$ & 2.0 & $-0.9200$ & 3 & 4.04 & 0.066(22) \\
        $D$ & 2.0 & $-0.8500$ & 3 & 43.08 & 0.117(10) \\
        $E$ &  2.1 & $-0.7700$ & 3 & 18.06 & 0.092(21) \\
        $F$ & 2.2 & $-0.6000$  & 3& 11.33 & 0.1284(100) \\
        \hline
        % 3NDelta
        $A$ &  1.8 & $-1.0890$ & 4 & $-$2.89 & --- \\
        $B$ & 2.0 & $-0.9490$ & 4 & $-$9.35 & --- \\
        $C$ & 2.0 & $-0.9200$ & 4 & 9.10 & 0.084(25) \\
        $D$ & 2.0 & $-0.8500$ & 4 & 13.78 & 0.121(19) \\
        $E$ &  2.1 & $-0.7700$ & 4 & 10.54 & 0.109(20) \\
        $F$ & 2.2 & $-0.6000$ & 4 & 9.81 & 0.16(13) \\
        \hline
        % 4NDelta
        $A$ &  1.8 & $-1.0890$ & 5 & $-$45.08 & --- \\
        $B$ & 2.0 & $-0.9490$ & 5 & $-$125.56 & --- \\
        $C$ & 2.0 & $-0.9200$ & 5 & 5.91 & 0.114(16) \\
        $D$ &  2.0 & $-0.8500$ & 5 & 7.83 & 0.101(26) \\
        $E$ &  2.1 & $-0.7700$ & 5 & $-$79.45 & --- \\
        $F$ &  2.2 & $-0.6000$ & 5 & 7.66 & 0.178(15) \\
      \end{tabular}
    \end{ruledtabular}
    \caption{The Bayes factor and extracted binding momenta for the
      fits to the $J^P=1^+$ nuclear states of baryon number
      $B=2,\ldots,5$. Dashes in the last column indicate cases where
      the model is likely not bound.}
    \label{ivrestab}
  \end{center}
\end{table}     

\begin{figure}
  \includegraphics[width=0.6\columnwidth]{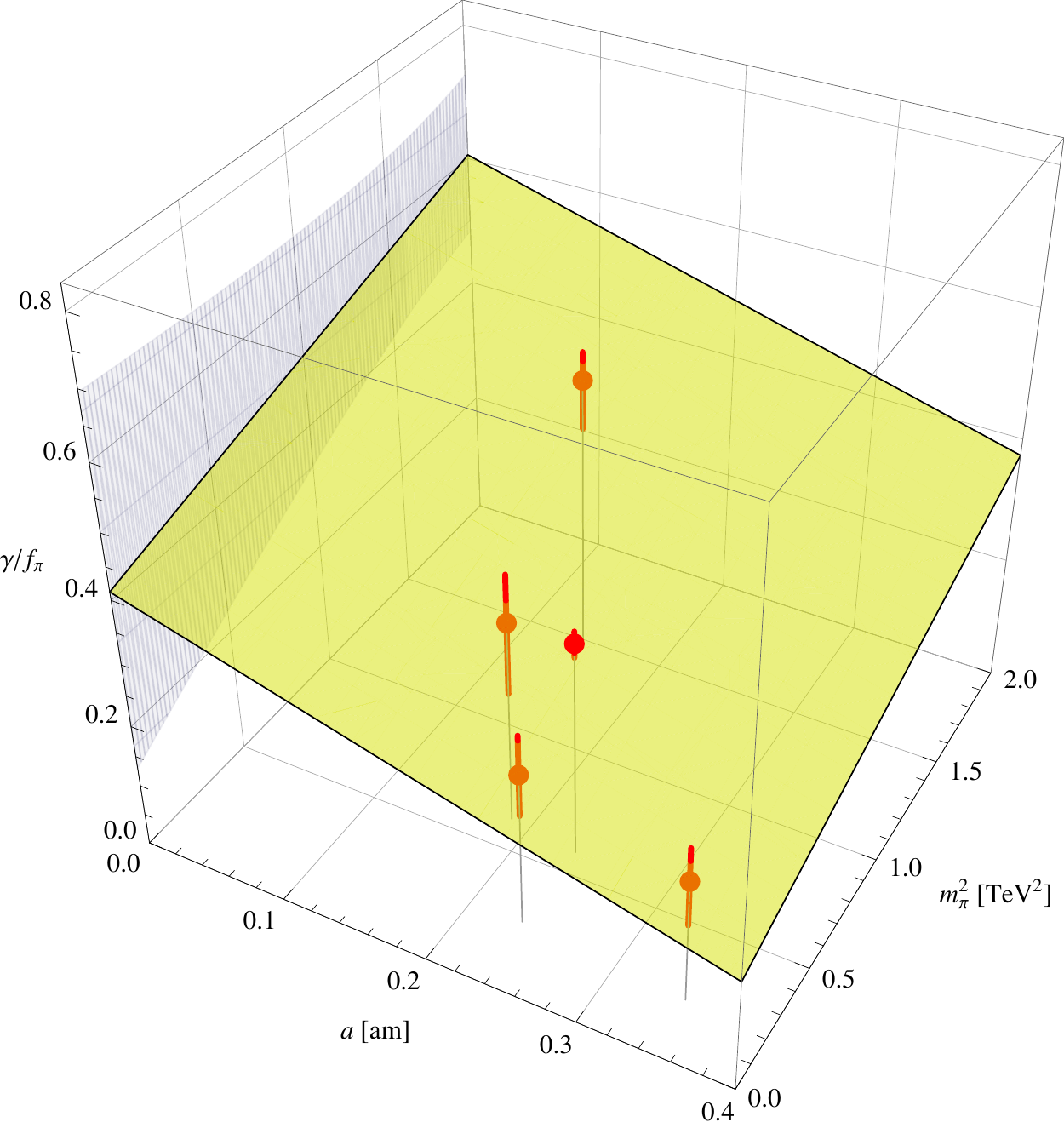}
  \caption{Continuum limit fit to the binding momentum of the $J^P=1^+$, $B=2$ nucleus
    as a function of $a$ (in attometers) and $m_\pi^2$ (in TeV${}^2$).
    The shaded region on the box wall corresponds to the uncertainty
    on the extrapolation.}
  \label{figJ1res2}
\end{figure}

\begin{figure}
  \includegraphics[width=0.6\columnwidth]{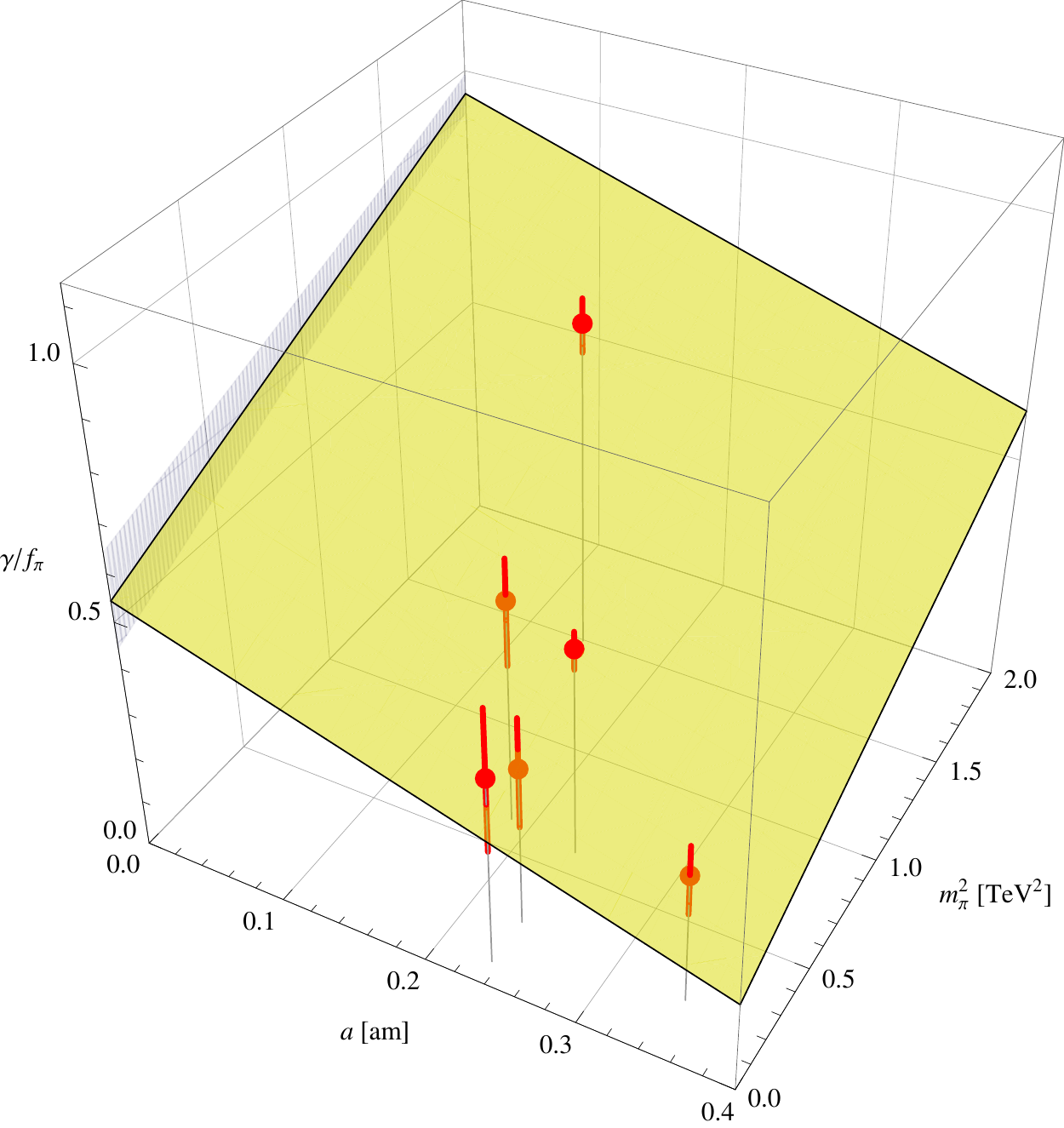}
  \caption{Continuum limit fit to the binding momentum of the $J^P=1^+$, $B=3$ nucleus
    as a function of $a$ (in attometers) and $m_\pi^2$ (in
    TeV${}^2$). The shaded region on the box wall corresponds to the
    uncertainty on the extrapolation.}
  \label{figJ1res3}
\end{figure}

\begin{figure}
  \includegraphics[width=0.6\columnwidth]{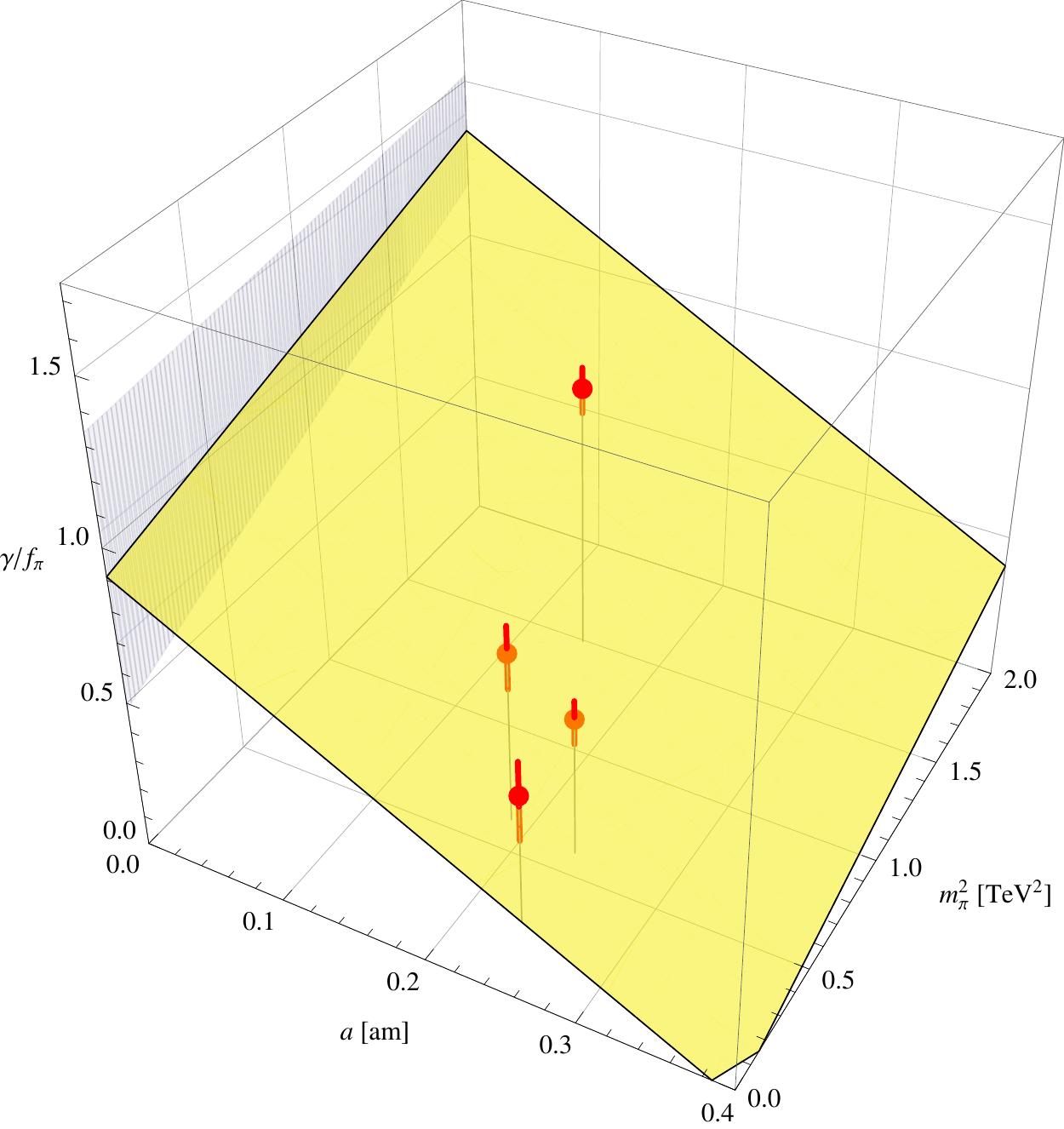}
  \caption{Continuum limit fit to the  binding momentum of the $J^P=1^+$ $B=4$ nucleus
    as a function of $a$ (in attometers) and $m_\pi^2$ (in
    TeV${}^2$). The shaded region on the box wall corresponds to the
    uncertainty on the extrapolation.}
  \label{figJ1res4}
\end{figure}

\begin{figure}
  \includegraphics[width=0.3\columnwidth]{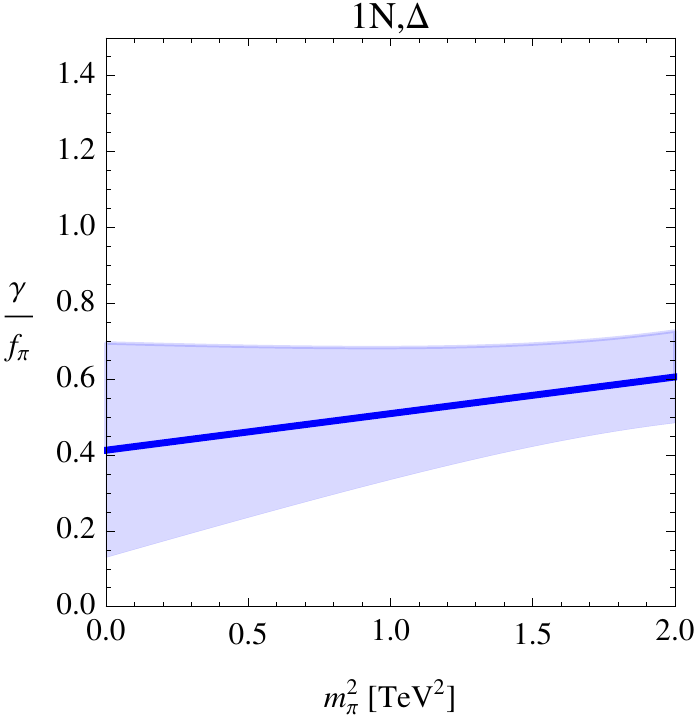}\qquad
  \includegraphics[width=0.3\columnwidth]{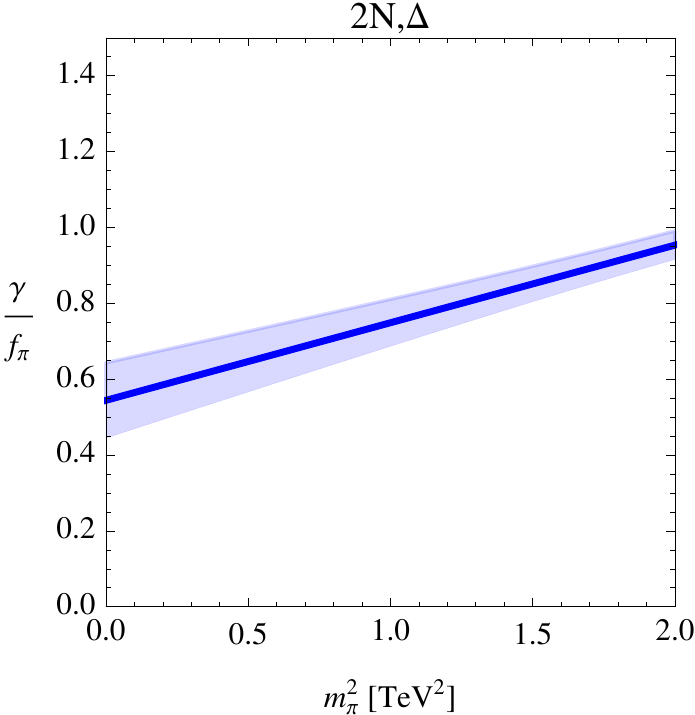}\qquad
  \includegraphics[width=0.3\columnwidth]{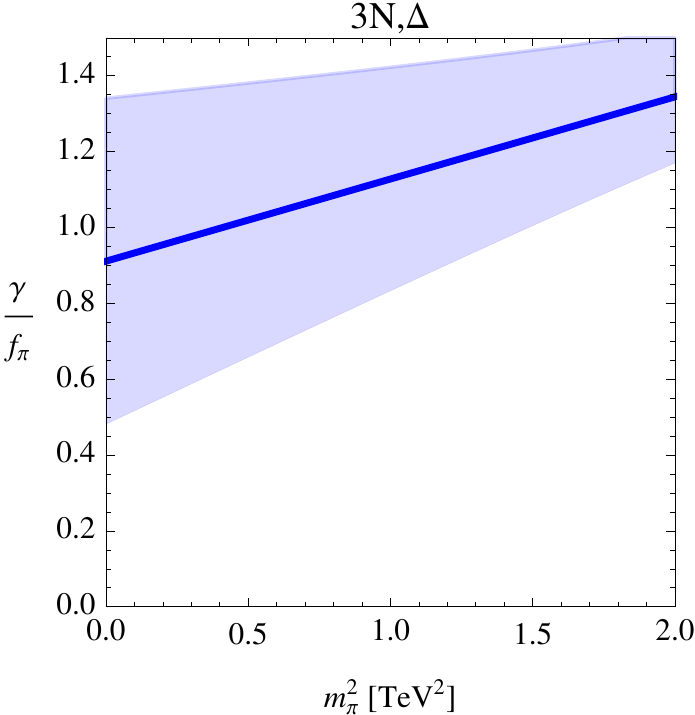}
  \caption{Behaviour of the extracted binding momenta of $J^P=1^+$, $B=2,3,4$
    nuclei as a function of $m_\pi^2$. The shaded regions correspond
    to the uncertainties on the fitted results.}
  \label{figJ1All}
\end{figure}

Having extracted the binding energies of these states on each
ensemble, we can investigate the continuum limit by comparing the
various ensembles.  We focus on the $B=2,3,4$, $J^P=1^+$ states ($B>4$
states are very likely unbound --- the scattering state fit is
preferred on most sets of ensembles) and assume a simple functional
form for the dependence of the binding momenta on the lattice spacing
and pion mass,
 \be
\label{contextrap}
\frac{\gamma_{nN,\Delta}}{f_\pi}(a, m) = \gamma^{(0)}_{nN,\Delta} + a\
\delta^{(a)}_{n} + {m_\pi}^2 \delta^{(m)}_n \,.
\ee 
The infinite
volume extrapolated binding momenta are fitted with this form using
least-squares minimisation (additional fits involving higher order
terms have also been investigated but were not well-constrained).  Note that the 
parameters $\delta_{n}^{(a,m}$ are dimensionful. We
find that the $B=2$ and $3$ states are clearly bound relative to
$(B-1)M_N+M_\Delta$ for a significant range of quark masses with the
binding momenta tending to decrease with the quark mass. For the
$B=3$ state at heavier masses, the significance of the binding is
particularly high.  These bound states are protected against decay
into $B$ nucleons by the combination of baryon number and baryonic
equivalent of G-parity (these nuclei are partners of the $(B-1)\pi +
\rho$ systems which differ in G-parity from $B\pi$ systems). At the
current level of statistics, we cannot cleanly determine if the $B=4$
state is bound or not in the continuum limit as, unlike $B=2,3$, the
$B=4$ extrapolation is very sensitive to removing a single data
point.\footnote{Note that for the higher body systems to be identified
  as bound states, they must also have energies that stabilise them
  against break up into any combination of sub-components, for example
  we would require $E_{3N,\Delta}< {\rm min}(E_{3N}+E_{0N,\Delta},
  E_{2N}+E_{1N,\Delta},E_{1N}+E_{2N,\Delta}$,\ldots.}  The $B=2,3,4$
$J^P=1^+$ fits, along with the projections to the continuum limit (in
which the $\delta^{(a)}_{n}$ are set to zero), are shown in
Fig.~\ref{figJ1res2} -- \ref{figJ1res4} as a function of $m_\pi$ and
$a$. We present the results using physical units, attometers for the
lattice spacing and TeV for the pion mass (these arise from the
arbitrary choice of $f_\pi=246$ GeV). For clarity, we again show the
continuum limit fits as functions of $m_\pi$ in Fig.~\ref{figJ1All}.
The central conclusion of our study is that multiple few-body bound
states appear for the range of quark masses investigated here.

\begin{figure}
  \includegraphics[width=0.9\columnwidth]{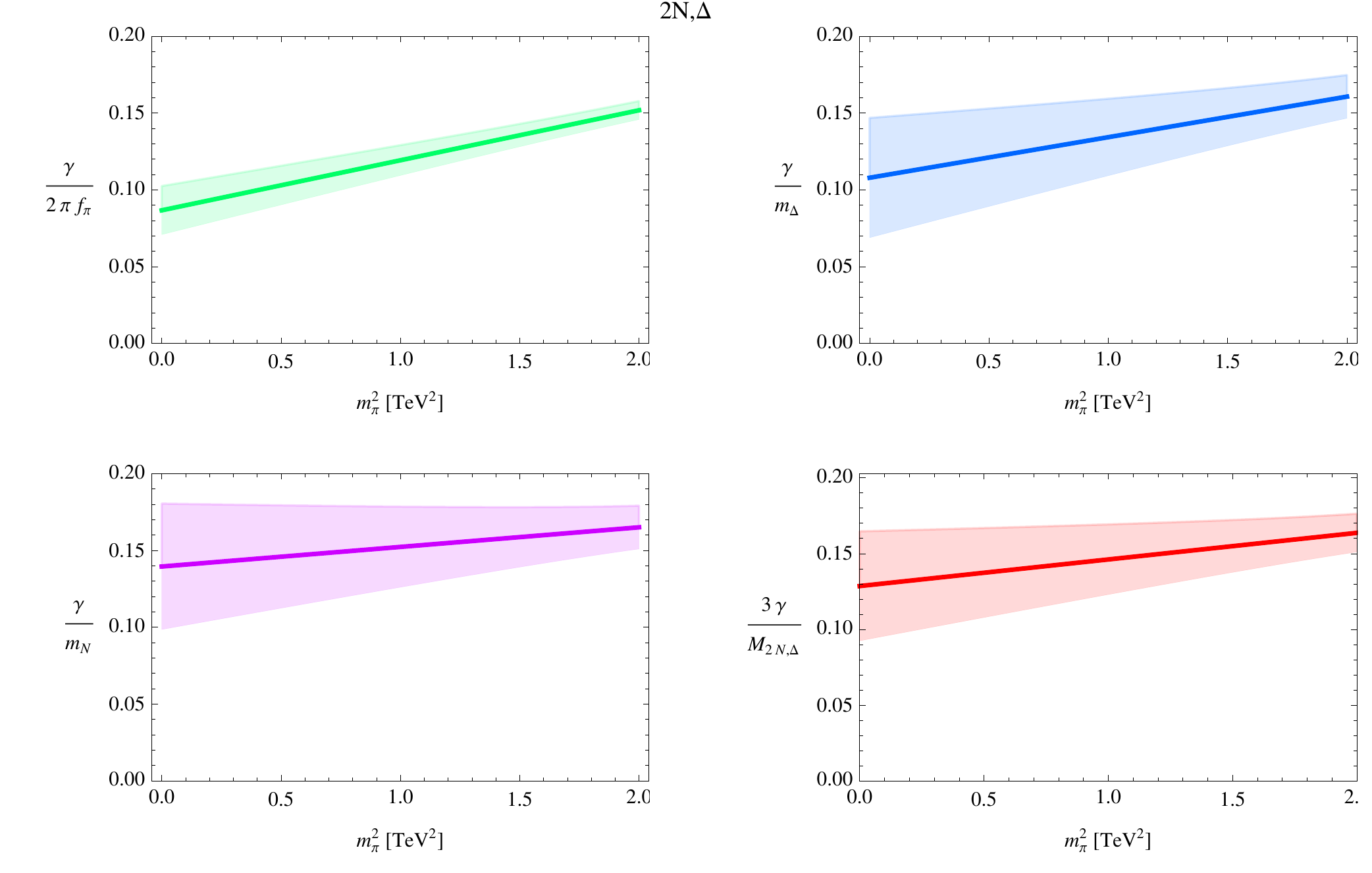}
  \caption{Behaviour of the binding momentum of the $J^P=1^+$, $B=3$
    nucleus as a function of $m_\pi^2$ for different choice of
    normalisation: $2\pi f_\pi$, $m_N$, $m_\Delta$,
    $\frac{1}{3}(2m_N+m_\Delta)$.  The shaded regions correspond to
    the uncertainties on the results.}
  \label{fig:extrapChoice}
\end{figure}

The choice of units in which to fit the binding momenta  has some
influence on the extrapolation because of the correlation between
measurements of the different quantities that enter the fit. 
The continuum limit results from
using four different alternative normalisations on the left hand 
side of Eq.~(\ref{contextrap})  are shown for the
$2N\Delta$ system in Fig.~\ref{fig:extrapChoice}; all are consistent
within their uncertainties, but the size of uncertainty varies.

\begin{figure}
  \includegraphics[width=0.7\columnwidth]{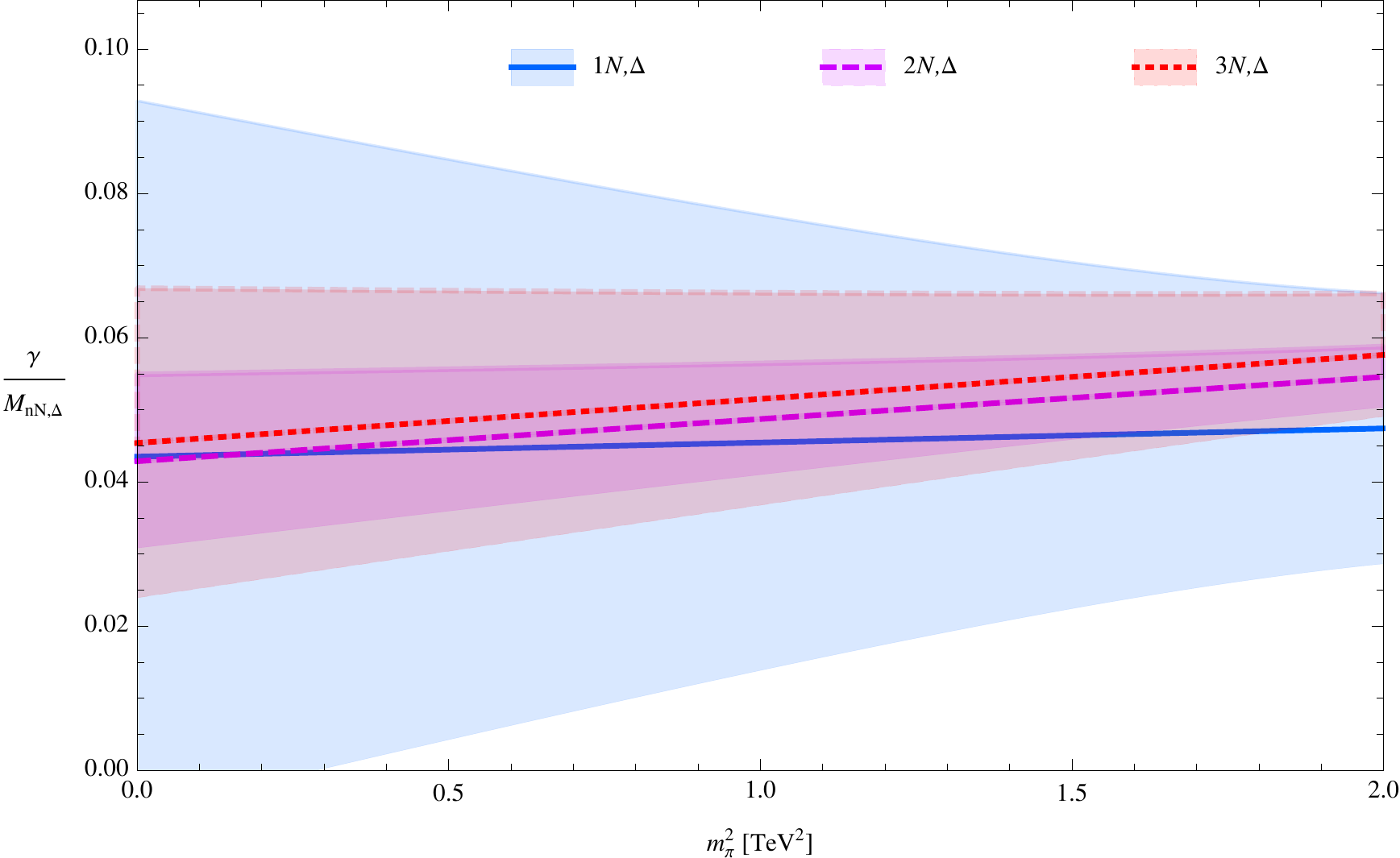}
  \caption{Binding momenta in units of the
    nuclear mass of the various $J^P=1^+$ nuclei as a function of
    $m_\pi^2$.  The shaded regions correspond to the uncertainties.}
  \label{fig:bindperBar}
\end{figure}

It is interesting to compare the binding momenta for differing
nuclei.  Figure \ref{fig:bindperBar} show the ratio of the binding
momentum (as defined through the fit) to the rest mass of the nucleus
for the $1N\Delta$, $2N\Delta$, and $3N\Delta$ systems.  Intriguingly,
this quantity appears to be insensitive to the baryon number for the
systems we have studied, although the uncertainties are too large to make 
definitive statements. 
%This is quite distinct from the behaviour 
%observed for light nuclei in the real world.
Despite this fact, the apparent absence of $B>4$ systems may suggest
that while the $N\Delta$ two-body interaction is attractive, the
$NN\Delta$ three-body interaction is repulsive and eventually
overcomes the two-body attraction as the number of $N$'s increases. It
is also possible that our interpolating operators are not sufficiently
close to the ground state eigenstates for larger $B$ but instead overlap
more strongly onto scattering states. More sophisticated choices of interpolating 
operators may be necessary to identify the bound states if they are present.

\subsection{Other nuclei}

In our investigations, we have focused on the $J=0,1$ systems of
the highest possible flavour symmetry. It is possible that $J\geq2$
systems, or states in other flavour representations could also be
bound nuclei. However, investigating this is beyond our current
scope. While a complete investigation of the nuclear spectrum of this
theory is far beyond the needs of current dark matter phenomenology, a
number of interesting questions could be investigated in this
direction.

\section{Nuclear properties and processes }
\label{sec:properties}
A quantitative understanding of the spectrum of nuclear states
provides important input into dark sector phenomenology based on this
model. However, this is by no means the only useful information that
can be extracted from lattice field theory calculations, and in this
section, we discuss further possibilities.  While we will not pursue
calculations of most of these properties in this initial study, they
can be investigated in the future if there are strong phenomenological
motivations.

\subsection{Scalar couplings}
The couplings of dark sector hadrons or nuclei to scalar currents can
be extracted from the quark mass variation of the masses of the
hadronic or nuclear states, making use of the Feynman-Hellman
theorem. For the case of single baryons, this approach has been used
to extract the relevant light- and strange-quark $\sigma$-terms in QCD
(see Ref.~\cite{Junnarkar:2013ac} for a recent overview), but also in
the dark matter context for SU($N_c=4$) baryons in
Ref.~\cite{Appelquist:2014dja}.  Following the standard
parameterisation of these quantities, we define the dimensionless,
renormalisation-scale invariant quantity
\begin{equation}
  \label{fsimga}
  f_{q}^{(H)} = \frac{\langle H | m_q \overline{q}q|H\rangle}{M_H}\,,
\end{equation}
for a quark flavour, $q$, and a hadron, $H$.  Through the
Feynman-Hellman theorem, this can be recast as
\begin{equation}
  \label{fsigma2}
  f_{q}^{(H)} = \frac{m_q}{M_H}\frac{\partial M_H}{\partial m_q}\,,
\end{equation}
which can then be evaluated by using lattice calculations of hadron
energies over a range of quark masses.

Ideally, precise calculations for many closely spaced quark masses,
volumes and lattice spacings would be performed, but this would be a
very computationally demanding task. Instead, we shall perform a less
intensive calculation and aim to understand the typical size of these
couplings rather than precise values.  To do so, we focus on a single
representative set of gauge configurations, the $16^3\times48$ $C$
ensemble, and perform partially-quenched measurements of the hadron
masses for many values of the valence quark mass around the single sea
quark mass and assume that the partial quenching effects are
small.\footnote{If the quark masses were light enough such that chiral
  perturbation theory were a controlled expansion, these partially
  quenched lattice calculations would determine a subset of the low
  energy constants of partially quenched chiral perturbation theory
  that govern the $\sigma$-terms.}  The PCAC quark mass for this
ensemble is $a m_q=0.0823(4)$ and we use valence masses $a m_{v} =
0.07,0.072,\ldots, 0.09$.  An important advantage of this approach is
that there are strong correlations between the measurements of the
hadron masses for the various valence quark masses, allowing for
precise estimates of the differences with considerably smaller
statistical sample sizes than would be needed if we were using
independent ensembles for each mass.

\begin{figure}
  \includegraphics[width=0.7\columnwidth]{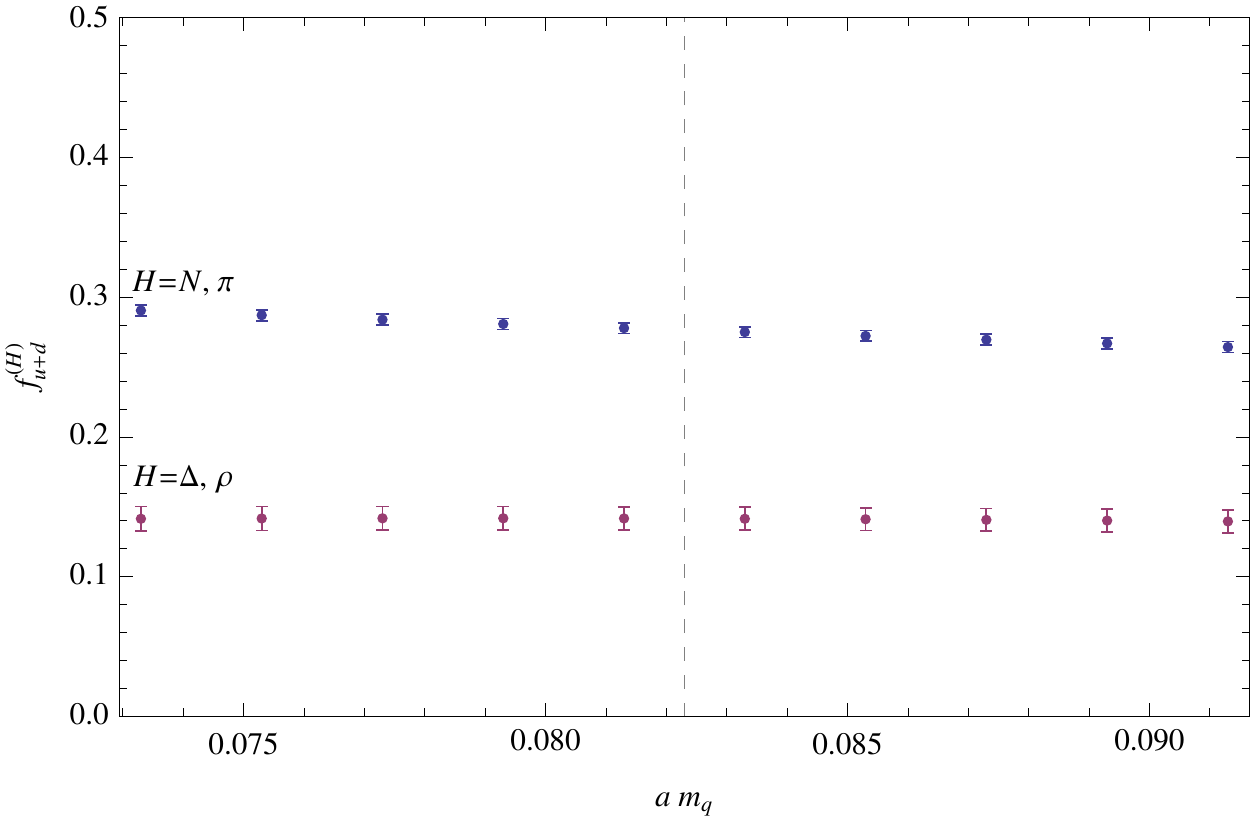}
  \caption{The numerical extractions of the quantities $f_{u+d}^{(N)}$
    and $f_{u+d}^{(\Delta)}$ as a function of the valence quark
    mass. The sea quark mass is indicated by the dashed line. }
  \label{fig:sigma}
\end{figure}

In Fig.~\ref{fig:sigma}, the extracted values of the quantities
$f_{u+d}^{(N)}$ and $f_{u+d}^{(\Delta)}$ are shown as a function of
the valence quark mass, using the finite difference approximation
$\frac{\partial M_H}{\partial m_q}\to [M_H(m_q)- M_H(m_q-\delta
m_q)]/\delta m_q$.  The extracted values of the couplings at the
unitary point are
\begin{equation}
  \label{fsigma3}
  f_{u+d}^{(N)} =f_{u+d}^{(\pi)} =0.276(4),\qquad
  f_{u+d}^{(\Delta)} =f_{u+d}^{(\rho)} =0.14(1),\qquad
\end{equation}
where only statistical uncertainties are shown. These values are
consistent with the expectations of naive dimensional analysis. 
As discussed above,
these values are only estimates and are subject to uncertainties from
the effects of partial quenching (and also from discretisation and
finite volume effects) which we estimate to be ${\cal O}(30\%)$. The
values of the couplings will also depend on the quark mass in a
non-trivial way.

For $N_c=3$ light nuclei, the nuclear $\sigma$-terms (scalar current
matrix elements in a nucleus) have also recently been studied for the
first time \cite{Beane:2013kca}. Because of correlated two- and
higher-body interactions, nuclear $\sigma$-terms will
differ from the sum of the $\sigma$-terms of their constituents, but
such effects were seen to be small in Ref.~\cite{Beane:2013kca}. 
Nuclear effects may be larger for $N_c=2$, but we leave such calculations 
for future work.

\subsection{Electroweak-analog interactions}
The couplings of single hadrons and tightly bound nuclei to additional
weakly-coupled gauge sectors through quark bilinear operators can be
straightforwardly determined using the same methods by which hadron
form factors \cite{Hagler:2009ni} and polarisabilities
\cite{Tiburzi:2011vk} are studied in QCD. In the current context, the
two-colour quarks could be charged under a U(1) symmetry, resulting in
either charged nuclei (depending on the U(1) that is gauged), or nuclei
whose internal structure gives rise to higher multiple moment
couplings, or higher order couplings (polarisabilities), to the U(1) interactions.

\subsection{Nuclear interactions}
Despite the Euclidean space formulation inherent in numerical studies
of lattice field theory, scattering processes below inelastic
thresholds can be investigated using lattice methods, as can two-body
decays/fusions induced by non-strong interaction dynamics (the
analogues of the weak current processes, $np\to d\gamma$ or $\nu d\to
n n e^{+}$, for example). Such determinations make use of a careful analysis
of the finite volume spectra of these systems which is modified by the
various interactions \cite{Luscher:1986pf,Luscher:1990ux,Lellouch:2000pv}. Recent
theoretical work has also focused on two-body systems with multiple
interaction channels
\cite{Hansen:2012tf,Briceno:2014oea,Briceno:2012yi,Wu:2014vma} and on
three-body interactions
\cite{Beane:2007qr,Detmold:2008gh,Polejaeva:2012ut,Briceno:2012rv,Hansen:2013dla}.
However, more complex interactions are currently beyond our ability to
investigate as the requisite formalism is not known.

\section{Discussion}
\label{sec:discussion}
The central conclusion of this work is that SU($N_c=2$) gauge theory
with $N_f=2$ degenerate flavours of quarks in the fundamental
representation exhibits a complex spectrum that includes bound
``nuclei'', states with baryon number, $B \geq2$. In combination with
the real-world nuclei that we observe in nature, and with studies of QCD at
heavier-than-physical quark masses, where deeply bound light nuclei are observed
\cite{Beane:2012vq,Yamazaki:2012hi,Beane:2013kca}, this leads to a
fascinating glimpse of ``nuclei'' in a more general context, and may
point to the pervasive nature of nuclei in strongly interacting
theories.\footnote{There have also been significant attempts to
  understand baryon-baryon
  \cite{Witten:1979kh,Kaplan:1995yg,Kaplan:1996rk,Banerjee:2001js,Belitsky:2002ni,
    Cohen:2002im,Bonanno:2011yr} and three-baryon
  \cite{Phillips:2013rsa} interactions in the large $N_{c}$
  limit. However, the existence of bound states arising from these
  interactions is not clear -- see Ref.~\cite{Beane:2002ab} for
  insightful discussions.}  Such a conclusion would have interesting
consequences for understanding how usual or unusual QCD is in the
space of similar theories. Also, if complex bound states are indeed
ubiquitous, then it is important to consider their contributions in
other strongly interacting dark matter scenarios.  For $N_c=2$ QCD as
a possible dark sector candidate, the existence of nuclei leads to a
range of interesting and novel phenomenology that we explore in a
companion paper \cite{Detmold:2014qqa}.

In the context of real world QCD ($N_c=3$), there is currently an
intense focus on investigating light nuclei from first principles,
both to understand how nuclei emerge from the underlying quark and
gluon degrees of freedom, and also to be able to make reliable
predictions for nuclear matrix elements of electroweak and other
currents that are important for a range of ongoing and future
experiments.  Performing a study analogous to the one presented here
for more complex theories such as SU($N_c=4$), while interesting, is
prohibitively expensive at the present time. As well as the naive
scaling of the cost with the number of degrees of freedom, SU($N_c=2$)
is special as there are nuclear states for which contractions can be
performed straightforwardly. The SU($N_c=4$) case, real QCD ($N_c=3$),
and most other theories require much larger resources in order to
study nuclei as they suffer from exponential signal-to-noise
degradation and the complexity of the requisite contractions for
multi-baryon systems \cite{Doi:2012xd,Detmold:2012eu,Gunther:2013xj}
presents a significant challenge.

\acknowledgments{We thank Stefan Meinel, Martin Savage and Jesse
  Thaler for discussions. MM is supported by a Simons Postdoctoral
  Fellowship, WD by a US Department of Energy Early Career Research
  Award DE-SC0010495 and the Solomon Buchsbaum Fund at MIT and AVP by
  Department of Energy grant DE-FG02-94ER40818.}

\appendix
% \section{Parameter values}
% \label{apptab}

\section{Extracted energies}
\label{fitresults}
\begin{table}[ht]
  \begin{center}
    \begin{ruledtabular}
      \begin{tabular}{llccc}
      	\multicolumn{2}{c}{State}&\multicolumn{3}{c}{Volume} \\
        $B$ & $J$ & $12^3 \times 72$ & $16^3 \times 72$ & $20^3 \times 72$ \\
        \hline
        2 & 0 & 0.007(1) & 0.0030(9) & 0.001(1) \\
        3 & 0 & 0.020(3) & 0.010(2) & 0.005(3) \\
        4 & 0 & 0.050(7) & 0.020(5) & 0.010(7) \\
        5 & 0 & 0.09(1) & 0.04(1) & 0.02(1) \\
        6 & 0 & 0.20(1) & 0.06(1) & 0.04(1) \\
        7 & 0 & 0.20(2) & 0.09(3) & 0.06(2) \\
        \hline
        2 & 1 & $-$0.010(3) & $-$0.008(3) & $-$0.006(3) \\
        3 & 1 & $-$0.002(8) & $-$0.007(5) & $-$0.006(8) \\
        4 & 1 & 0.03(1) & 0.00(1) & $-$0.00(1) \\
        5 & 1 & 0.07(1) & 0.02(1) & 0.01(2) \\
        6 & 1 & 0.10(2) & 0.05(3) & 0.03(2) \\
        7 & 1 & 0.20(3) & 0.08(5) & 0.06(3) \\
      \end{tabular}
    \end{ruledtabular}
    \caption{ The fitted energy shifts (in lattice units) on each
      volume for ensemble $A$.  }
    \label{Arestab}
  \end{center}
\end{table}     

\begin{table}[ht]
  \begin{center}
    \begin{ruledtabular}
      \begin{tabular}{llcccc}
      	\multicolumn{2}{c}{State}&\multicolumn{4}{c}{Volume} \\
        $B$ & $J$ & $12^3 \times 48$ & $16^3 \times 48$ & $20^3 \times 48$ & $24^3 \times 48$ \\
        \hline
        2 & 0 & 0.030(5) & 0.010(3) & 0.003(2) & 0.006(1) \\
        3 & 0 & 0.10(1) & 0.040(7) & 0.020(6) & 0.020(4) \\
        4 & 0 & 0.20(3) & 0.10(1) & 0.05(1) & 0.050(7) \\
        5 & 0 & 0.40(5) & 0.20(2) & 0.10(1) & 0.09(1) \\
        6 & 0 & 0.70(6) & 0.30(2) & 0.20(3) & 0.10(1) \\
        7 & 0 & 1.00(6) & 0.50(3) & 0.30(5) & 0.20(3) \\
        \hline
        2 & 1 & $-$0.04(1) & $-$0.020(7) & $-$0.010(5) & $-$0.009(4) \\
        3 & 1 & $-$0.01(2) & $-$0.02(1) & $-$0.01(1) & $-$0.004(7) \\
        4 & 1 & 0.09(5) & 0.03(1) & 0.01(1) & 0.02(1) \\
        5 & 1 & 0.30(6) & 0.10(2) & 0.06(2) & 0.05(1) \\
        6 & 1 & 0.60(8) & 0.20(3) & 0.10(3) & 0.10(2) \\
        7 & 1 & 0.8(1) & 0.40(5) & 0.20(4) & 0.20(4) \\
      \end{tabular}
    \end{ruledtabular}
    \caption{ The fitted energy shifts (in lattice units) on each
      volume for ensemble $B$.  }
    \label{Brestab}
  \end{center}
\end{table}     

\begin{table}[ht]
  \begin{center}
    \begin{ruledtabular}
      \begin{tabular}{llccc}
      	\multicolumn{2}{c}{State}&\multicolumn{3}{c}{Volume} \\
        $B$ & $J$ & $12^3 \times 48$ & $16^3 \times 48$ & $20^3 \times 48$ \\
        \hline
        2 & 0 & 0.007(1) & 0.003(1) & 0.002(1) \\
        3 & 0 & 0.030(6) & 0.010(3) & 0.007(3) \\
        4 & 0 & 0.07(1) & 0.030(6) & 0.020(6) \\
        5 & 0 & 0.10(1) & 0.06(1) & 0.04(1) \\
        6 & 0 & 0.30(2) & 0.10(1) & 0.06(1) \\
        7 & 0 & 0.40(3) & 0.20(2) & 0.09(1) \\
        \hline
        2 & 1 & $-$0.030(3) & $-$0.020(2) & $-$0.010(1) \\
        3 & 1 & $-$0.030(8) & $-$0.020(5) & $-$0.010(4) \\
        4 & 1 & $-$0.01(1) & $-$0.01(1) & $-$0.010(8) \\
        5 & 1 & 0.07(2) & 0.01(1) & 0.00(1) \\
        6 & 1 & 0.20(3) & 0.05(2) & 0.02(1) \\
        7 & 1 & 0.30(4) & 0.10(3) & 0.05(2) \\
      \end{tabular}
    \end{ruledtabular}
    \caption{ The fitted energy shifts (in lattice units) on each
      volume for ensemble $C$.  }
    \label{Crestab}
  \end{center}
\end{table}     

\begin{table}[ht]
  \begin{center}
    \begin{ruledtabular}
      \begin{tabular}{llccccc}
      	\multicolumn{2}{c}{State}&\multicolumn{5}{c}{Volume} \\
        $B$ & $J$ & $12^3 \times 48$ & $16^3 \times 48$ & $16^3 \times 72$ & $20^3 \times 48$ & $24^3 \times 48$ \\
        \hline
        2 & 0 & 0.001(1) & 0.001(1) & 0.001(1) & $-$0.0005(9) & $-$0.001(1) \\
        3 & 0 & 0.004(5) & 0.003(5) & 0.005(3) & $-$0.000(3) & $-$0.000(3) \\
        4 & 0 & 0.01(1) & 0.010(9) & 0.010(5) & 0.002(6) & 0.004(7) \\
        5 & 0 & 0.04(2) & 0.03(1) & 0.030(8) & 0.01(1) & 0.01(1) \\
        6 & 0 & 0.08(2) & 0.05(1) & 0.05(1) & 0.02(1) & 0.03(1) \\
        7 & 0 & 0.10(3) & 0.09(2) & 0.08(1) & 0.04(2) & 0.06(2) \\
        \hline
        2 & 1 & $-$0.010(1) & $-$0.010(2) & $-$0.010(2) & $-$0.010(2) & $-$0.010(1) \\
        3 & 1 & $-$0.020(6) & $-$0.020(6) & $-$0.010(5) & $-$0.020(5) & $-$0.020(4) \\
        4 & 1 & $-$0.02(1) & $-$0.01(1) & $-$0.009(8) & $-$0.020(9) & $-$0.020(8) \\
        5 & 1 & 0.00(2) & 0.00(1) & 0.01(1) & $-$0.01(1) & $-$0.01(1) \\
        6 & 1 & 0.05(3) & 0.03(1) & 0.03(1) & 0.00(1) & 0.01(1) \\
        7 & 1 & 0.10(4) & 0.07(2) & 0.06(2) & 0.02(2) & 0.03(2) \\
      \end{tabular}
    \end{ruledtabular}
    \caption{ The fitted energy shifts (in lattice units) on each
      volume for ensemble $D$.  }
    \label{Drestab}
  \end{center}
\end{table}     

\begin{table}[ht]
  \begin{center}
    \begin{ruledtabular}
      \begin{tabular}{llccc}
      	\multicolumn{2}{c}{State}&\multicolumn{3}{c}{Volume} \\
        $B$ & $J$ & $12^3 \times 72$ & $16^3 \times 72$ & $20^3 \times 72$ \\
        \hline
        2 & 0 & 0.007(1) & 0.002(1) & 0.000(1) \\
        3 & 0 & 0.030(6) & 0.009(3) & 0.004(4) \\
        4 & 0 & 0.06(1) & 0.020(6) & 0.010(7) \\
        5 & 0 & 0.10(1) & 0.050(9) & 0.03(1) \\
        6 & 0 & 0.20(2) & 0.08(1) & 0.05(1) \\
        7 & 0 & 0.40(2) & 0.10(1) & 0.07(3) \\
        \hline
        2 & 1 & $-$0.020(3) & $-$0.010(1) & $-$0.009(2) \\
        3 & 1 & $-$0.010(8) & $-$0.010(4) & $-$0.010(5) \\
        4 & 1 & 0.01(1) & $-$0.005(8) & $-$0.008(9) \\
        5 & 1 & 0.09(1) & 0.02(1) & 0.01(1) \\
        6 & 1 & 0.20(2) & 0.06(1) & 0.03(2) \\
        7 & 1 & 0.30(2) & 0.10(1) & 0.05(3) \\
      \end{tabular}
    \end{ruledtabular}
    \caption{ The fitted energy shifts (in lattice units) on each
      volume for ensemble $E$.  }
    \label{Erestab}
  \end{center}
\end{table}

\begin{table}[ht]
  \begin{center}
    \begin{ruledtabular}
      \begin{tabular}{llcccc}
      	\multicolumn{2}{c}{State}&\multicolumn{4}{c}{Volume} \\
        $B$ & $J$ & $12^3 \times 72$ & $16^3 \times 72$ & $20^3 \times 72$ & $24^3 \times 72$ \\
        \hline
        2 & 0 & $-$0.002(2) & $-$0.004(1) & $-$0.004(1) & $-$0.001(2) \\
        3 & 0 & $-$0.003(6) & $-$0.010(4) & $-$0.010(4) & $-$0.004(5) \\
        4 & 0 & 0.01(1) & $-$0.010(9) & $-$0.01(1) & $-$0.01(1) \\
        5 & 0 & 0.07(1) & 0.00(1) & $-$0.01(2) & $-$0.01(1) \\
        6 & 0 & 0.20(1) & 0.03(2) & $-$0.00(3) & $-$0.00(3) \\
        7 & 0 & 0.30(2) & 0.07(2) & 0.01(5) & 0.01(5) \\
        \hline
        2 & 1 & $-$0.020(2) & $-$0.010(1) & $-$0.010(1) & $-$0.008(2) \\
        3 & 1 & $-$0.030(7) & $-$0.030(5) & $-$0.020(5) & $-$0.020(6) \\
        4 & 1 & $-$0.03(1) & $-$0.03(1) & $-$0.03(1) & $-$0.03(1) \\
        5 & 1 & 0.04(1) & $-$0.01(1) & $-$0.03(2) & $-$0.03(2) \\
        6 & 1 & 0.10(2) & 0.01(2) & $-$0.01(4) & $-$0.02(3) \\
        7 & 1 & 0.20(2) & 0.06(2) & $-$0.00(6) & $-$0.01(5) \\
      \end{tabular}
    \end{ruledtabular}
    \caption{ The fitted energy shifts (in lattice units) on each
      volume for ensemble $F$.  }
    \label{Frestab}
  \end{center}
\end{table}

\bibliography{DarkNucLQCD}

\end{document}